# The representation of physical motions
# by various types of quaternions

D. H. Delphenich [†]

**Abstract:** It is shown that the groups of Euclidian rotations, rigid motions, proper, orthochronous Lorentz transformations, and the complex rigid motions can be represented by the groups of unit-norm elements in the algebras of real, dual, complex, and complex dual quaternions, respectively. It is shown how someof the physically-useful tensors and spinors can be represented by the various kinds of quaternions. The basic notions of kinematical states are described in each case, except complex dual quaternions, where their possible role in describing the symmetries of the Maxwell equations is discussed.

[†]     E-mail: david_delphenich@yahoo.com, Website: neo-classical-physics.info.

# CONTENTS

## V.  COMPLEX DUAL QUATERNIONS

# INTRODUCTION

According to Wilhelm Blaschke [**1**], the first vestiges of the concept of a quaternion went to back to the work of Leonhard Euler on rigid-body rotations in 1776 [**2**]. Olinde Rodrigues seems to have developed much of the basics of the subject in 1840, without actually introducing the term "quaternion." However, the first major attempt to develop the concept in its own right, along with its applications to physics, seems to have the posthumously-published book of Sir William Rowan Hamilton that he called *Elements of Quaternions* [**3**] that came almost a century after Euler in 1866, although the work was done starting in 1843.

Hamilton envisioned quaternions as a geometric algebra that would extend the three-dimensional algebra of the vector cross product to a four-dimensional algebra in which the Euclidian scalar product also played a role in defining the product of its elements. Thus, the algebra would be closely related to rotations by the fact that the vector cross product defines a Lie algebra on $\mathbb{R}^3$ that is isomorphic to $\mathfrak{so}(3; \mathbb{R})$, which consists of infinitesimal generators of Euclidian rotations, as well as the fact that rotations preserve the Euclidian scalar product.

In that monumental treatise, the coefficients of quaternions were treated as if they could be either real or complex, as it suited the purpose, although Hamilton suggested that the complex kind might be referred to as "biquaternions." However, since the work predated the theory of relativity, no mention was made of the role that the complex quaternions might play in regard to relativistic motions.

Apparently, the subsequent history of the theory of quaternions was somewhat marginal to the mainstream of both mathematics and physics for quite some time. However, from 1899 to 1913 there was an international Quaternion Society that formed around Alexander MacFarlane that was dedicated to keeping the concept alive. For the most part, their main acceptance in the early years seems to have been with the algebraists (cf., e.g., Shaw [**4**]), who saw them as a useful example of an algebra. However, later researchers in physics and mechanical engineering continued to expand upon the usefulness of quaternions in applications that involved the representation of three-dimensional Euclidian rotations.

For instance, to this day, the inertial navigation community regards the representation of kinematics by real quaternions as being their most computationally efficient algorithm for the propagation of rotating frames. Although the method of Euler angles requires one less coordinate to describe a rotation, nevertheless, the differential equations that one must integrate involve products of trigonometric functions, which tends to slow down the computational speed considerably. Although the method of direction cosine matrices only involves basic arithmetic operations on the components of the matrices, nonetheless, they are matrices with nine components, which also tends to slow things down. Thus, the four components of a unit quaternion seem to provide an ideal compromise.

In a different part of geometry and mechanics, geometers such as Louis Poinsot [**5**] and Michel Chasles [**6**] were exploring the geometry of the larger group of rigid motions, which also includes translations, in addition to the rotations; both two and three dimensional rigid motions were dealt with in detail. The geometry that they were considering was actually projective geometry, not affine Euclidian geometry. Chasles found that the concept of an infinitesimal center of rotation for a planar rigid motion could be extended to a "central axis" for any three-dimensional rigid motion, which allowed one to decompose the motion into a rotation about the axis and a translation along it. Previously, in the context of statics, Poinsot found that a finite set of spatially-distributed force vectors that acted on a rigid body could be replaced with an equivalent force-moment about a central axis and a force along it. The German geometers Julius Plücker [**7**] and his illustrious student Felix Klein [**8**] expanded upon the projective-geometric nature of these constructions, and introduced the term "Dyname" for a finite spatial distribution of force vectors. (The French were using the term "torseur," which eventually became the more modern term "torsor.") In the meantime, Sir Robert Ball [**9**] had introduced the term "screw" to describe the canonical form of the rigid motion and "wrench" to describe the canonical form of a force distribution. To some extent, the German school of geometrical kinematics culminated in the 1903 treatise [**10**] of another student of Plücker named Eduard Study that was entitled *Die Geometrie der Dynamen*.

One of the innovations that Study introduced in that work was the algebra of "dual numbers," which he applied to study of quaternions to produce "dual quaternions." Actually, William Kingdon Clifford had previously sketched out a theory [**11**] of what he was calling "biquaternions," although his usage of that word was inconsistent with that of Hamilton, since Clifford's biquaternions were quaternions whose components were dual numbers, while Hamilton's biquaternions had complex components. However, the treatise of Study developed the concepts in much more detail than that of Clifford. Since dual quaternions represent an algebra over $\mathbb{R}^8$, some authors (e.g., MacAulay [**12**]) referred to them as "octonions." Unfortunately, that usage is inconsistent with the modern usage of that term to refer to Cayley numbers, which defines a division algebra over $\mathbb{R}^8$, unlike dual quaternions, which have divisors of zero, and therefore cannot be a division algebra.

The dual numbers represent an algebra over $\mathbb{R}^2$, just as the complex numbers do, but the difference is that the basic object of the dual number algebra is a symbol $\varepsilon$ that is nilpotent – viz., $\varepsilon^2 = 0$ – while the basic object of the complex algebra is a symbol $i$ with the property that $i^2 = -1$. The effect of introducing $\varepsilon$ is to produce an algebra that is not a division algebra – in particular, it has divisors of zero – with the property that the product of two dual numbers $a + \varepsilon b$ and $c + \varepsilon d$ is $ac + \varepsilon(ad + bc)$, which then combines the usual multiplication of real numbers with their addition. Although it is not obvious at this stage of the discussion, when one uses such numbers for the components of quaternions, it allows on to represent three-dimensional rigid motions by means of the group of unit dual quaternions, just as the unit quaternions carry a representation of the group of three-dimensional Euclidian rotations; in fact, it is the spin representation that is isomorphic to $SU(2)$.

Apparently, the main acceptance of the methods of dual quaternions was by the mechanical engineering community. One of the earliest applications of screws and dual quaternions to mechanics was by the Russian Zanichevskiy in 1889 [**13**], although no copies of that paper remain, as it was destroyed by the Bolsheviks during the revolution of 1917, along with some later work of Kotjelnikoff [**14**]. Richard von Mises developed the application to mechanics in a pair of papers in 1924 [**15**] that were widely cited to this day. Another classic of the Russian school was by Dimentberg [**16**], which also contained a bibliography of Russian work that was not widely known outside of the Soviet Union. The general fields of application of dual quaternions to mechanics that is being discussed today seem to be in the theory of mechanisms [**17–19**], and especially robot manipulators [**20-22**]. Just as real quaternions give a computationally-efficient algorithm for dealing with rigid-body rotations in real time, dual quaternions give a computationally-efficient algorithm for dealing with rigid-body motions that also include translations, such as the motion of joints in manipulators.

Nonetheless, in the physics community itself, one finds the observation of Herbert Goldstein in a footnote to the 1980 edition [**23**] of his standard textbook *Classical Mechanics* in the context of rigid-body kinematics:

> "…Such a combination of translation and rotation is called a *screw motion*. There seems to be little present use for this version of Chasles' theorem, nor for the elaborate mathematics of screw motions as developed in the nineteenth century…"

Meanwhile, in a different part of the physics community, the complex quaternions that Hamilton had only alluded to were being applied to the emerging physics of special relativity. One of the earliest researchers to develop that application was the Polish physicist Ludwik Silberstein [**24**], who showed how the Lorentz transformations could be represented by the action of unit complex quaternions on the quaternions. Because the complex quaternions, like the dual quaternions, are also an algebra – but not a division algebra – over $\mathbb{R}^8$, and they admit more automorphisms than the real quaternions, due to the possibility of complex conjugation, they also admit more ways of defining the action of unit complex quaternions on the quaternions, and more types of invariant subspaces for the actions, which correspond to the different types of tensors that one can represent by quaternions.

As for the acceptance of the methods of complex quaternions into relativity theory, one should note that according to Silberstein, in the cited reference, Minkowski felt that quaternions were "too narrow and clumsy for the purpose." However, various papers on the subject of quaternions and relativity followed, just the same. (See, e.g., Weiss [**25**] and Rastall [**26**].)

It is the fact that the representation of various tensors (scalar, vector, bivector, spinor) by complex quaternions corresponds, not to differing numbers of component indices, but to differing *actions* that makes these representations somewhat more esoteric than the more conventional tensor representations, but it is in the fact that one can represent different types of tensors by the same basic algebra that one finds the power of the methodology. In particular, various attempts were made – notably, by Lanczos [**27**] – to apply the methods of complex quaternions to the modeling of both the Maxwell equations for electromagnetism and the Dirac equation for the wave function of the electron. The big problem that emerged was the unification problem of finding a field (or wave) equation that would include both Maxwell and Dirac as special cases.

This unification problem was closely related to the Einstein-Maxwell problem, which concerned finding a field equation that would imply both Maxwell's equations of electromagnetism and Einstein's equations of gravitation as consequences. One of the many attempts [**28**] that Einstein, together with Mayer, made along those lines involved the use of what he was calling "semi-vectors," which were later showed by Blaton [**29**] to be a slight generalization of the spinors that were gradually being introduced into quantum physics due to the discovery of the magnetic moment of the electron and the Uhlenbeck-Goudsmit hypothesis that it was due to some form of intrinsic angular momentum.

One of the intriguing aspects of the Lanczos equations was that they involved an even higher-dimensional field space than the complex quaternions, namely, since they seemed to involve pairs of complex quaternions, one might think of the wave function as taking its values in an algebra of real dimension sixteen. Interestingly, both Albert Proca [**30**] and Sir Arthur Stanley Eddington [**31**] were suggesting that the correct form of the quantum wave function should take its values in the full sixteen-dimensional Clifford algebra of Minkowski space, and not just $\mathbb{C}^4$. However, as is usually the case, the Lanczos equation did not attract widespread attention, mostly due to the problem of the physical interpretation of the pairs of complex quaternions.

One suggestion that is worthy of consideration was made more recently by Gsponer and Hurni [**32**], who conjectured that perhaps the pairing of wave functions is simply due to the more modern notion of isospin symmetry, such as one finds in the proton-neutron doublet or the electron-neutrino doublet. Certainly, such a concept was not known to Lanczos at the time of his paper in 1927, so it would not have been considered back then.

Another sixteen-real-dimensional algebra that has a natural place in physical mechanics is the algebra of complex dual quaternions, which have the same multiplication table for the basis elements as real quaternions, but their components come from the four-real-dimensional algebra of complex dual numbers. These numbers look essentially the same as real dual numbers, except that the components are complex numbers. Thus, the complex dual quaternions can be regarded as a complexification of the real dual quaternions, so they have some of the features of both the dual and the complex numbers. As a complexification of the real dual quaternions, the group of unit complex dual quaternions naturally carries a representation of the group $ISO(3; \mathbb{C})$ of complex rigid motions, which are then the semi-direct product of the three-dimensional complex Euclidian rotation group with the three-dimensional complex translation group. Although the complex rotation group is actually isomorphic to the identity component of the Lorentz group – i.e., the proper, orthochronous Lorentz group – and the translation group $\mathbb{C}^3$ includes $\mathbb{R}^4$ as a subgroup, nonetheless, the nature of the semi-direct product does not permit one to find a Poincaré subgroup of $ISO(3; \mathbb{C})$. However, this does not mean that there are no physically useful applications of the group, since it acts quite naturally on $\mathbb{C}^3$, which can be used as a model for the field spaces of electromagnetism, namely, the spaces of bivectors and 2-forms over $\mathbb{R}^4$.

The representation of electromagnetic fields as fields with values in $\mathbb{C}^3$, such as in the form $\mathbf{E} + i\mathbf{B}$, goes back at least as far as Riemann's lectures on partial differential equations (see Weber [**33**]), although it only got a brief passing mention at the time.  It was next discussed by Silberstein [**34a, b**] in 1907 and independently by Arthur Conway [**35**] in 1911.  Later, it was notably employed by Ettore Majorana in order to put Maxwell's and Dirac's equations into a common formalism, although that work took the form of notes that were not published until much later in a compilation volume [**36**].  Independently of him, J. Robert Oppenheimer [**37**] also employed the complex representation in order to discuss the problem of finding a wave function for the photon that would be analogous to the one that the Dirac equation gives for the electron.  Interestingly, that problem is still open, since the statistical interpretation of the wave function assumes that one can localize it to a point particle, which is impossible for the photon, although one can still speak of a momentum-space wave function for it.  The complex representation had also been developed in the context of general relativity theory, as well, and some researchers referred to it as the method of "3-spinors"  (See, e.g., [**38, 39**].)

Although the main application of quaternions in this monograph is to kinematics, nevertheless, for the sake of completeness, we shall also discuss complex dual quaternions, which have not been the subject of as much discussion in physics as real, dual, and complex quaternions.  As we see it, the main application of complex dual quaternions to the symmetries of electromagnetic field equations, so the kinematics of complex rigid frames will not be pursued at the moment.  Since the complex dual quaternions provide a sixteen-real-dimensional field space, that fact might explain the Lanczos equations in a physically reasonable way as an alternative to isospin doublets.

The basic structure of this monograph is straightforward:

In chapter I, we define the general notion of an algebra and discuss some of the elementary notions concerning them that will be applied in the later chapters.  The fact that an algebra is a special type of ring is emphasized, so some of the material is general to rings, while other material is specific to algebras.  In particular, the notion of the tensor product of algebras will be crucial to the discussion of dual, complex, and complex dual quaternions.

In chapter II we introduce the real quaternions, which are at the root of all of the other variants that we will subsequently discuss.  In particular, we show that the group of unit real quaternions is a Lie group that is isomorphic to $SU(2)$ by a straightforward association of quaternions with 2×2 complex matrices.  We then show how the unit real quaternions act on the three-dimensional vector subspace of pure – or "vector" – quaternions in a manner that represents the action of $SO(3; \mathbb{R})$ on three-dimensional real Euclidian space, although diametral pairs of unit quaternions get associated with the same proper rotation in the manner of the spin representation of the orthogonal group.  We then show how one derives the basic kinematical objects from such an action when the unit quaternion varies differentiable in time.  In particular, we discuss both velocity and acceleration for vectors and frames and show the forms that take in both inertial and co-moving frames.

In chapter III, we introduce the algebra of real dual numbers and show how one does some of the usual trigonometry and linear algebra using them instead of real numbers.  We then define the algebra of dual quaternions to be the tensor product of the real quaternions with the dual numbers and go over the same basic topics that we did in the chapter on real quaternions. We then see that the unit dual quaternions carry a representation of the Lie group of rigid motions in three-dimensional real Euclidian space and examine how that group acts on dual vectors, which are the analogue of pure quaternions in this case. We then show how many of the kinematical expressions that were derived before are obtained essentially by changing the ring of coefficients for the vectors from real numbers to dual numbers.

In chapter IV, we discuss complex quaternions. The basic flow of ideas is the same as in the previous two chapters, although we will see that there are more automorphisms that one can introduce in the complex case, which define more actions of the unit complex quaternions on the complex quaternions and more invariant subspaces. Since that group is easily seen to be isomorphic to $SL(2; \mathbb{C})$, one sees that the complexification of real rotations gives transformations that have fundamental significance in special relativity, as does the group $SO(3; \mathbb{C})$. Furthermore, one finds that that when the scalar are complex one can have non-trivial "null quaternions," which are related to the isotropic vectors of Minkowski space. They also make it possible to find actions of the group of unit complex quaternions whose invariant subspaces could represent $SL(2; \mathbb{C})$ spinors. We then examine the kinematics of frames in the various invariant subspaces.

Finally, in chapter V, we attempt follow the same basic template for exploring the algebra of complex dual quaternions and examine what issues become relevant when one complexifies the algebra of real dual quaternions, or similarly, "dualizes" the algebra of complex quaternions. We conclude by suggesting that the most immediate application of complex dual quaternions is to the symmetries of the Maxwell equations, which then becomes a matter for a later study of the role of quaternions in physical field theories.

A certain familiarity with the basic definitions of differentiable manifolds and Lie groups will be assumed in what follows, such as one might learn from Frenkel [**40**]. However, most of the discussion is algebraic in character, and assumes just the rudiments of linear algebra, while concepts from abstract algebra, such as rings, modules, and fields, will be introduced as necessary for the benefit of physicists who are not familiar with them.

# CHAPTER I

# ALGEBRAS.

**1. General notions [1-3].** If $V$ is an $n$-dimensional vector space whose scalars come from a field $\mathbb{K}$, which will always be either the field $\mathbb{R}$ of real numbers or the field $\mathbb{C}$ of complex numbers in what follows, then an *algebra* over $V$ is defined by a $\mathbb{K}$-bilinear map $V{\times}V \to V$, $(a, b) \mapsto ab$ that one regards as a multiplication of vectors to produce another vector. Thus, for any elements $a$, $b$, $c \in V$ and any scalar $\lambda \in \mathbb{K}$, one must have:

$$(a + b)c = ac + bc, \qquad a(b + c) = ab + ac, \qquad (\lambda a)b = a(\lambda b) = \lambda(ab).$$

Since the vector space $V$ has an Abelian group structure that is defined by vector addition and the first two conditions that were just stated represent right and left distributivity, any algebra can be regarded as a *ring* algebraically (cf., e.g., Jacobson [**4**], which also includes a chapter on algebras.). However, the last set of conditions specializes the definition to vector spaces.

The multiplication that defines an algebra does not have to be associative or commutative, not does it have to admit a unity element or multiplicative inverses. Indeed, it might admit *divisors* of zero, which would be non-zero elements $a$ and $b$ such that:

$$ab = 0.$$

If $\{\mathbf{e}_i, i = 1, \ldots, n\}$ is a basis for the vector space that underlies an algebra $A$ then one can obtain all of the important structure of the algebra from the multiplication table for the basis elements, since any other elements are linear combinations of them and the product is assumed to be bilinear. One can then summarize the multiplication table for the basis elements in the form of a set of linear equations that express the various products as linear combination of the basis elements again:

$$\mathbf{e}_i \, \mathbf{e}_j = a_{ij}^k \mathbf{e}_k \,, \tag{1.1}$$

in which the component array $a_{ij}^k$ is referred to as the set of *structure constants* for the algebra in that basis. These constants can also be regarded as the components of a third-rank tensor of mixed type over $A$, since the algebra multiplication is a bilinear map from $A{\times}A$ to $A$, and thus defines an element of $A^* \otimes A^* \otimes A$.

There is an essential difference between defining a basis for the vector space that underlies an algebra $A$ and defining a set of *generators* for the algebra. A set $S \subset A$ consists of generators for $A$ if every element of $A$ can be expressed as a linear combination of *products* of elements of $S$. Thus, although any basis will generate the algebra, often, as we shall see, a subset of a basis might generate the other basis elements by way of products of the basis elements.

If $v = v^i \, \mathbf{e}_i$ and $w = w^i \, \mathbf{e}_i$ are arbitrary elements of $A$ then one sees that their product $vw$ has components with respect to $\mathbf{e}_i$ that can be obtained from:

$$(ab)^k = a_{ij}^k v^i w^j. \qquad (1.2)$$

An especially important class of non-associative algebras is given by *Lie algebras*, for which the multiplication of $a$ and $b$ is written $[a, b]$. The Lie bracket is then required to be anti-symmetric [1]) and satisfy the Jacobi identity:

$$[a, b] = -[b, a], \qquad [a, [b, c]] + [b, [c, a]] + [c, [a, b]] = 0.$$

The Jacobi identity can be regarded as a measure of the non-associativity of the Lie bracket. However, if one has an associative algebra over a vector space $V$ then one can define a Lie algebra over $V$ by the commutator bracket:

$$[a, b] = ab - ba. \qquad (1.3)$$

The associativity of the product is necessary in order to make the bracket satisfy the Jacobi identity.

A particular example of a Lie algebra that will recur in what follows is the algebra over $\mathbb{R}^3$ that is defined by the vector cross product [2]):

$$[\mathbf{a}, \mathbf{b}] = \mathbf{a} \times \mathbf{b} = (\varepsilon_{ijk} \, a^j \, b^k) \, \mathbf{e}_i, \qquad (1.4)$$

in which $\{\mathbf{e}_1, \mathbf{e}_2, \mathbf{e}_3\}$ constitutes the canonical basis $\{(1, 0, 0), (0, 1, 0), (0, 0, 1)\}$ for $\mathbb{R}^3$, while $a^i$ and $b^i$ are the components of $\mathbf{a}$ and $\mathbf{b}$ with respect to that basis. Thus, the Levi-Cività symbol $\varepsilon_{ijk}$ also gives one the structure constants for that Lie algebra.

This Lie algebra is isomorphic to the Lie algebra $\mathfrak{so}(3; \mathbb{R})$ of infinitesimal three-dimensional Euclidian rotations. We regard the latter as defined by anti-symmetric real 3×3 matrices, when given the commutator bracket. A useful basis for the Lie algebra $\mathfrak{so}(3; \mathbb{R})$ is defined by the elementary infinitesimal rotation matrices:

$$J_1 = \begin{bmatrix} 0 & 0 & 0 \\ 0 & 0 & -1 \\ 0 & 1 & 0 \end{bmatrix}, \qquad J_2 = \begin{bmatrix} 0 & 0 & 1 \\ 0 & 0 & 0 \\ -1 & 0 & 0 \end{bmatrix}, \qquad J_3 = \begin{bmatrix} 0 & -1 & 0 \\ 1 & 0 & 0 \\ 0 & 0 & 0 \end{bmatrix}, \qquad (1.5)$$

for infinitesimal rotations around the $x$, $y$, and $z$ axes, respectively.

---

[1])   For some authors, if one wishes to admit vector spaces over $\mathbb{Z}_2$, the anti-symmetry is replaced by the requirement that $[a, a] = 0$ in any case. We shall not, however, need such generality for our purposes.

[2])   We shall adhere to the notational convention that lower-case Latin indices always range from 1 to 3 and lower-case Greek indices range from 0 to 3.

The isomorphism of $(\mathbb{R}^3, \times)$ with $\mathfrak{so}(3; \mathbb{R})$ is then defined by the *adjoint representation*, ad: $(\mathbb{R}^3, \times) \to \mathfrak{so}(3; \mathbb{R})$, $\mathbf{a} \mapsto [\mathrm{ad}(\mathbf{a})]$, where:

$$\mathrm{ad}(\mathbf{a})\,\mathbf{b} = \mathbf{a} \times \mathbf{b} \qquad (1.6)$$

and $[\mathrm{ad}(\mathbf{a})]$ is the matrix of the linear map $\mathrm{ad}(\mathbf{a})$: $\mathbb{R}^3 \to \mathbb{R}^3$ with respect to some basis. If that basis is the canonical basis then one finds that:

$$J_i = [\mathrm{ad}(\mathbf{e}_i)], \quad (i = 1, 2, 3). \qquad (1.7)$$

One can then say that:

$$\mathrm{ad}(\mathbf{a}) = a^i\,J_i = \begin{bmatrix} 0 & -a^3 & a^2 \\ a^3 & 0 & -a^1 \\ a^3 & a^1 & 0 \end{bmatrix}. \qquad (1.8)$$

Interestingly, when one defines the vector product of complex 3-vectors in $\mathbb{C}^3$, the resulting Lie algebra $\mathfrak{so}(3; \mathbb{C})$ is isomorphic to the Lie algebra $\mathfrak{so}(1, 3)$ of infinitesimal Lorentz transformations of Minkowski space, which, for us, will be $\mathbb{R}^4$ given the scalar product of signature type $(+1, -1, -1, -1)$. These infinitesimal Lorentz transformations will be defined by real 4×4 matrices $l$ with property that:

$$\eta l + l\eta = 0,$$

in which $\eta = \mathrm{diag}[+1, -1, -1, -1]$ is the matrix of the scalar product.

If the basis for the complex vector space $\mathbb{C}^3$ is given by the canonical basis then one can give a basis for the complex Lie algebra $\mathfrak{so}(3; \mathbb{C})$ by way of $J_i = [\mathrm{ad}(\mathbf{e}_i)]$, as before. If one regards $\mathfrak{so}(3; \mathbb{C})$ as a real Lie algebra then can give a basis for it by way of $J_i$ and $K_i = iJ_i$, and one sees that, from the $\mathbb{C}$-bilinearity of the Lie bracket, the commutation rules for the basis are:

$$[J_i, J_j] = \varepsilon_{ijk}\,J_k, \qquad [J_i, K_j] = \varepsilon_{ijk}\,K_k, \qquad [K_i, K_j] = -\,\varepsilon_{ijk}\,J_k, \qquad (1.9)$$

which are then isomorphic to those of $\mathfrak{so}(1, 3)$, when one replaces $J_i$ and $K_i$ with the real 4×4 matrices:

$$\hat{J}_i = \begin{bmatrix} 0 & 0 \\ 0 & J_i \end{bmatrix}, \quad \hat{K}_1 = \begin{bmatrix} 0 & 1 & 0 & 0 \\ 1 & 0 & 0 & 0 \\ 0 & 0 & 0 & 0 \\ 0 & 0 & 0 & 0 \end{bmatrix}, \qquad \hat{K}_2 = \begin{bmatrix} 0 & 0 & 1 & 0 \\ 0 & 0 & 0 & 0 \\ 1 & 0 & 0 & 0 \\ 0 & 0 & 0 & 0 \end{bmatrix}, \qquad \hat{K}_1 = \begin{bmatrix} 0 & 0 & 0 & 1 \\ 0 & 0 & 0 & 0 \\ 0 & 0 & 0 & 0 \\ 1 & 0 & 0 & 0 \end{bmatrix}, \qquad (1.10)$$

which then describe the three elementary infinitesimal rotations and boosts, respectively.

Thus, one can see that, in a sense, an infinitesimal boost is like an imaginary infinitesimal rotation.

We have just given two examples of a fundamental class of associative algebras in the form of *matrix algebras*, which are defined by vector spaces of square matrices under matrix multiplication. We will use the notation $M(n, \mathbb{R})$ for the algebra of real $n \times n$ matrices and $M(n, \mathbb{C})$ for complex $n \times n$ matrices. Both of these algebras have a unity in the form of the identity matrix. An early theorem of Cayley stated that any associative algebra can be represented by a matrix algebra. In many cases, this representation is an isomorphism.

A vector subspace $S$ of an algebra $A$ is called a *subalgebra* if the product of any two elements of $S$ is another element of $S$; from the bilinearity of the product, the product of any linear combination of elements of $S$ with any other linear combination of elements of $S$ will then belong to $S$. For instance, 0 and $A$ are (improper) subalgebras, and in the case of the Lie algebra $\mathfrak{so}(3; \mathbb{R})$, the set of all anti-symmetric 3×3 real matrices $\omega^i_j$ such that $\omega^i_j v^j = 0$ is a subalgebra, and, in fact, a one-dimensional subalgebra that is isomorphic to $\mathfrak{so}(2; \mathbb{R})$. Similarly, $M(n, \mathbb{R})$ can be represented as a subalgebra of $M(n, \mathbb{C})$; for instance, by representing any matrix in $M(n, \mathbb{C})$ as the sum of a real matrix, which then belongs to $M(n, \mathbb{R})$ and an imaginary one. Note that although the imaginary matrices in $M(n, \mathbb{C})$ take the form of $i$ times a real matrix, nonetheless, the product of two imaginary matrices is real, so the imaginary matrices do not form a subalgebra.

An important example of a subalgebra is the *center* of any algebra $A$. This ideal $Z(A)$ would then consist of all elements of $A$ that commute with every other element; i.e., $Z(A) = \{z \in A \mid az = za \text{ for all } a \in A\}$. At the very least, 0 and 1 will have this property (if there is a unity), so if $A$ has a unity all scalar multiples of 1 will be contained in the center. For instance, the Schur lemma says that the only square matrices that commute with all other square matrices are scalar multiples of the identity matrix, which says that the center of those algebras will be defined by those scalar multiples.

One can easily see that $Z(A)$ must be a subalgebra of $A$, since if $z$ and $z'$ are elements of $Z(A)$ and $a$ is an arbitrary element of $A$ then:

$$a \, (zz') = \, z \, a \, z' = (zz') \, a$$

so $zz'$ commutes with any element of $A$, as well.

Any algebra product can be polarized into a sum of a commutator and an anti-commutator:

$$ab = \tfrac{1}{2} \, [a, b] + \tfrac{1}{2} \, \{a, b\},$$

where:

$$\{a, b\} = ab + ba.$$

This results in a corresponding polarization of the structure constants with respect to any choice of basis:

$$a_{ij}^k = b_{ij}^k + c_{ij}^k, \qquad\qquad b_{ij}^k = b_{ji}^k, \qquad c_{ij}^k = -c_{ji}^k.$$

For a Lie algebra, by anti-symmetry, only the $c_{ij}^k$ will be non-vanishing.

A *Clifford algebra* $\mathcal{C}(n, <.,.>)$ over an $n$-dimensional orthogonal space $(V, <.,.>)$ has the property that:

$$\{a, b\} = 2<a, b>, \qquad a, b \in V. \tag{1.11}$$

A Clifford algebra over an $n$-dimensional vector space will be $2^n$-dimensional and can be represented by some matrix algebra. Any orthonormal frame $\{\mathbf{e}_a, a = 1, .., n\}$ for $V$ will then define a set of generators for the algebra, since any element of the algebra can be expressed as a linear combination of products of the generators. One then sees that a basis for $\mathcal{C}(n, <.,.>)$ is defined by all of the linearly independent products $\{1, \mathbf{e}_a, \mathbf{e}_a\,\mathbf{e}_b, \dots, \mathbf{e}_1 \dots \mathbf{e}_n\}$, when one takes into account the basic defining relation (1.11).

The Clifford algebra that will be of interest to us is the one $\mathcal{C}(3, \delta_{ij})$ that is defined over real, three-dimensional Euclidian space $(\mathbb{R}^3, \delta_{ij})$. Relative to the canonical basis, which is also orthogonal, one then has:

$$\mathbf{e}_i\,\mathbf{e}_j + \mathbf{e}_j\,\mathbf{e}_i = 2\delta_{ij}. \tag{1.12}$$

This algebra is then eight-dimensional as a real algebra and has a basis that is defined by the set of products $\{1, \mathbf{e}_i, \mathbf{e}_2\mathbf{e}_3, \mathbf{e}_3\mathbf{e}_1, \mathbf{e}_1\mathbf{e}_2, \mathbf{e}_1\mathbf{e}_2\mathbf{e}_3\}$.

From (1.12), the set $\{1, \mathbf{e}_2\mathbf{e}_3, \mathbf{e}_3\mathbf{e}_1, \mathbf{e}_1\mathbf{e}_2\}$ closes under multiplication and thus defines a four-dimensional subalgebra of $\mathcal{C}(3, \delta_{ij})$ that one calls the *even subalgebra*. The four-dimensional vector space complement that is spanned by the set $\{\mathbf{e}_i, \mathbf{e}_1\mathbf{e}_2\mathbf{e}_3\}$ does not close, so it is not a subalgebra, and one calls the elements of this space the *odd* elements of $\mathcal{C}(3, \delta_{ij})$.

Although one might think that we might also use the Clifford algebra $\mathcal{C}(4, \eta_{\mu\nu})$ over Minkowski space, nonetheless, we shall find that $\mathcal{C}(3, \delta_{ij})$ is still fundamental to the complex quaternions, which relate to Lorentz transformations, when one complexifies it.

An algebra that has no divisors of zero is an *integral domain*. An algebra with unity becomes a *division algebra* when every element then admits a multiplicative inverse. According to Adams's theorem, the only real division algebras, up to isomorphism, are $\mathbb{R}, \mathbb{C}, \mathbb{H}, \mathbb{O}$, which are algebras of real numbers, complex numbers, real quaternions, and octonions – or Cayley algebras – respectively. A commutative division algebra is then a *field*. The division algebras $\mathbb{R}$ and $\mathbb{C}$ are both associative and commutative, while $\mathbb{H}$ is associative, but not commutative, and $\mathbb{O}$ is neither associative nor commutative. The only complex division algebra is $\mathbb{C}$ itself (see Dickson [**2**], in the section on complex algebras).

**2. Ideals in algebras.** If $U$ and $V$ are subsets of an algebra $A$ then we shall define their product to be the linear subspace $UV = \text{span}\{uv \mid u \in U, v \in V\}$; that is, it consists of all finite linear combinations of products of elements in the two sets. Thus, $UU \subset U$ iff $U$ is a subalgebra of $A$.

A subset $\mathcal{I}$ of an algebra $A$ is called a *left-sided ideal* of $A$ iff $A\mathcal{I} \le \mathcal{I}$. Clearly, $A$ and 0 are always left-ideals in any algebra. The latter situation is, however, distinct from the notion of a *zero left-ideal*; for such an ideal, $A\mathcal{I} = 0$. However, if $A$ is a division algebra then there would no zero left-ideals in it. If $A$ has a unity $e$ then the only left-ideal that contains $e$ is $A$ itself.

If $S$ is a subset of $A$ then the *left-ideal generated by $S$* is defined to be the set $\mathcal{I}(S)$, which can be characterized as the smallest (with respect to inclusion) left-ideal that contains $S$. For instance, if the set consists of only the unity $e$ then $\mathcal{I}(e) = A$.

As we said above, by definition, a left-ideal is a linear subspace of an algebra $A$. One can easily show that any left-ideal $\mathcal{I}$ must also be a sub-algebra; i.e., $\mathcal{I}\mathcal{I} \le \mathcal{I}$. One simply takes two representative elements $u$ and $u'$ in $\mathcal{I}$ and notes that their products $uu'$ and $u'u$ both belong to $\mathcal{I}$ since the element on the left is a general element in $A$, while the elements on the right are presumed to be elements of $\mathcal{I}$. However, it is not always true that a sub-algebra must be a left-ideal. For instance, any line through the origin in an algebra will be a subalgebra, but it will not generally be fixed by left-multiplication by every element of the algebra unless the algebra is one-dimensional to begin with. Complex multiplication gives a familiar example in which lines through the origin can get rotated by multiplication.

There are two important classes of special elements in any left-ideal $\mathcal{I}$: An element $n \in \mathcal{I}$ is called *nilpotent* if $n^p = 0$ for some positive integer $p$, and the minimum such $p$ is called the *degree of nilpotency*. For instance, in the algebra of 2×2 real matrices, the matrix:

$$\begin{bmatrix} 0 & 1 \\ 0 & 0 \end{bmatrix}$$

is nilpotent of degree 2. It is clear that a division algebra can have no non-zero nilpotent elements, since $nn^{p-1} = 0$ would then define a pair of divisors of zero.

An element $\varepsilon \in A$ is called *idempotent* if $\varepsilon^2 = \varepsilon$. For instance, if $A$ has a unity element then it will always be an idempotent. Similarly, 0 is always an idempotent, and we will treat this as the trivial case.

If one considers matrix algebras then one sees that projection operators behave like idempotents. Basically, the first application projects all of the elements of the vector space that the matrix acts on onto a subspace, while the second application of the projection acts like a unity on the subspace.

For a finite-dimensional vector space $V$ one can always find (non-unique) supplementary subspace $S^c$ to any given subspace $S$ such that $V = S \oplus S^c$. Once a supplement has been chosen for $S$, any element $\mathbf{v} \in V$ can be uniquely expressed in the form $\mathbf{s} + \mathbf{s}^c$, where $\mathbf{s} \in S$ and $\mathbf{s}^c \in S^c$.

This then defines projection operators $P\colon V \to S$, $\mathbf{v} \mapsto \mathbf{s}$, $P^c\colon V \to S^c$, $\mathbf{v} \mapsto \mathbf{s}^c$. Furthermore, one must have:

$$P^c = I - P. \tag{1.13}$$

There is then a corresponding unique decomposition of the identity transformation $I$ into the sum:

$$I = P + P^c. \tag{1.14}$$

For instance, in two dimensions one can express the identity matrix as the sum:

$$\begin{bmatrix} 1 & 0 \\ 0 & 1 \end{bmatrix} = \begin{bmatrix} 1 & 0 \\ 0 & 0 \end{bmatrix} + \begin{bmatrix} 0 & 0 \\ 0 & 1 \end{bmatrix}$$

of the projection onto the $x$ axis and the projection onto the $y$ axis.

One also notes that the projection operators have the property that $PP^c = P^cP = 0$, which derives from the fact that $S \cap S^c = 0$.

One can reverse the logic and say that if one is given two linear transformations $P$, $P^c$ of $A$ to itself such that:

1. $P^2 = P$, $P^{c2} = P^c$        (idempotency),
2. $PP^c = P^cP = 0$        (orthogonality),
3. $I = P + P^c$        (decomposition of the identity)

then there are subspaces $S$ and $S^c$ of $A$ such that $A = S \oplus S^c$ and $S \cap S^c = 0$; the subspace are defined simply by the images of the linear transformations.

One similarly defines two idempotent elements $\varepsilon_1$ and $\varepsilon_2$ in an algebra $A$ to be *orthogonal* iff $\varepsilon_1\varepsilon_2 = \varepsilon_2\varepsilon_1 = 0$. If one has, moreover, that $e = \varepsilon_1 + \varepsilon_2$ then one can express $A$ as the direct sum $\mathcal{I}(\varepsilon_1) \oplus \mathcal{I}(\varepsilon_1) = A\varepsilon_1 + A\varepsilon_2$. There is always the question of reducibility to address at a time like this, so we define an idempotent $\varepsilon$ to be *primitive* iff is can be expressed as the sum of two orthogonal non-trivial idempotents and *imprimitive* otherwise. For instance, in the present case, if $e = \varepsilon_1 + \varepsilon_2$ then $e$ would not be a primitive idempotent.

Furthermore, from the definition of an idempotent, one will have:

$$(e - \varepsilon)(e - \varepsilon) = e - 2\varepsilon + \varepsilon^2 = e - \varepsilon, \qquad \varepsilon(e - \varepsilon) = 0.$$

Thus, $\varepsilon^{\hat{c}} = e - \varepsilon$ is also an idempotent and is orthogonal to $\varepsilon$. One can then obtain a decomposition of the unity element:

$$e = \varepsilon + \varepsilon^{\hat{c}}.$$

Along with left-ideals, one can also define a *right-ideal* $\mathcal{I}$ in $A$ to be a linear subspace such that $\mathcal{I}A \leq A$; $\mathcal{I}$ is therefore also a sub-algebra of $A$. There are then corresponding definitions for the right-ideal generated by a subset and the decomposition of the unity element by right-ideals of orthogonal idempotents.

Finally, one can define a *two-sided ideal* – or simply, *ideal* – in an algebra $A$ to be a subspace $\mathcal{I}$ that is both a left-ideal and a right ideal. This can also be expressed by saying that $A\mathcal{I}A \leq \mathcal{I}$. A two-sided ideal of an algebra (i.e., ring) is more closely analogous to a normal subgroup of a group than the previous two types of one-sided ideals, since one finds that the difference vector space $A - \mathcal{I}$, which is composed of equivalence classes of elements in $A$ that differ by an element of $\mathcal{I}$, is an algebra iff $\mathcal{I}$ is a two-sided ideal. Note that if $A$ is a commutative algebra then all ideals will be two-sided.

**3. Automorphisms of algebras.** An *automorphism* of an algebra $A$ is a linear isomorphism $a: A \rightarrow A$, $v \mapsto v^{\alpha}$ that respects the order of multiplication:

$$(vw)^{\alpha} = v^{\alpha} w^{\alpha}.$$

For instance, complex conjugation of complex numbers has this property:

$$(z_1 z_2)^{*} = z_1^{*} z_2^{*}.$$

One calls $\alpha$ an *anti-automorphism* when the order of multiplication is reversed:

$$(vw)^{\alpha} = w^{\alpha} v^{\alpha}.$$

The transposition of square matrices and the inversion of invertible matrices fall into this category:

$$(AB)^{\mathrm{T}} = B^{\mathrm{T}} A^{\mathrm{T}}, \qquad (AB)^{-1} = B^{-1} A^{-1}.$$

All of the examples that were given so far are also examples of *involutions*; i.e., $\alpha^2 = I$.

When one composes two automorphisms, the result is an automorphism. However, the composition of two anti-automorphisms is an automorphism, while the composition of an automorphism and an anti-automorphism – in either order – is an anti-automorphism. For instance, the Hermitian conjugation of square complex matrices is the composition of complex conjugation and transposition:

$$\dagger = \mathrm{T}^* = {}^*\mathrm{T},$$

which then becomes an anti-automorphism:

$$(AB)^\dagger = B^\dagger A^\dagger.$$

As long as the sum or difference $v \pm v^\alpha$ is still a member of $A$, one can define the *polarization* of any element $v \in A$ into a part $v^+$ that is fixed by the automorphism (or anti-automorphism) $\alpha$ and a part $v^-$ that goes to its negative under $\alpha$:

$$v = v^+ + v^-, \qquad (v^+)^\alpha = v^+, \qquad (v^-)^\alpha = -v^-,$$

by setting:

$$v^\pm = \tfrac{1}{2}(v \pm v^\alpha).$$

When $\alpha$ is complex conjugation, one polarizes a complex number into a sum of a real and an imaginary part. When $\alpha$ is the transposition of square matrices, the result is the sum of a symmetric matrix and a skew-symmetric one, while if $\alpha$ is Hermitian conjugation, the result is the sum of a Hermitian matrix and a skew-Hermitian one. However, inversion of invertible matrices would not admit this decomposition, since the sum or difference of an invertible matrix does not have to be invertible.

Polarization induces a decomposition of the identity operator on $A$ into the sum of two projections:

$$I = P^+ + P^-,$$

with:

$$P^\pm(v) = v^\pm.$$

There is a corresponding direct sum decomposition of $A$, as a vector space, into $A^+ \oplus A^-$, where:

$$P^\pm(A) = A^\pm.$$

These subspaces do not have to be sub-algebras, though. For instance, the product of imaginary numbers is a real number, while the product of symmetric or skew-symmetric matrices does not have to be symmetric or skew-symmetric, respectively.

When $\alpha$ is an involutory anti-automorphism, the operation of $A$ on itself by $\alpha$-*conjugation:*

$$A \times A \to A, \ (a, b) \mapsto aba^\alpha$$

has the useful property that it always has $A^\pm$ for invariant subspaces. That is, if $b \in A^\pm$ then $aba^\alpha \in A^\pm$, as well. This follows from the fact that if $b^\alpha = \pm b$ then:

$$(aba^\alpha)^\alpha = ab^\alpha a^\alpha = \pm aba^\alpha.$$

This fact will prove repeatedly useful in our discussions of the action of unit quaternions of various types on the quaternions, more generally.

**4. Representations of algebras**. A *representation* of an $n$-dimensional algebra $A$ over a field $\mathbb{K}$ is a homomorphism $\rho\colon A \to M(m; \mathbb{K})$. That is, it is a linear map that associates every element $a \in A$ with a unique matrix $\rho(a)$ in the algebra of $m \times m$ matrices with elements in the field $\mathbb{K}$ that also has the property that:

$$\rho(ab) = \rho(a)\,\rho(b);$$

i.e., it respects the products.

The image $\rho(A)$ will be a sub-algebra of $M(m; \mathbb{K})$, but not necessarily one that is isomorphic to $A$. For instance, the trivial map that takes every element of $A$ to the 0 matrix is still a homomorphism, but not a very fascinating one. Indeed, $\rho(A)$ will be isomorphic to $A$ iff $\rho$ is also injective, which is true iff ker $\rho = 0$. In such a case, one will call the representation *faithful*, and if $m = n$, in addition, then the representation is an isomorphism, and one says that $A$ is itself a matrix algebra.

Any algebra admits at least two non-trivial representations, which are defined by left-multiplication and right-multiplication. If the algebra is a division algebra then one can also define another representation by means of conjugation.

If $a \in A$ then left-multiplication by $a$ defines a linear map $L(a): A \rightarrow A$, $b \mapsto ab$. However, the linear map does not have to be invertible. In fact, $L(a)$ is invertible iff $a$ is invertible. Thus, if $A$ is a division algebra then its non-zero elements will be represented in $GL(n; \mathbb{K})$.

If one chooses a basis $\mathbf{e}_i$ for $A$, which we assume to be $n$-dimensional, then the linear transformation $L(a)$ can be represented by an $n \times n$ matrix $[L(a)]_i^j$ with entries in the field of scalars for $A$ by way of:

$$L(a)\mathbf{e}_i = \mathbf{e}_j [L(a)]_i^j. \tag{1.15}$$

Thus, one can define $L: A \rightarrow M(n; \mathbb{K})$, $a \mapsto [L(a)]_i^j$, and one finds that it is, in fact, a representation of $A$. If $A$ has a unity $e$ then the representation must be faithful, since otherwise there would be distinct elements $a \neq a'$ in $A$ such that $L(a) = L(a')$. That would mean that one would have to have $ab = a'b$ for all $b \in A$. In particular, this would have to be true for $b = e$, which would imply that $a = a'$.

However, one immediately sees that the representation $L$ is not usually likely to be an isomorphism, since the dimension of $M(n; \mathbb{K})$ is $n^2$, as opposed to $n$ for $A$. Thus, the only possible dimensions in which this might happen are 0 and 1.

If one chooses a basis for the algebra $A$ then the components of the product $vw$ of two elements $v$ and $w$ can be expressed in terms of the structure constants $a_{jk}^i$ as in (1.2), and this gives us the matrix $[L(a)]_i^j$ in terms of the structure constants, as well; namely:

$$(vw)^i = a_{jk}^i\, v^j\, w^k = [L(v)]_j^i\, w^j,$$

with:

$$[L(v)]_j^i = a_{kj}^i\, v^k.$$

Thus, the matrix of the map $L$ itself is given by the structure constants in that basis.

The representation by right-multiplication is entirely analogous to the case of left-multiplication. First one defines $R(a)$: $A \to A$, $b \mapsto ba$ and then $R$: $A \to M(n; \mathbb{K})$, $a \mapsto [R(a)]$. From the previous argument, one sees that if one has chosen a basis for $A$ such that its structure constants are $a^i_{jk}$ then the matrix of $[R(v)]$ in that basis will be:

$$[R(v)]^i_j = a^i_{jk} \, v^k.$$

Hence, depending upon the symmetry of the product, left and right multiplication might or might nor be closely related processes. In particular, for Lie algebras, whose structure constants are then anti-symmetric, it is only necessary to examine the left multiplication, which then gives the adjoint representation of the Lie algebra, which allows one to define the "roots" of the Lie algebra as eigenvalues of the matrices that one associates with elements in a "Cartan subalgebra."

Because the structure of the algebra is essentially contained in the structure constants, one can determine much of that structure by looking at the properties of the general matrix $[L(v)]^i_j$. In particular, looking for its eigenvectors and eigenvalues gives one the characteristic and minimal polynomials with coefficients that depend upon $a^i_{jk}$ and the components of the general element $v$. The roots and factorizability of these polynomials then have much to say about the structure of the algebra itself. Note, furthermore, that although the matrix $[L(v)]^i_j$ will change with a change of basis, the characteristic polynomial will not.

Since we will have no immediate need for this approach to the structure of algebras, we simply refer the interested readers to some of the earlier literature (e.g., Shaw [**1**], Dickson [**2**], or Albert [**3**]).

The representation of an $n$-dimensional division algebra $A$ over $\mathbb{K}$ in $M(n; \mathbb{K})$ by conjugation is called the *adjoint representation* (although this a different usage from the one that relates to Lie algebras). First, if $a \in A$ then one defines conjugation by $a$ as the linear map ad($a$): $A \to A$, $b \mapsto aba^{-1}$. Once again, since $A$ is a division algebra, as long as $a \neq 0$ the linear transformation ad($a$) will be invertible. One then defines the adjoint representation of $A$ by ad: $A \to M(n; \mathbb{K})$, $a \mapsto [\mathrm{ad}(a)]^i_j$, where:

$$a \, \mathbf{e}_i \, a^{-1} = \mathbf{e}_j [\mathrm{ad}(a)]^j_i. \tag{1.16}$$

The three types of representations that we just defined are closely related to the previous kinds of ideals, since both concepts are related to the multiplication of elements. Thus, one can think of a left-ideal $\mathcal{I}$ in $A$ as an invariant subspace of the representation $L$ since $L(a)\mathcal{I} \leq \mathcal{I}$ for every $a \in A$; i. e., the representation on $\mathcal{I}$ is *irreducible*. If $\mathcal{I} = \mathcal{I}(\varepsilon)$ for some idempotent element $\varepsilon \in A$ then the representation is irreducible iff $\varepsilon$ is primitive.

Analogous statements apply to the case of right-ideals and right-multiplication.

One also sees the adjoint representation of any division algebra $A$ has two-sided ideals for invariant subspaces. There is also closely-related "chiral" representation of $A \times A$ on $A$ that takes any $((a, b), c)$ to $acb$. This representation would also have two-sided ideals for its invariant subspaces.

**5. Tensor products of algebras**. Although the representations of physical fields in the various types of quaternions are essentially an alternative to the tensor and spinor product representations that are customarily used in theoretical physics, nevertheless, we shall still have to clarify what we mean by saying that the various types of quaternion algebras are obtained by tensoring the algebra of real quaternions by various coefficients rings. We mean that those coefficient rings can all be regarded as real algebras of varying dimensions that are obtained by taking the tensor product of the two algebras as real vector spaces.

If $A$ and $B$ are both $\mathbb{K}$-algebras of dimensions $n$ and $m$, respectively, then their *tensor product algebra* is a $\mathbb{K}$-algebra $A \otimes_{\mathbb{K}} B$ of dimension $nm$ that is defined over the corresponding tensor product of vector spaces by also accounting for the products on $A$ and $B$ to give a product on $A \otimes_{\mathbb{K}} B$:

$$(a \otimes b)(a' \otimes b') = aa' \otimes bb'.$$

One can also think of $A \otimes_{\mathbb{K}} B$ as consisting of linear combinations of elements in $A$ with coefficients in $B$.

If $\{\mathbf{e}_i, i = 1, \ldots, n\}$ is a basis for $A$ and $\{\mathbf{f}_\alpha, \alpha = 1, \ldots, m\}$ is a basis for $B$ then $\{\mathbf{e}_i \otimes \mathbf{f}_\alpha, i = 1, \ldots, n, \alpha = 1, \ldots, m)$ is a basis for $A \otimes_{\mathbb{K}} B$, when the components of any element of $A \otimes_{\mathbb{K}} B$ are taken from $\mathbb{K}$. Not all elements of $A \otimes_{\mathbb{K}} B$ are of the form $a \otimes b$, but only the decomposable ones. In that case, the components of an element of $A \otimes_{\mathbb{K}} B$ with respect to the basis $\mathbf{e}_i \otimes \mathbf{f}_a$ are of the form $a^i b^\alpha$; a more general element simply has a component matrix $\beta^{i\alpha}$ with elements in $\mathbb{K}$.

Since the main use that we will have for tensor products of algebras will involve tensoring the real quaternions with various coefficient algebras, we shall now show how the above remarks simplify somewhat in such a case. For example, we consider the complexification of a real algebra.

Let $A$ be an $n$-dimensional real algebra with a basis defined by $\{\mathbf{e}_i, i = 1, \ldots, n\}$ and a multiplication table that is defined by

$$\mathbf{e}_i \, \mathbf{e}_j = a_{ij}^k \mathbf{e}_k \, .$$

We regard $\mathbb{C}$ as a division algebra over $\mathbb{R}^2$ with a basis defined by $\{1, i\}$ whose multiplication table is:

$$11 = 1, \; 1i = i1 = i, \qquad ii = -1.$$

When one takes the tensor product $A \otimes_{\mathbb{R}} \mathbb{C}$, the basis that one can define on it from the given ones consists of $2n$ members $\{\mathbf{e}_i \otimes 1, \mathbf{e}_i \otimes i\}$, which we abbreviate to $\{\mathbf{e}_i, i\mathbf{e}_i\}$. Thus, in order extend the two given multiplication tables, we only need to account for the products that involve $i$, which we do by way of:

$$\mathbf{e}_i \, (i\mathbf{e}_j) = (i\mathbf{e}_j) \, \mathbf{e}_i = i \, (\mathbf{e}_i \, \mathbf{e}_j) = a_{ij}^k \, (i\mathbf{e}_k) = (ia_{ij}^k)\mathbf{e}_k \, , \qquad (1.17)$$

$$(i\mathbf{e}_i)(i\mathbf{e}_j) = i^2 \, \mathbf{e}_i \, \mathbf{e}_j = - \, \mathbf{e}_i \, \mathbf{e}_j = - \, a_{ij}^k \mathbf{e}_k \, . \qquad (1.18)$$

If one regards $A \otimes_{\mathbb{R}} \mathbb{C}$ as a real algebra then a typical element $v$ can be represented in the form:

$$v = v_{\text{Re}}^i \mathbf{e}_i + v_{\text{Im}}^i (i\mathbf{e}_i) \, ,$$

but if one regards it as a complex algebra with a basis given by $\mathbf{e}_i$ then the same elements takes the form:

$$v = (v_{\text{Re}}^i + iv_{\text{Im}}^i)\mathbf{e}_i \, .$$

Similarly, if one regards $A \otimes_{\mathbb{R}} \mathbb{C}$ as a real algebra then the new set of structure constants is now $a_{IJ}^K$, where $I$, $J$, $K$ run from 1 to $2n$, the structure constants $a_{ij}^k$ are unchanged, and from (1.17), (1.18), the missing ones are:

$$a_{i+n,j}^{k+n} = a_{i,j+n}^{k+n} = a_{ij}^k \, , \qquad\qquad a_{i+n,j+n}^k = - \, a_{ij}^k \, ,$$

all other being zero.

However, if we regard $A \otimes_{\mathbb{R}} \mathbb{C}$ as a complex algebra then the structure constants remain unchanged. In fact, more generally, when one tensors a given algebra with various other coefficient algebras, the structure constants for a real basis remain the same, while the character of the multiplication is solely due to the character of the multiplication in the coefficient algebra $B$, since one assumes that products are $B$-bilinear. Therefore, one can always deal with the products of coefficients and products of basis elements separately. In particular, all of the algebras that we will be dealing with are obtained by tensoring the real quaternions with various other algebras, namely, the complex numbers, the dual numbers, and the complex dual numbers. Hence, the basic multiplication table for the basis elements will not change fundamentally, while the idiosyncrasies of the coefficient algebra will affect the products of coefficients.

# CHAPTER II

# REAL QUATERNIONS

The algebra of real quaternions is fundamental to the extensions that follow, so we first introduce the formalism at that level, and then show how one extends it in the subsequent chapters.

**1. The group of Euclidian rotations.** Three-dimensional Euclidian space, for now, will be $\mathbb{R}^3$ when it is given the Euclidian scalar product $<.,.>$. A scalar product on a $\mathbb{K}$-linear space $V$ is, of course, a bilinear functional $V \times V \to \mathbb{K}$, $(\mathbf{v}, \mathbf{w}) \mapsto <\mathbf{v}, \mathbf{w}>$ that is symmetric and non-degenerate. Thus, one always has:

$$<\mathbf{v}, \mathbf{w}> = <\mathbf{w}, \mathbf{v}>, \tag{2.1}$$

and for every $\mathbf{v} \in V$ the linear map $\mathbf{v}^* : V \to V^*$, $\mathbf{w} \mapsto \mathbf{v}^*(\mathbf{w}) = <\mathbf{v}, \mathbf{w}>$ is an isomorphism.

If $\{\mathbf{e}_i, i = 1, 2, 3\}$ is a basis for $\mathbb{R}^3$ then it will be said to be *orthonormal* iff:

$$<\mathbf{e}_i, \mathbf{e}_j> = \delta_{ij} . \tag{2.2}$$

We shall also refer to an orthonormal basis for $E^3 = (\mathbb{R}^3, <.,.>)$ as an *orthonormal frame* for that vector space.

The scalar product of any two vectors $\mathbf{v} = v^i \mathbf{e}_i$ and $\mathbf{w} = w^j \mathbf{e}_j$ can be obtained from the scalar products of the frame members using bilinearity:

$$<\mathbf{v}, \mathbf{w}> = \delta_{ij} v^i w^j = v^1 w^1 + v^2 w^2 + v^3 w^3 . \tag{2.3}$$

In particular, the scalar product of any vector with itself takes the form:

$$\| \mathbf{v} \|^2 = <\mathbf{v}, \mathbf{v}> = \delta_{ij} v^i v^j = \sum_{i=1}^{3} (v^i)^2 , \tag{2.4}$$

and we will refer to $\| \mathbf{v} \|^2$ as the *norm-squared* of $\mathbf{v}$ and its square root as the *norm* of $\mathbf{v}$.

Because we are dealing with real numbers, the only way that $\| \mathbf{v} \|$ can vanish is if $\mathbf{v} = 0$. Thus, one refers to the scalar product as *positive-definite*. (Of course, this property does not apply to the Minkowski space scalar product.)

A map $R: \mathbb{R}^3 \to \mathbb{R}^3$ will be called an *orthogonal transformation* – or simply, a *rotation* – if it preserves the scalar product; i.e.:

$$<R\mathbf{v}, R\mathbf{w}> = <\mathbf{v}, \mathbf{w}> \qquad \text{for all } \mathbf{v}, \mathbf{w}. \tag{2.5}$$

Since the scalar product is bilinear and symmetric, one can then show:

**Theorem:**

Any orthogonal map $R$ must be an invertible linear transformation.

Proof:

Linearity:

$$<R(\alpha\mathbf{v}), R\mathbf{w}> = \alpha <\mathbf{v}, \mathbf{w}> = \alpha <R\mathbf{v}, R\mathbf{w}> = <\alpha R\mathbf{v}, R\mathbf{w}>,$$

$$<R(\mathbf{v} + \mathbf{v}'), R\mathbf{w}> = <\mathbf{v} + \mathbf{v}', \mathbf{w}> = <\mathbf{v}, \mathbf{w}> + <\mathbf{v}', \mathbf{w}>$$
$$= <R\mathbf{v}, R\mathbf{w}> + <R\mathbf{v}', R\mathbf{w}> = <R\mathbf{v} + R\mathbf{v}', R\mathbf{w}>,$$

and as these relations must be true for all vectors, one can conclude the linearity:

$$R(\alpha\mathbf{v}) = \alpha R\mathbf{v}, \qquad R(\mathbf{v} + \mathbf{v}') = R\mathbf{v} + R\mathbf{v}'.$$

Invertibility:

If $R\mathbf{v} = 0$ then:

$$<R\mathbf{v}, R\mathbf{v}> = <\mathbf{v}, \mathbf{v}> = 0,$$

which is only possible if $\mathbf{v} = 0$ in the positive-definite case; thus, ker $R = 0$, which makes $R$ injective. The fact that it is also surjective then follows from the nullity-rank theorem. Q.E.D.

Thus, $R$ can be represented by a matrix, which we shall denote by either $R$ or $R_j^i$, with respect to a chosen basis $\mathbf{e}_i$, since:

$$R\mathbf{e}_i = \mathbf{e}_j R_i^j. \tag{2.6}$$

The matrix can also be said to act on the components of a vector $\mathbf{v} = v^i\mathbf{e}_i$ since:

$$R\mathbf{v} = v^i(\mathbf{e}_j R_i^j) = (R_i^j v^i)\mathbf{e}_j; \tag{2.7}$$

i.e., if $\mathbf{v}' = R\mathbf{v}$ then:

$$v'^i = R_j^i v^j. \tag{2.8}$$

Note that the same matrix acts on the column vector $[v^i]$ directly on the left and on the row vector $[\mathbf{e}_i]$ by its transpose on the right.

Because of (2.3) and (2.5), the matrix of any rotation $R$ has the property that:

$$\delta_{kl} R_i^k R_j^l = \delta_{ij}. \tag{2.9}$$

Thus, one can say that:

$$R^{-1} = R^{\mathrm{T}}, \tag{2.10}$$

where the T refers to the transpose operator.

Since this means that $RR^T = R^T R = I$, if one takes the determinant of both sides then one finds that:

$$\det(R) = \pm 1. \tag{2.11}$$

The positive sign refers to *proper rotations*, while the negative sign gives *improper* ones, which are the product of a proper rotation with a reflection through the origin, whose matrix is then $– I$.  Only the proper rotations are regarded as physical motions.

If one looks at the eigenvalues of a typical rotation $R$ then the characteristic polynomial will take the form:

$$\det(R − \lambda I) = a\lambda^3 + b\lambda^2 + c\lambda + d,$$

with all real coefficients.  Although the roots do not have to all be real, nonetheless, the complex roots must only occur in complex conjugate pairs ([3]).  Thus, at least one of them must be real.  From orthogonality, however, one sees that for a real eigenvalue $\lambda$ with a corresponding eigenvector $\mathbf{v}$, one must have:

$$<R\mathbf{v}, R\mathbf{v}> = \lambda^2 <\mathbf{v}, \mathbf{v}> = <\mathbf{v}, \mathbf{v}>,$$

which makes $\lambda = \pm 1$ unless $\mathbf{v} = 0$ (again, in the positive-definite case), although the negative sign refers to an improper rotation.

Thus a three-dimensional rotation will always have an *axis*; i.e., a line through the origin whose points are all fixed by the rotation.  The effect of the rotation on any plane perpendicular to that axis will be a planar rotation, whose matrix relative to an orthonormal frame in that plane can be given the form:

$$R = \begin{bmatrix} \cos\theta & −\sin\theta \\ \sin\theta & \cos\theta \end{bmatrix}, \tag{2.12}$$

where the angle of rotation $\theta$ is measured positive clockwise.

One sees that the characteristic polynomial of such a matrix is:

$$\lambda^2 − 2\cos\theta\,\lambda + 1,$$

and its eigenvalues will take the form:

$$\lambda = e^{i\theta} = \cos\theta + i\sin\theta, \tag{2.13}$$

which makes all of them complex numbers on the unit circle.  We shall see formulas that are analogous to this one show up in all of the following sections.

The corresponding eigenvectors for any $e^{i\theta}$ can then be chosen to take the form:

_________________

([3])  Caveat: This will no longer be true when we get to complex rotations.

$$\begin{bmatrix} 1 \\ \pm i \end{bmatrix},$$

independently of $\theta$; in particular, they are the same for the complex conjugate of $\lambda$, which amounts to a rotation through an angle of $-\theta$. Since they are clearly complex vectors, this will only be of interest when we go on to complex rotations.

As a result of all of this, the characteristic polynomial for a proper three-dimensional rotation factors into either of two forms:

$$(\lambda - 1)(\lambda^2 - 2\cos\theta\,\lambda + 1) \qquad \text{or} \qquad (\lambda - 1)^3,$$

depending upon whether one root is real and the other two are complex or whether all of them are real ( = 1), which can only be true for the identity matrix.

There is a useful formula for the effect of a rotation on a vector $\mathbf{v}$ that is discussed in theoretical kinematics [**1**] and is called *Rodrigues's formula:*

$$\mathbf{v}' = \cos\theta\,\mathbf{v} + (1 - \cos\theta) <\mathbf{v}, \mathbf{u}> \mathbf{u} + \sin\theta\,\mathbf{u} \times \mathbf{v}, \tag{2.14}$$

in which axis of rotation is described by a unit vector $\mathbf{u}$ and the angle of rotation is $\theta$. This formula will also recur in the sequel in various analogous forms.

Any general rotation can be expressed uniquely as a product of elementary rotations about the three orthonormal axes of a frame. Their matrices take the form:

$$R(\theta, 0, 0) = \begin{bmatrix} 1 & 0 & 0 \\ 0 & \cos\theta & -\sin\theta \\ 0 & \sin\theta & \cos\theta \end{bmatrix}, \qquad R(0, \phi, 0) = \begin{bmatrix} \cos\psi & 0 & \sin\phi \\ 0 & 1 & 0 \\ -\sin\phi & 0 & \cos\phi \end{bmatrix},$$

$$\tag{2.15}$$

$$R(0, 0, \psi) = \begin{bmatrix} \cos\psi & -\sin\psi & 0 \\ \sin\psi & \cos\psi & 0 \\ 0 & 0 & 1 \end{bmatrix},$$

once one has chosen a particular order for the product, since the product of rotations does not generally commute, unless they are both performed about the same axis.

The real numbers $\theta$, $\phi$, $\psi$ are called the *Euler numbers* for the sequence of rotations, and are sometimes referred to as the *roll*, *pitch*, and *yaw* angles, respectively. They then define a local coordinate chart for the differentiable manifold $O(3; \mathbb{R})$, which can then be seen to be three-dimensional. Since it also has a group structure under the composition of rotations, and the group operations of product and inversion are differentiable, one then sees that $O(3; \mathbb{R})$ is a real, three-dimensional, non-Abelian Lie group, while $O(2; \mathbb{R})$ is a real, one-dimensional, compact, Abelian one that is diffeomorphic to a pair of circles.

Since the determinant function is continuous, the level sets corresponding to $\pm 1$ are disjoint connected components, and the connected component that contains the identity matrix is a group $SO(3; \mathbb{R})$, which is then composed of proper Euclidian rotations.  It can be shown to be diffeomorphic to $\mathbb{RP}^3$ as a manifold, so it is compact.  The proof follows easily using quaternions, as we shall see.

The Lie algebra $\mathfrak{so}(3; \mathbb{R})$ of $SO(3; \mathbb{R})$, which represents the infinitesimal generators of one-parameter subgroups of proper rotations, can be obtained by differentiating the basic property of orthogonal matrices when one assumes that the rotations define a differentiable curve through the identity in $SO(3; \mathbb{R})$:

$$\frac{d}{ds}\bigg|_{s=0} [R(s)R^{\mathrm{T}}(s)] = \dot{R}(s)R^{\mathrm{T}}(s) + R(s)\dot{R}^{\mathrm{T}}(s) = 0.$$

When one sets $\dot{R}(0) = \omega$, so $\dot{R}^{\mathrm{T}}(0) = \omega^{\mathrm{T}}$, one gets the defining property of infinitesimal rotation matrices:

$$\omega + \omega^{\mathrm{T}} = 0; \tag{2.16}$$

i.e., they are anti-symmetric.

The Lie algebra $\mathfrak{so}(3; \mathbb{R})$ can also be conveniently represented by the vector cross product that is defined on $\mathbb{R}^3$:

$$[\mathbf{v}, \mathbf{w}] = \mathbf{v} \times \mathbf{w} = \varepsilon_{ijk}\, v^i\, w^j\, \mathbf{e}_k\,. \tag{2.17}$$

In order to show the isomorphism, one needs only to define the adjoint action of $\mathfrak{so}(3; \mathbb{R})$, in the present form, on itself, which was discussed in Chapter I.  The isomorphism of these two representations of $\mathfrak{so}(3; \mathbb{R})$ can be obtained by associating the basis elements $\mathbf{e}_i$ of $\mathbb{R}^3$ with the elementary matrices $J_i$, which define a basis for the matrix representation of that Lie algebra.

The structure constants – i.e., the commutation relations – for the Lie algebra can be obtained from the Lie brackets of the orthonormal basis vectors:

$$[\mathbf{e}_i, \mathbf{e}_j] = \varepsilon_{ijk}\, \mathbf{e}_k\,. \tag{2.18}$$

One sees that the eigenvalues of the elementary matrices are $0$, $\pm i$, while the eigenvectors are unchanged from those of the corresponding finite rotations.  Thus, the axis of a finite rotation can also be obtained from then zero eigenvector of its infinitesimal generator.  If the latter is expressed in the form ad($\mathbf{v}$) then its zero eigenspace will be the line through $\mathbf{v}$ itself, since $\mathbf{v} \times \mathbf{w} = 0$ iff $\mathbf{v}$ is collinear with $\mathbf{w}$.

Three-dimensional, Euclidian rotations can also be represented by 2×2 complex unitary matrices with unity determinant, which defines a group that is usually denoted by $SU(2)$. Thus, the space of its defining representation is $\mathbb{C}^2$, when it is given the Hermitian inner product. Such an inner product is not symmetric in the same sense as the usual scalar product, but must satisfy:

$$(\mathbf{v}, \mathbf{w}) = (\mathbf{w}, \mathbf{v})^*.$$

As a result, the norm-squared of any vector will be real.

Moreover, a complex basis $\{\mathbf{e}_1, \mathbf{e}_2\}$ for $\mathbb{C}^2$ is called *unitary* iff:

$$(\mathbf{e}_a, \mathbf{e}_b) = \delta_{ab},$$

and one finds that the general component expression for the inner product becomes:

$$(\mathbf{v}, \mathbf{w}) = \delta_{ab}\, v^a\, w^{b*}.$$

A $\mathbb{C}$-linear transformation $U$ of $\mathbb{C}^2$ is then called *unitary* when it preserves the Hermitian inner product:

$$(U\mathbf{v}, U\mathbf{w}) = (\mathbf{v}, \mathbf{w}), \qquad \text{for all } \mathbf{v}, \mathbf{w} \in \mathbb{C}^2.$$

As a result, the matrix of a unitary transformation must satisfy:

$$UU^\dagger = U^\dagger U = I, \qquad\qquad \text{i.e.,} \quad U^{-1} = U^\dagger,$$

which implies that every unitary transformation is invertible, moreover.

This also implies that the modulus of det $U$ must be unity for any unitary matrix $U$, since:

$$\det(UU^\dagger) = \det U\, (\det U)^* = \|\det U\|^2 = 1.$$

One finds that because of the unitarity constraint on its elements a typical 2×2 unitary matrix $U$ does not need four independent complex numbers to specify it uniquely, but only two:

$$U = \begin{bmatrix} \alpha & \gamma \\ \beta & \delta \end{bmatrix} = \begin{bmatrix} \alpha & -\beta^* \\ \beta & \alpha^* \end{bmatrix}.$$

The complex numbers $\alpha$, $\beta$, $\gamma$, $\delta$ are then the *Cayley-Klein parameters* ([4]), which go back to Klein's work on the theory of tops, and when one expresses $\alpha$ and $\beta$ in terms of real and imaginary components:

$$\alpha = e^0 + ie^3, \quad \beta = e^2 + ie^1,$$

---

([4]) A standard reference of Cayley-Klein parameters and Euler parameters, as well as their relationship to Euler angles, is Goldstein [**2**].

the four real parameters $e^0$, …, $e^3$ that one introduces are sometimes referred to as the *Euler parameters* (as distinct from the Euler *angles*). We shall soon see that the Euler parameters were essentially the components of a real unit quaternion.

When one further imposes the constraint that $U$ have unity determinant, this implies that:

$$\alpha\alpha^* + \beta\beta^* = 1,$$

and one sees that if one represents the two column vectors of $U$ as a pair of vectors $\{\mathbf{U}^1, \mathbf{U}^2\}$ in $\mathbb{C}^2$ then the conditions that were imposed on a matrix in $SU(2)$ say that this pair of vectors must constitute a special unitary frame. Since the association of a pair $(U, -U)$ of matrices in $SU(2)$ with a rotation matrix $R$ in $SO(3)$ – a process that will become quite straightforward when we have introduced quaternions – is often referred to as defining the "spin" covering group of $SO(3)$, we will call a special unitary frame in $\mathbb{C}^2$ a *spin frame*. Thus, any oriented, orthonormal frame in $E^3$ is associated with two spin frames.

Just as an oriented, orthonormal frame in $\mathbb{R}^2$ is defined by specifying one of the two frame members, similarly, a spin frame in $\mathbb{C}^2$ is defined by specifying one of the two complex vectors. For instance, one can take the column $\mathbf{U}^1 = [\alpha, \beta]^{\mathrm{T}}$ in $U$ to be the first frame member and then define the other by:

$$\mathbf{U}^2 = \begin{bmatrix} -\beta^* \\ \alpha^* \end{bmatrix} = \begin{bmatrix} 0 & -1 \\ 1 & 0 \end{bmatrix} \begin{bmatrix} \alpha \\ \beta \end{bmatrix}^* = J\,\mathbf{U}^{1*},$$

in which we have introduced $J$ for the matrix of a clockwise rotation through $\pi/2$ radians.

Because of this, and the fact that $\mathbf{U}^1$ has unit norm, one sees that a matrix in $SU(2)$ can just as well be described by a unit vector in $\mathbb{C}^2$. This one-to-one correspondence between $SU(2)$ matrices and unit vectors in $\mathbb{C}^2$ is at the heart of the description of spin by Pauli spinors, and is also quite elegantly incorporated into the theory of real quaternions, as we shall see.

## 2. The algebra of real quaternions [3-6].

The algebra $\mathbb{H}$ of real quaternions is defined over the vector space $\mathbb{R}^4$ by giving the multiplication table for the canonical basis $\{\mathbf{e}_0, …, \mathbf{e}_3\}$

$$\mathbf{e}_0\,\mathbf{e}_\mu = \mathbf{e}_\mu\,\mathbf{e}_0 = \mathbf{e}_\mu, \qquad \mathbf{e}_i\,\mathbf{e}_j = -\,\delta_{ij}\,\mathbf{e}_0 + \varepsilon_{ijk}\,\mathbf{e}_j\,\mathbf{e}_k\ . \tag{3.1}$$

One immediately notes that since products of some of the basis elements can produce other basis elements, the basis in question does not constitute a minimal set of generators.

For instance, since $\mathbf{e}_1\mathbf{e}_1 = -\mathbf{e}_0$ and $\mathbf{e}_1\mathbf{e}_2 = \mathbf{e}_3$ , one could use $\{\mathbf{e}_1, \mathbf{e}_2\}$ to generate $\mathbb{H}$, since the given basis then consists of $\{-\mathbf{e}_1\mathbf{e}_1, \mathbf{e}_1, \mathbf{e}_2, \mathbf{e}_1\mathbf{e}_2\}$.

One can read off the structure constants for $\mathbb{H}$ relative to the canonical basis from (3.1) directly:

$$a_{0\mu}^{\kappa} = a_{\mu0}^{\kappa} = \delta_{\mu}^{\kappa}, \qquad a_{ij}^0 = -\delta_{ij}, \qquad a_{ij}^k = \varepsilon_{ijk} . \tag{3.2}$$

A typical quaternion then takes the form:

$$q = q^{\mu}\,\mathbf{e}_{\mu} , \tag{3.3}$$

and if $p = p^{\mu}\,\mathbf{e}_{\mu}$ is another quaternion then their product $pq$ can be expressed in the explicit form:

$$\begin{aligned} pq = \quad &(p^0q^0 - p^1q^1 - p^2q^2 - p^3q^3)\,\mathbf{e}_0 \\ &+ (p^1q^0 + p^0q^1 - p^3q^2 + p^2q^3)\,\mathbf{e}_1 \\ &+ (p^2q^0 + p^3q^1 + p^0q^2 - p^1q^3)\,\mathbf{e}_2 \\ &+ (p^3q^0 - p^2q^1 + p^1q^2 + p^0q^3)\,\mathbf{e}_3 . \end{aligned} \tag{3.4}$$

The element $\mathbf{e}_0$ then represents the unity element of the algebra $\mathbb{H}$, and, in fact, it generates the *center* of $\mathbb{H}$, which is defined all elements that commute with all other elements, so it consists of all scalar multiples $q^0\mathbf{e}_0$ . Hence, it will often be convenient to simply abbreviate $\mathbf{e}_0$ by 1. However, when we get to orthogonality, it is important to remember that 1 is still a *vector*, so, in particular $<1, \mathbf{e}_i> = 0$, as we will see.

The algebra $\mathbb{H}$ contains an infinitude of subalgebras that are $\mathbb{R}$-isomorphic to $\mathbb{C}$, such as the subalgebras generated by $\{1, \mathbf{e}_1\}$, $\{1, \mathbf{e}_2\}$, and $\{1, \mathbf{e}_3\}$. In fact, more generally, if $\mathbf{q} = q^i\mathbf{e}_i$ satisfies $<\mathbf{q}, \mathbf{q}> = 1$, with a definition of the scalar product that we will give shortly, then $\{1, \mathbf{q}\}$ generates a subalgebra that is $\mathbb{R}$-isomorphic to $\mathbb{C}$.

Any quaternion can be expressed in the *scalar-plus-vector* form:

$$q = q^0 + \mathbf{q}, \qquad\qquad (\mathbf{q} = q^i\,\mathbf{e}_i), \tag{3.5}$$

where one calls $S(q) = q^0$ the *scalar* part of $q$ and $V(q) = \mathbf{q}$, the *vector* or *pure quaternion* part of $q$. Thus, the scalars represent the center of $\mathbb{H}$.

One can introduce the *conjugation* automorphism: If $q = q^0 + \mathbf{q}$ then:

$$\overline{q} = q^0 - \mathbf{q}. \tag{3.6}$$

In fact, this is an anti-automorphism, since:

$$\overline{p\,q} = \overline{q}\,\overline{p} . \tag{3.7}$$

One sees that polarizing the identity operator with respect to conjugation expresses it as the sum:

$$I = S + V, \qquad (3.8)$$

of two complementary projections $S : \mathbb{H} \to S\mathbb{H}$ and $V : \mathbb{H} \to S\mathbb{H}$ that are defined by:

$$Sq = \tfrac{1}{2}(q + \overline{q}), \qquad Vq = \tfrac{1}{2}(q - \overline{q}). \qquad (3.9)$$

One then has a corresponding direct sum decomposition $\mathbb{H} = S\mathbb{H} \oplus V\mathbb{H}$.

The product of any two quaternions $q$ and $r$ can be expressed in the scalar-plus-vector form:

$$qr = (q^0 r^0 - <\mathbf{q}, \mathbf{r}>) + r^0 \mathbf{q} + q^0 \mathbf{r} + \mathbf{q} \times \mathbf{r}, \qquad (3.10)$$

since:

$$\mathbf{q}\mathbf{r} = -<\mathbf{q}, \mathbf{r}> + \mathbf{q} \times \mathbf{r}. \qquad (3.11)$$

Note that the scalar part of the product then behaves like the Minkowski scalar product, even though we are still only talking about Euclidian geometry.  We then define:

$$(q, r) = S(qr) = \tfrac{1}{2}(qr + \overline{rq}) = q^0 r^0 - <\mathbf{q}, \mathbf{r}>. \qquad (3.12)$$

Of particular interest in what follows will be the general expression for the square of any quaternion:

$$q^2 = (q^0)^2 - <\mathbf{q}, \mathbf{q}> + 2q^0 \mathbf{q}. \qquad (3.13)$$

Although the algebra $\mathbb{H}$ is associative, it is not commutative.  One notes that, in fact:

$$[\mathbf{q}, \mathbf{r}] = 2\,\mathbf{q} \times \mathbf{r}, \qquad (3.14)$$

so although the Lie algebra that is defined by the commutator bracket is isomorphic to $\mathfrak{so}(3; \mathbb{R})$, there is a factor of 2 involved that relates to the fact that the three-dimensional Euclidian rotations will be represented by half-angle rotations.

The algebra $\mathbb{H}$ is, as we mentioned before, a division algebra; in particular, it has no divisors of zero.  In order to find the multiplicative inverse to any non-zero quaternion $q$, we can go back to the expression (3.10) for $qr$ and set it equal to 1.  This gives the following conditions on $r$:

$$q^0 r^0 - <\mathbf{q}, \mathbf{r}> = 1, \qquad r^0 \mathbf{q} + q^0 \mathbf{r} = -\mathbf{q} \times \mathbf{r}.$$

In the second equation, we see that a linear combination of $\mathbf{q}$ and $\mathbf{r}$ can lie in the plane of $\mathbf{q} \times \mathbf{r}$ only if it equals zero.  Thus, $\mathbf{r} = \lambda \mathbf{q}$ for some real number $\lambda$.  But, from the left-hand side this makes $r^0 = -\lambda q^0$.  From the first equation, we see that one must then have $(q^0)^2 - <\mathbf{q}, \mathbf{q}> = 1/\lambda$.  We shall now see that, in fact, $\lambda = -\| q \|^{-2}$.

From (3.10), one sees that:

$$q\bar{q} = \bar{q}q = q^0 q^0 + <\mathbf{q}, \mathbf{q}> \equiv \| q \|^2. \tag{3.15}$$

Thus, we are now looking at the Euclidian norm over $\mathbb{R}^4$, as well as over $\mathbb{R}^3$. The level surfaces of that norm are then real 3-spheres of radius $\| q \|$.

More generally, we can define another scalar product on $\mathbb{H}$ by way of:

$$<q, r> = S(q\bar{r}) = q^0 r^0 + <\mathbf{q}, \mathbf{r}>. \tag{3.16}$$

One then sees that this scalar product amounts to the Euclidian scalar product for quaternions of vector type; in particular, one sees that one also has $\| q \|^2 = <q, q>$.

Since $\| q \|$ is positive-definite for real $q$, as long as $q$ itself is non-zero, one can define the multiplicative inverse to $q$ by:

$$q^{-1} = \frac{\bar{q}}{\| q \|^2}. \tag{3.17}$$

The non-zero quaternions $Q^*$ then define a multiplicative group that is also a four-dimensional real Lie group. It contains the subgroup $Q_1$ of all unit quaternions, and since any non-zero quaternion $q$ can be expressed in "polar" form $\| q \| \, \hat{q}$, where $\hat{q} = q / \| q \|$ is a unit quaternion, one sees that the group $Q^*$ is the product $\mathbb{R}^* \times Q_1$ of the group of non-zero real numbers under multiplication and the group of unit quaternions.

This polar form of any $q$ can then be expressed in the form:

$$q = \| q \| \, (\cos \tfrac{1}{2} \alpha + \sin \tfrac{1}{2} \alpha \, \hat{\mathbf{q}}), \tag{3.18}$$

in which $\hat{\mathbf{q}}$ is a unit vector that generates an axis of rotation, so:

$$\mathbf{q} = \| q \| \sin \tfrac{1}{2} \alpha \, \hat{\mathbf{q}}, \tag{3.19}$$

and the angle $\alpha$, which is then defined by:

$$\cos \tfrac{1}{2} \alpha = \frac{q^0}{\| q \|}, \tag{3.20}$$

represents one-half an angle of rotation about that axis. The appearance of the factor 1/2 will become more necessary when we see that the group of unit quaternions is isomorphic to $SU(2)$, which doubly covers the group $SO(3)$.

One easily verifies that $\mathbb{H}$ has no non-trivial nilpotents of degree two by setting the expression (3.13) for $q^2$ equal to 0, which would make:

$$(q^0)^2 = <\mathbf{q}, \mathbf{q}>, \qquad q^0 \mathbf{q} = 0.$$

From the second equation, we know that either $q^0$ or $\mathbf{q}$ is 0.  In the former case, this would make $<\mathbf{q}, \mathbf{q}> = 0$, and for real vectors this would imply that $\mathbf{q} = 0$, which is the trivial case $q = 0$.  Similarly, in the latter case, if $\mathbf{q} = 0$ then $<\mathbf{q}, \mathbf{q}> = 0$ vanishes, and with it $q^0$, which again gives the trivial case.

In order to find the non-trivial idempotents in $\mathbb{H}$, one goes back to the expression (3.13) for $q^2$ and sets it equal to $q = q^0 + \mathbf{q}$.  This implies that one must have:

$$q^0 = (q^0)^2 - <\mathbf{q}, \mathbf{q}>, \qquad \mathbf{q} = 2q^0\mathbf{q}.$$

If we address the second one first then we see that either $\mathbf{q} = 0$ or $\mathbf{q} \neq 0$.  In the former case, from the first equation, one must have $q^0 = 1$, which gives the trivial idempotent $q = 1$.  If $\mathbf{q} \neq 0$ then $q^0 = 1/2$ , which implies that $<\mathbf{q}, \mathbf{q}> = -1/4$.  As long as we are dealing with real vectors this is impossible, although it will be possible when we go on to complex vectors.

We conclude that there are no non-trivial idempotents in $\mathbb{H}$.

Although $\mathbb{H}$ is not a complex algebra, nonetheless, for some purposes – such as $SU(2)$ spinors – one can represent $\mathbb{H}$ as a *real* algebra over $\mathbb{C}^2$.  In order to see this, one first reverts to the classical notation 1, $i$, $j$, $k$ for the principal units of $\mathbb{H}$.  Since $k = ij$, one can rearrange the terms in the expansion of a typical quaternion as follows:

$$q = q^0 + q^1 i + q^2 j + q^3 ij = (q^0 + q^1 i) + (q^2 + q^3 i)\, j = z^1 + z^2 j, \qquad (3.21)$$

in which we have introduced the complex components:

$$z^1 = q^0 + q^1 i, \qquad z^2 = q^2 + q^3 i. \qquad (3.22)$$

One should be careful about regarding $j$ as another version of $i$, since even though $j^2 = -1$, nonetheless, $ij = -ji$.  As a result, one sees that left scalar multiplication and right scalar multiplication are distinct in this case, since, for example, if $z = u + iv$ is a complex number then:

$$zj = (u + iv)j = uj + vij = uj - ji = j(u - iv) = jz^*.$$

This explains the sense in which this algebra over $\mathbb{C}^2$ is not really a complex algebra, since one finds that the product is $\mathbb{R}$-bilinear, but not $\mathbb{C}$-bilinear, namely:

$$
\begin{aligned}
(z^1 + z^2 j)(w^1 + w^2 j) &= z^1 w^1 + z^2 j\, w^1 + z^1 w^2 j + z^2 j\, w^2 j \\
&= z^1 w^1 + z^2\, w^{1*} j + z^1 w^2 j + z^2 w^{2*} jj \\
&= (z^1 w^1 - z^2 w^{2*}) + (z^1 w^2 + z^2\, w^{1*})\, j.
\end{aligned}
$$

If the product were $\mathbb{C}$-bilinear then the complex conjugates would not appear. However, complex conjugation is still $\mathbb{R}$-linear, so the product is $\mathbb{R}$-bilinear. This situation is closely related to the fact that $SU(2)$ is not a complex Lie group, although it is defined in terms of complex 2×2 matrices, since its real dimension – viz., 3 – is not even.

In order to be consistent with the real case, one defines the conjugate of $q$ as:

$$\overline{q} = z^{1*} - z^2 j. \tag{3.23}$$

This makes:

$$\| q \|^2 = q\overline{q} = z^1 z^{1*} + z^2 z^{2*}, \tag{3.24}$$

which then defines a Hermitian scalar product on $\mathbb{H}$. In particular, $\| q \|^2$ is still a real number.

The inverse of $q$ is still $\overline{q} / \| q \|^2$, although the definition of conjugate and norm-squared have changed. Similarly, one can still define unit quaternions by $\| q \| = 1$, although the explicit form for $\| q \|$ is Hermitian complex, now.

Although we have put $j$ to the right of the complex component $z^2$ in the above expressions, one can also represent quaternions by putting $i$ to the left and using $j$ as the imaginary unit, as well:

$$q = q^0 + q^1 i + q^2 j + q^3 ij = (q^0 + q^2 j) + i(q^1 + q^3 j) = z^1 + i\, z^2, \tag{3.25}$$

in which:

$$z^1 = q^0 + q^2 j, \qquad z^2 = q^1 + q^3 j, \tag{3.26}$$

this time.

One still has $iz = z^* i$, so the product of two quaternions takes the form:

$$zw = (z^1 + i\, z^2)(w^1 + i\, w^2) = z^1 w^1 - z^2 w^{1*} + i(z^{1*} w^2 + z^2 w^1). \tag{3.27}$$

The conjugate of $q$ takes the form:

$$\overline{q} = z^{1*} - iz^2, \tag{3.28}$$

which differs from the previous expression only by the placement of the $i$. Thus, the inverse of a non-zero quaternion still has the same form, as does the product $q\overline{q} = \| q \|^2$.

The distinction between these two ways of representing a real quaternion as a pair of complex numbers will become essential when we discuss the action of $SU(2)$ on Pauli spinors in the next section. For now, we observe that the pair $(z^1, z^2)$ of complex numbers that we associated with the tetrad $(q^0, ..., q^3)$ of real numbers that define a real quaternion are essentially two of the four Cayley-Klein parameters that one associates with the four Euler parameters in order to represent a rotation, as discussed in the first section of this chapter.

**3. The action of rotations on quaternions**. We shall first prove that the Lie group $Q_1$ is isomorphic to $SU(2)$ and then go on to show how one can isometrically represent vectors and orthonormal frames in $E^3$ by quaternions of vector type in such a way that a certain action of unit quaternions on the latter space becomes two-to-one equivalent to the action of rotations on $E^3$. We will discuss the representation of spin frames and Pauli spinors by real quaternions, shortly.

The algebra $\mathbb{H}$ can be represented isomorphically as a real subspace of the complex algebra $M(2; \mathbb{C})$ by simply defining the association of basis elements.

One defines a basis $\{ \tau_\mu , \mu = 0, \ldots, 3 \}$ for the complex four-dimensional vector space $M(2; \mathbb{C})$ by way of:

$$\tau_0 = \begin{bmatrix} 1 & 0 \\ 0 & 1 \end{bmatrix}, \qquad \tau_1 = i \begin{bmatrix} 1 & 0 \\ 0 & -1 \end{bmatrix}, \quad \tau_2 = \begin{bmatrix} 0 & 1 \\ -1 & 0 \end{bmatrix}, \quad \tau_3 = i \begin{bmatrix} 0 & 1 \\ 1 & 0 \end{bmatrix}, \qquad (4.1)$$

which are then seen to verify the multiplication table:

$$\tau_0 \tau_\mu = \tau_\mu \tau_0 = \tau_\mu, \qquad \tau_i \tau_j = - \delta_{ij} \tau_0 + \varepsilon_{ijk} \tau_j \tau_k . \qquad (4.2)$$

This is formally identical to (3.1), so the linear map defined by taking $\mathbf{e}_\mu$ to $\tau_\mu$ and extending to the other elements by $\mathbb{R}$-linearity represents $\mathbb{H}$ isomorphically as a real subalgebra of the complex algebra $M(2; \mathbb{C})$. The typical quaternion $q^\mu \mathbf{e}_\mu$ then goes to the matrix:

$$[q] = q^\mu \, \tau_\mu = \begin{bmatrix} q^0 + iq^1 & q^2 + iq^3 \\ -q^2 + iq^3 & q^0 - iq^1 \end{bmatrix} = q^0 \tau_0 + \begin{bmatrix} iq^1 & q^2 + iq^3 \\ -q^2 + iq^3 & -iq^1 \end{bmatrix}. \quad (4.3)$$

The $\tau$ matrices have the property that $\tau_0$ represents the unity of the algebra and the other three matrices $\tau_i$, $i = 1, 2, 3$ are anti-Hermitian:

$$\tau_i^\dagger = - \tau_i . \qquad (4.4)$$

In fact, they relate to the usual Pauli $\sigma$ matrices, which are Hermitian, by the rule:

$$\tau_1 = i\sigma_3, \qquad \tau_2 = i\sigma_2, \qquad \tau_3 = i\sigma_1, \qquad (4.5)$$

which also involves a permutation of the axes. The reason that we shall use anti-Hermitian matrices, instead of Hermitian ones, is that the Lie algebra that is generated by the $\tau_i$ is $\mathfrak{su}(2)$, which then represents infinitesimal rotations directly, and when we extend this to $\mathfrak{sl}(2; \mathbb{C})$ by adding the Hermitian matrices as infinitesimal boosts, it will not be so confusing as to what roles are played by the two types of matrices.

The matrix that represents the conjugate of $q$ takes the form:

$$[\overline{q}] = \begin{bmatrix} q^0 - iq^1 & -q^2 - iq^3 \\ q^2 - iq^3 & q^0 + iq^1 \end{bmatrix} = q^0 \tau_0 - \begin{bmatrix} iq^1 & q^2 + iq^3 \\ -q^2 + iq^3 & -iq^1 \end{bmatrix} = [q]^\dagger. \quad (4.6)$$

Thus, conjugation corresponds to the Hermitian adjoint operation in this real case.

The determinant of $[q]$ is:

$$\det [q] = \| q \|^2, \quad (4.7)$$

and one sees that the multiplicative group $Q^*$ of non-zero quaternions then corresponds to a real subgroup $GL(2; \mathbb{C})$ that is isomorphic to $GL(4; \mathbb{R})$ and the subgroup $Q_1$ of unit quaternions then corresponds to a real subgroup of $GL(2; \mathbb{C})$ that is isomorphic to $SU(2)$, since the inverse of the matrix $[q]$ for any unit quaternion $q$ is

$$[q]^{-1} = \begin{bmatrix} q^0 - iq^1 & -q^2 - iq^3 \\ q^2 - iq^3 & q^0 + iq^1 \end{bmatrix} = [q]^\dagger. \quad (4.8)$$

This association of unit quaternions with elements of $SU(2)$ gives a concise way of showing that the manifold of the Lie group $SU(2)$ is diffeomorphic to a real 3-sphere.

When one defines $\mathbb{H}$ as a real algebra over $\mathbb{C}^2$, as we did at the end of the last section, the association of a quaternion $q = z^1 + z^2 j$ with a 2×2 complex matrix becomes:

$$[q] = \begin{bmatrix} z^1 & z^2 \\ -z^{2*} & z^{1*} \end{bmatrix}. \quad (4.9)$$

We recognize that this amounts to the association of the complex Cayley-Klein parameters, which are now $z^1$ and $z^2$, with the four real Euler parameters, which are now the components $q^\mu$ of a real quaternion, before one imposes the unitarity constraint.

The determinant of this matrix then becomes:

$$\det [q] = z^1 z^{1*} + z^2 z^{2*} = \| q \|^2, \quad (4.10)$$

and:

$$[\overline{q}] = \begin{bmatrix} z^{1*} & -z^2 \\ z^{2*} & z^1 \end{bmatrix} = [q]^\dagger. \quad (4.11)$$

This once more shows the Hermitian nature of $\mathbb{H}$ in this formulation.

Since the matrices $[q^i \tau_i]$ that represent pure quaternions define a Lie algebra under Lie bracket that is isomorphic to $\mathfrak{su}(2)$, which is, in turn, isomorphic to $\mathfrak{so}(3; \mathbb{R})$, one sees that the Lie algebra of infinitesimal rotations can be represented by the Lie algebra of pure quaternions.

Having established the isomorphism of the Lie group $Q_1$ with $SU(2)$, one then defines an action of $Q_1$ on $\mathbb{H}$ as follows:

$$Q_1 \times \mathbb{H} \to \mathbb{H}, \ (u, q) \mapsto uq\overline{u} \ . \tag{4.12}$$

This action has the spaces $S\mathbb{H}$ and $V\mathbb{H}$ of quaternions of scalar and vector type, respectively, as invariant subspaces. Clearly, if $q$ is a scalar then $q$ goes to itself under the above action, so the action is trivial on scalars. However, if $q$ is a vector then $\overline{q} = -q$, which means that:

$$\overline{uq\overline{u}} \ = u \, \overline{q} \, \overline{u} = - \, uq\overline{u} \ ,$$

so $q' = uq\overline{u}$ is also a vector. If two vectors $\mathbf{x}$ and $\mathbf{y}$ go to the same vector under the action of $u$ then $u\mathbf{x}\overline{u} = u\mathbf{y}\overline{u}$, which makes $\mathbf{x} = \mathbf{y}$, by cancellation on the left and right, so the action is injective. Similarly, it is surjective, since the equation $\mathbf{x}' = u\mathbf{x}\overline{u}$ can be solved by way of $\mathbf{x} = \overline{u}\mathbf{x}'u$ for any $\mathbf{x}'$. It is also linear since:

$$u(\alpha\mathbf{x} + \beta\mathbf{y})\overline{u} = \alpha(u\mathbf{x}\overline{u}) + \beta(u\mathbf{y}\overline{u}) \ .$$

One notes that the restriction of scalar product $<\cdot,\cdot>$ on quaternions to the three-dimensional vector space $V\mathbb{H}$ of pure quaternions, namely:

$$<\mathbf{x}, \mathbf{y}> = S(\mathbf{x}\overline{\mathbf{y}}) = \delta_{ij} x^i y^j, \tag{4.13}$$

makes $V\mathbb{H}$ isometric to $E^3$. The action in question is then an isometry of that Euclidian structure, since if $q' = uq$ then:

$$q'\overline{q'} \ = uq\overline{u} \, u\overline{q} \, \overline{u} = u \, q\overline{q} \, \overline{u} \ = u\overline{u} \cdot q\overline{q} = q\overline{q} \ .$$

If $\mathbf{e}_i$ is an orthonormal frame in $V\mathbb{H}$ then the action of a unit quaternion $u$ on $\mathbf{e}_i$ can also be defined by a matrix $R_i^{\ j}$ by way of:

$$u\mathbf{e}_i\overline{u} = \mathbf{e}_j R_i^{\ j} \ . \tag{4.14}$$

Since the action is an isometry, the matrix $R_i^{\ j}$ is orthogonal. It also becomes clear that both $u$ and $-u$ can be associated with the same rotation matrix $R_i^{\ j}$, due to the quadratic nature of the action of the unit quaternions on vectors.

The pairs of unit quaternions $\{q, -q\}$ that are, in fact, antipodal points on the unit 3-sphere in $\mathbb{H}$. The line through the origin of $\mathbb{R}^4$ that connects the two antipodal points can then represent the rotation in $SO(3)$, which is, in fact, diffeomorphic to $\mathbb{R}P^3$ as a manifold. Thus, in a sense, the components of quaternions behave like the homogeneous coordinates of points in $\mathbb{R}\mathrm{P}^3$, and the two-fold covering map $SU(2) \rightarrow SO(3)$ can be regarded as the association of $\{u, -u\}$ to $R_i^{\ j}$, and topologically this is the association of antipodal points on a real 3-sphere to points in $\mathbb{R}\mathrm{P}^3$ by way of the line through the antipodal points.

One can show that the association of $u$ with $R$ is an order-reversing homomorphism by noting that the action of $uu'$ on $\mathbf{x}$ is:

$$uu'\mathbf{x}\overline{uu'} = u(u'\mathbf{x}\overline{u'})\overline{u} = u(\mathbf{x}R')\overline{u} = \mathbf{x}\,R'R.$$

If one puts $u$ into polar form then the action (4.12) on vectors $\mathbf{v}$ takes the form:

$$(\cos\tfrac{1}{2}\theta + \sin\tfrac{1}{2}\theta\ \mathbf{u})\,\mathbf{v}\,(\cos\tfrac{1}{2}\theta - \sin\tfrac{1}{2}\theta\ \mathbf{u})$$

$$= \cos\theta\,\mathbf{v} + (1 - \cos\theta)<\mathbf{u},\mathbf{v}>\mathbf{u} + \sin\theta\,\mathbf{u}\times\mathbf{v}, \qquad (4.15)$$

which is again Rodrigues's formula.

In summation, if $E^3$ is represented isometrically by the vector space $V\mathbb{H}$, when it is given the scalar product that is the restriction of the one defined on $\mathbb{H}$, then the action of $Q_1 \cong SU(2)$ on $V\mathbb{H}$ that was defined in (4.12) is isometric and maps to the action of $SO(3)$ on $E^3$ by right matrix multiplication in such a way that antipodal unit quaternions go to the same proper rotation.

One can also use the left and right multiplication of a quaternion $q$ by a unit quaternion $u$, when expressed in the complex form, to account for $SU(2)$ spinors. However, one finds that in order to get the most intuitively appealing results, one must represent real quaternions in the form $z^1 + iz^2$ when dealing with left multiplication and in the form $z^1 + z^2 j$ when dealing with right multiplication.

Namely, since $\{1, i\}$ defines a $\mathbb{C}$-basis for $\mathbb{C}^2$, which we temporarily denote by $\{\mathbf{c}_1, \mathbf{c}_2\}$ and left multiplication by a unit quaternion $u$ is an invertible $\mathbb{R}$-linear map, the image of the basis $\mathbf{c}_a$ by $L(u)$ can be expressed in terms of $\mathbf{c}_a$ by means of an invertible 2×2 complex matrix:

$$u\mathbf{c}_a = \mathbf{c}_b[L(u)]_a^b. \qquad (4.16)$$

One can derive an explicit expression for the matrix $[L(u)]_a^b$ by considering the expression for the product $uq$, although one must multiply the unit $i$ on the *left* in order to represent the quaternions. Thus, if $u = u^1 + iu^2$ and $q = q^1 + iq^2$ then:

$$uq = u^1 q^1 - u^{2*} q^2 + i(u^{1*} q^2 + u^2 q^1). \tag{4.17}$$

This makes:

$$[L(u)]_a^b = \begin{bmatrix} u^1 & -u^{2*} \\ u^2 & u^{1*} \end{bmatrix}, \tag{4.18}$$

which is a matrix in $SU(2)$.

If $q' = L(u)q = uq$ then:

$$q' \overline{q'} = uq\overline{q}\,\overline{u} = q\overline{q}, \tag{4.19}$$

so left multiplication is an isometry of the Hermitian structure, which is consistent with our previous statement about $[L(u)]_a^b$.

Therefore, if one represents an element of $\mathbb{C}^2$, such as an $SU(2)$ spinor, by $q$ in its complex form and an element of $SU(2)$ by a unit quaternion $u$ then the left action $qu$ corresponds to the multiplication of a matrix in $SU(2)$ times the element of $\mathbb{C}^2$.

Similar statements are true for right multiplication, which defines a matrix $[R(u)]_a^b$ by way of:

$$\mathbf{c}_a\, u = \mathbf{c}_b [R(u)]_a^b, \tag{4.20}$$

and this matrix is also an element of $SU(2)$. However, this time, the basis $\mathbf{c}_a$ refers to $\{1, j\}$.

From considering the product $qu$, when one multiplies the $j$ on the *right*, this time:

$$qu = q^1 u^1 - q^2 u^{2*} + (q^1 u^2 + q^2 u^{1*})\, j, \tag{4.21}$$

one derives:

$$[R(u)]_a^b = \begin{bmatrix} u^1 & u^2 \\ -u^{2*} & u^{1*} \end{bmatrix}, \tag{4.22}$$

which is seen to be a unitary matrix with determinant 1, as well as the transpose of the matrix $[L(u)]_a^b$, which one sees immediately from comparing (4.22) to (4.18)

However, the right multiplication of a quaternion $q$ by $u$ now corresponds to the right multiplication of a row vector in $\mathbb{C}^{2*}$ by the matrix $[R(u)]_a^b$. Thus, left and right multiplication of a quaternion $q$ by a unit quaternion $u$ correspond to the action of $SU(2)$ on $\mathbb{C}^2$ vectors and covectors, respectively, as long as one represents a $\mathbb{C}^2$ vector $(z^1, z^2)$ in the form $z^1 + jz^2$ and a covector in $\mathbb{C}^{2*}$ in the form $z^1 + z^2 j$.

If one takes the tensor product $z^a w_b$ of a $\mathbb{C}^2$ vector and a $\mathbb{C}^2$ covector then one obtains a second-rank mixed tensor whose components define a complex 2×2 matrix, which can also represent a vector in $\mathbb{R}^3$. The simultaneous left action of $u$ and right action of $\overline{u}$ on that tensor then corresponds to the conjugation of its component matrix by the $SU(2)$ matrices that represent $u$ and $\overline{u}$, which is the usual action that one encounters in spinor algebra for vectors.

### 4. The kinematics of fixed-point rigid bodies [7].

The mechanics of fixed-point rigid bodies (i.e., tops) is usually first treated as being based in differentiable curves in $SO(3)$. That is because the assumption of rigidity allows one to represent the state of a rigid body at each point in time by some chosen orthonormal frame at a chosen point.

For instance, flight mechanics often defines a body frame that is centered at the center of mass of the vehicle and has a positive $x$-axis along the longitudinal axis, pointing forward, a positive $y$-axis that points out the left wing, and a $z$-axis that points vertically upward; one might invert for the $z$ and $z$ directions for some purposes. We shall use the term *attitude* that is used in that discipline to refer to the angular position of a rigid body, rather than *orientation*, which has an unrelated, but established, usage in the context of frames.

The assumption of a fixed point for the rotational motion then means that there is no translation of the frame through the course of time. All of the other possible attitudes of the rigid body are then in one-to-one correspondence with points of $SO(3)$. Hence, when one initial frame is chosen, the time evolution of frames in the rigid body can be associated with a curve $R: \mathbb{R} \to SO(3)$, $t \mapsto R(t)$, such that $R(0) = I$, since the initial frame is the reference for the other ones.

If one assumes sufficient differentiability conditions for the curve $R(t)$ then the first time derivative $\dot{R}(t)$ can be regarded as the angular velocity of the motion relative to the initial frame, which then represents the angular velocity in an "inertial" frame. If one translates the tangent vector $\dot{R}(t)$ in the tangent space to $SO(3)$ at $R(t)$ back to the identity $I$ then one gets a corresponding element [5]):

$$\omega(t) = \dot{R}(t)\, R(t)^{-1} \tag{5.1}$$

of $T_I SO(3)$, which represents the angular velocity in a co-moving – i.e., non-inertial – frame. Since $T_I SO(3)$ can be identified with the Lie algebra $\mathfrak{so}(3)$, when it is defined by the right-invariant vector fields on $SO(3)$, one can think of angular velocity as a curve in $\mathfrak{so}(3)$.

If one goes to the next time derivative $\ddot{R}(t)$ then that takes the form of a curve in $T^2 SO(3)$ that takes its value in the vector space $T_{\dot{R}(t)} T_{R(t)} SO(3)$ at each $t$. This curve then represents the angular acceleration of the motion relative to the initial frame, which must be erected in both the tangent space $T_{R(t)} SO(3)$ and the tangent space to the point $\dot{R}(t)$ in that space, as well.

In order to get the angular acceleration relative to the co-moving frame, one can differentiate (5.1):

---

$$\frac{d\omega}{dt} = \ddot{R}R^{-1} + \dot{R}\dot{R}^{-1} = \ddot{R}R^{-1} - \dot{R}R^{-1}\dot{R}R^{-1} = \ddot{R}R^{-1} - \omega\omega,$$

and define the angular acceleration relative to the co-moving frame by:

$$\alpha(t) = \ddot{R}R^{-1} = \frac{d\omega}{dt} + \omega\omega. \qquad (5.2)$$

This then defines a curve in $T_\omega T_I SO(3; \mathbb{R}) = T_\omega \mathfrak{so}(3; \mathbb{R})$.

Higher derivatives of $R(t)$ can then be right-translated back to the identity to give the corresponding values in the co-moving frame.

The kinematical state of the rigid body relative to the initial frame can then be described by the "$k$-jet;"

$$j^k(t) = (t, R(t), \dot{R}(t), \ldots, \overset{(k)}{R}(t)),$$

while the kinematical state relative to the co-moving frame can take the form:

$$j^k(t)R(t)^{-1} = (t, I, \dot{R}(t)R(t)^{-1}, \ldots, \overset{(k)}{R}(t)R(t)^{-1}) = (t, I, \alpha(t), \ldots, \overset{(k-1)}{\omega}(t)).$$

In order to represent the kinematical state of a rigid body by quaternions, one maps the vector space $\mathbb{R}^3$ to the vector space $V\mathbb{H}$ of pure quaternions by taking every vector $\mathbf{x} = (x^1, x^2, x^3)$ to the quaternion $x = x^i \mathbf{e}_i$. Thus, every frame $\mathbf{f}_i = \mathbf{e}_i R$ in $\mathbb{R}^3$ becomes a frame $\mathbf{f}_i$ in $V\mathbb{H}$. Similarly, every sufficiently differentiable curve $\mathbf{x}(t)$ in $\mathbb{R}^3$ becomes a sufficiently differentiable curve $x(t) = x^i(t) \mathbf{e}_i$ in $V\mathbb{H}$, so every moving frame $\mathbf{f}_i(t)$ in $\mathbb{R}^3$ also becomes a moving frame $f_i(t)$ in $V\mathbb{H}$.

However, since the action of the rotation group on pure quaternions is by conjugation, not right-multiplication, in order to represent the frame $\mathbf{f}_i$ in terms of $\mathbf{e}_i$ one must first represent the rotation matrix $R$ by a unit quaternion $q$, and then let $q$ act on $\mathbf{e}_i$ by conjugation:

$$\mathbf{f}_i = q\mathbf{e}_i\overline{q}.$$

In order to represent a moving frame $\mathbf{f}_i(t)$, one then represents the curve $R(t)$ as a curve $q(t)$ in $Q_1$:

$$\mathbf{f}_i(t) = q(t)\mathbf{e}_i\overline{q}(t).$$

One then differentiates this curve directly to get:

$$\mathbf{v}_i \equiv \frac{d\mathbf{f}_i}{dt} = \dot{q}\mathbf{e}_i\overline{q} + q\mathbf{e}_i\dot{\overline{q}}. \qquad (5.3)$$

This system of equations expresses the time derivative of the rotation of the frame in terms of the original frame $\mathbf{e}_i$, which is then the way that things appear in an inertial frame. In the co-moving frame, one substitutes $\mathbf{e}_i = \overline{q}\mathbf{f}_i q$ and gets:

$$\frac{d\mathbf{f}_i}{dt} = \dot{q}\overline{q}\mathbf{f}_i + \mathbf{f}_i q\dot{\overline{q}} = \dot{q}\overline{q}\mathbf{f}_i - \mathbf{f}_i \dot{q}\overline{q} \, ,$$

in which we have used the fact that since $q\overline{q} = 1$, one gets $\dot{q}\overline{q} = -q\dot{\overline{q}}$ by differentiation.

If one introduces the quaternion:

$$\omega = \dot{q}\overline{q} \, , \tag{5.4}$$

which is analogous to (5.1), to play the role of angular velocity then we can summarize the previous formula as:

$$\frac{d\mathbf{f}_i}{dt} = [\omega, \mathbf{f}_i] = 2\omega \times \mathbf{f}_i \, . \tag{5.5}$$

One again, there is a factor of 2 compared to the Euclidian expression.

If one differentiates (5.3) again then one gets the acceleration of the moving frame relative to the initial frame as:

$$\mathbf{a}_i \equiv \frac{d\mathbf{v}_i}{dt} = \ddot{q}\mathbf{e}_i\overline{q} + 2\dot{q}\mathbf{e}_i\dot{\overline{q}} + q\mathbf{e}_i\ddot{\overline{q}} \, . \tag{5.6}$$

When one substitutes for $\mathbf{e}_i$ one then gets the corresponding expression relative to the co-moving frame:

$$\mathbf{a}_i = \ddot{q}\overline{q}\mathbf{e}_i + 2\dot{q}\overline{q}\mathbf{e}_i q\dot{\overline{q}} + \mathbf{e}_i q\ddot{\overline{q}} = \ddot{q}\overline{q}\mathbf{e}_i - 2\dot{q}\overline{q}\mathbf{e}_i\dot{q}\overline{q} + \mathbf{e}_i q\ddot{\overline{q}} \, .$$

If one differentiates (5.4) then one sees that:

$$\dot{\omega} = \ddot{q}\,\overline{q} + \dot{q}\,\dot{\overline{q}} = \ddot{q}\,\overline{q} + \|\dot{q}\|^2 \, .$$

One also finds that:

$$\|\omega\|^2 = \dot{q}\overline{q}q\dot{\overline{q}} = \|\dot{q}\|^2 \, ,$$

which makes:

$$\ddot{q}\,\overline{q} = \dot{\omega} - \|\omega\|^2 .$$

After substituting this in the last equation for $\mathbf{a}_i$, one gets:

$$\mathbf{a}_i = -\|\omega\|^2 \mathbf{e}_i - <\omega, \mathbf{e}_i> \omega + [\dot{\omega}, \mathbf{e}_i] \tag{5.7}$$

for the acceleration of the moving frame relative to itself.

When one goes on to the spin representation of Euclidian rotations, one sees that there are two ways of representing the motion of spin frames: by elements of $SU(2)$ and by unit vectors in $\mathbb{C}^2$, when it is given the Hermitian inner product, which we call $H^2$.

When one chooses a reference spin frame on $H^2$, any other spin frame can be described by a unique matrix in $SU(2)$. Thus, a time-parameterized family of spin frames is just a sufficiently-differentiable curve $U(t)$ in $SU(2)$. We shall then denote its first and second derivatives with respect to $t$ by $\dot{U}$ and $\ddot{U}$, respectively.

Of course, these matrices represent the time evolution of the frame with respect to – say – the initial frame, which amounts to the inertial description of the motion. If one wishes to describe it with respect to the moving spin frame $U(t)$ then one must right-translate $U(t)$ back to the identity matrix by means of $U(t)^{\dagger}$, and in so doing, translate $\dot{U}$ back to a tangent vector to $I$, namely:

$$\Omega = \dot{U}U^{\dagger},$$

and $\ddot{U}$, to a tangent vector to $(I, \Omega)$:

$$\alpha = \ddot{U}U^{\dagger}.$$

Thus, $\Omega$ is an element of the Lie algebra $\mathfrak{su}(2)$ of infinitesimal generators of special unitary transformations, while $\alpha$ is tangent to that vector space at $\Omega$.

If one differentiates $\Omega$ with respect to $t$, one gets:

$$\dot{\Omega} = \ddot{U}U^{\dagger} + \dot{U}\dot{U}^{\dagger} = \ddot{U}U^{\dagger} + \Omega U U^{\dagger}\Omega^{\dagger} = \ddot{U}U^{\dagger} - \Omega\Omega,$$

since $\Omega$ is skew-Hermitian; i.e.:

$$\alpha = \dot{\Omega} + \Omega\Omega.$$

Of course, when one associates the 2×2 skew-Hermitian matrix $\Omega$ with a 3×3 real orthogonal matrix $R$, a factor of 1/2 must be introduced into $\Omega$ in order to account for the double-valuedness of the covering map, which makes the rotational angle $\theta$ in $E^3$ correspond to the angle $\theta/2$ in $H^2$, and therefore, under differentiation, the angular velocity $\omega$ and angular acceleration $\alpha$ in $E^3$ will correspond to the angular velocity $\omega/2$ and $\alpha/2$ in $H^2$, as well.

The non-relativistic description of spinning matter was introduced into quantum mechanics by Wolfgang Pauli [**8**] in order to properly account for the spin of an electron in the Schrödinger equation, as demanded by the Uhlenbeck-Goudsmit hypothesis that the newly-discovered magnetic dipole moment of the electron might be proportional to an intrinsic angular moment – or "spin" – that the electron possessed. Of course, various theoreticians emphasized that this spin did not have to be represent an actual kinematical state of the electron, such as the way that the Earth rotates about its own axis while orbiting around the Sun. Indeed, nowadays, spin is seen to be more related to the dimension of the space in which the quantum wave function takes its values.

Pauli's modification of the Schrödinger equation involved first replacing the wave function that took its values in $\mathbb{C}$ with one that took its values in $\mathbb{C}^2$, which one then called a *Pauli spinor*. Just as the real Euclidian rotation group acts on rigid bodies in $E^3$, its covering group $SU(2)$ acts on rigid bodies when they are represented in $H^2$.

We have already observed in the first section of this chapter that the attitude of a rigid body can be equivalently described by a matrix in $SU(2)$ or a column vector in $H^2$ with unit Hermitian norm. We have also pointed out that $\mathbb{H}$ can be re-organized into a real algebra over $\mathbb{C}^2$. Thus, we see that since $Q_1$ is isomorphic to $SU(2)$ and the unit sphere in $H^2$, the multiplication of two real unit quaternions also describes the kinematics of a Pauli spinor when one assumes that its time evolution can be described by the action of a sufficiently-differentiable curve in $SU(2)$ on an initial spinor $\psi_0$ in $H^2$:

$$\psi(t) = U(t)\,\psi_0\,.$$

Thus, by successive differentiations, we get:

$$\dot{\psi} = \dot{U}\psi_0 = \Omega\,\psi, \qquad \ddot{\psi} = \ddot{U}\psi_0 = \alpha\,\psi,$$

with $\Omega$ and $\alpha$ defined as above in the case of $SU(2)$.

The interpretation of the Pauli equation for the spinning electron in terms of the so-called "hydrodynamical" intepretation of wave mechanics, which went back to Madelung, and was expanded upon by Takabayasi, Schönberg, and others, was discussed in the pair of papers by Bohm, Schiller, and Tiomno [**9**]. This interpretation of wave mechanics is not the same thing as examining its classical limit, since one does not take $h$ to zero in the process, but only gives the complex wave equations of quantum theory a real tensorial form. This is, in fact, the process by which one associates physical observables with the wave function when one wishes to represent those observables as tensor fields of various ranks, instead of Hermitian operators. The basic construction that emerges – perhaps by starting with the field Lagrangian and deriving the Noether currents that follow from the basic symmetries of the action functional – is that of *bilinear covariants*, which we now briefly discuss.

When one represents a Pauli spinor as $\psi = [z^1, z^2]^{\mathrm{T}}$, the basic physical observables generally follow from considering expressions of the form $\psi^{\dagger}\sigma^{\mu}\psi$, where $\sigma^0 = I$ and the $\sigma^i$, $i = 1, 2, 3$ are the Pauli matrices, with their indices raised using the Euclidian metric. One thus obtains four real functions from the four real functions $q^{\mu}$, $\mu = 0, \ldots, 3$ that define $z^1$ and $z^2$. A straightforward calculation gives:

$$\psi^{\dagger}\psi = z^1 z^{1*} + z^2 z^{2*} = \sum_{\mu}(q^{\mu})^2\,, \tag{5.8}$$

$$\psi^{\dagger}\sigma^1\psi = z^{1*}z^2 + z^{2*}z^1 = 2(q^0\,q^2 + q^1\,q^3), \tag{5.9}$$

$$\psi^{\dagger}\sigma^2\psi = i(z^1 z^{2*} - z^2 z^{1*}) = 2(q^0\,q^2 + q^1\,q^3), \tag{5.10}$$

$$\psi^{\dagger}\sigma^3\psi = z^1 z^{1*} - z^2 z^{2*} = (q^0)^2 + (q^1)^2 - (q^2)^2 - (q^3)^2. \tag{5.11}$$

The first number $\psi^{\dagger}\psi$ that one obtains is usually taken to represent a scalar density, such as a number density, or, when normalized to have unity for its integral over all space, a probability density function. The other three numbers $\psi^{\dagger}\sigma_i\psi$, $i = 1, 2, 3$ are taken to be proportional to the spin density of the particle that is described by $\psi$.

The four scalars can be assembled into a four-dimensional vector $(\psi^\dagger \sigma^\mu \psi)\sigma_\mu$ with real coefficients, and since we know now that the $\sigma_\mu$ represent the basis for $\mathbb{H}$ in a subalgebra of the matrix algebra $M(2; \mathbb{C})$, this suggests that the four-vector $(\psi^\dagger \sigma^\mu \psi)\,\mathbf{e}_\mu$, when the coefficients are expressed in terms of the $q^\mu$, is an element of $\mathbb{H}$.  If we examine the expressions $\overline{q}\mathbf{e}_\mu q$ then we see that:

$$\overline{q}\mathbf{e}_0 q = \overline{q}q = \psi^\dagger \psi, \tag{5.12}$$

$$\overline{q}\mathbf{e}_1 q = (\psi^\dagger \sigma^3 \psi)\,\mathbf{e}_1 + (\psi^\dagger \sigma^2 \psi)\,\mathbf{e}_1 + (\psi^\dagger \sigma^1 \psi)\,\mathbf{e}_1\,. \tag{5.13}$$

If you recall that in (4.5) we were permuting 1 and 3 in our association of the Pauli matrices with the $\tau_i$ that followed most naturally from our quaternion basis then one sees that, in effect, the quaternion $\overline{q}\mathbf{e}_0 q + \overline{q}\mathbf{e}_1 q$ is essentially the one that corresponds to $(\psi^\dagger \sigma^\mu \psi)\sigma_\mu$.  One can evaluate the other products $\overline{q}\mathbf{e}_2 q$ and $\overline{q}\mathbf{e}_3 q$, but all that one finds is expressions for the coefficients that resemble those of $\overline{q}\mathbf{e}_1 q$ with various permutations and sign changes of the components $q^\mu$.  One then presumes that the definition of the Pauli matrices essentially "favors" $\sigma^1$, in that sense, much as one usually singles out the $z$ component of angular momentum in quantum mechanics.

One also obtains physical observables from bilinear expressions that involve the differentials $d\psi$ and $d\psi^\dagger$.  In particular, the Noether current that is associated with the $U(1)$ phase invariance of the Lagrangian is the vector field that corresponds to the 1-form:

$$J = \frac{\hbar}{2mi}(\psi^\dagger d\psi - d\psi^\dagger \psi) = \frac{\hbar}{2mi}(z^{1*}\,dz^1 - z^1\,dz^{1*} + z^{2*}\,dz^2 - z^2\,dz^{2*})\,. \tag{5.14}$$

When we substitute the quaternion component expressions for $z^1$ and $z^2$, the current takes the form:

$$J = \frac{\hbar}{m}(q^0\,dq^1 - q^1\,dq^0 + q^2\,dq^3 - q^3\,dq^2). \tag{5.15}$$

If we form the quaternion expression that corresponds to $\psi^\dagger d\psi - d\psi^\dagger \psi$ then we get:

$$\overline{q}dq - d\overline{q}q = 2(q^0\,dq^i - q^i\,dq^0 + \varepsilon_{ijk}\,q^j\,dq^k)\,\mathbf{e}_i\,. \tag{5.16}$$

Once again, we see that the component of $\mathbf{e}_1$ is the desired expression, while the components of the other two basis elements are obtained by permuting the indices on the spatial components.

# CHAPTER III

# DUAL QUATERNIONS

Now that we have established the basic structure of the real quaternions, in order to discuss the dual quaternions, which belong to the real algebra $\mathbb{H} \otimes \mathbb{D}$, where $\mathbb{D}$ is the algebra of dual numbers, we mostly need to introduce that algebra and then observe what would change as a result of the tensoring operation. We will then see that the effect of introducing the dual numbers as components is to make it possible to represent translations in Euclidian space, as well as rotations, by means of quaternions.

**1. The group of rigid motions.** Since rigid motions are special type of affine transformation, we must now regard $E^3$ as a three-dimensional affine space $A^3$ that has been given a Euclidean scalar product on its tangent spaces. Hence, one can no longer form scalar combinations of points in the space $A^3$ and it has no uniquely-defined origin. One only knows that there is a transitive action of the translation group $(\mathbb{R}^3, +)$ that allows one to associate any pair of points $x, y \in A^3$ with a unique vector $\mathbf{s} \in \mathbb{R}^3$ that one interprets as the *displacement vector from $x$ to $y$*. Therefore the opposite displacement vector from $y$ to $x$ will be $-\mathbf{s}$. One can write this association in either form:

$$y - x = \mathbf{s}, \qquad y = x + \mathbf{s}.$$

One can then regard the action of the translation group as equivalent to a two-point map $s : A^3 \times A^3 \to \mathbb{R}^3$, $(x, y) \mapsto y - x$. This also allows us to define a *position vector field* $\mathbf{x}: A^3 \to \mathbb{R}^3$, $x \mapsto \mathbf{x}(x)$ relative to any choice of reference point $O$ by way of:

$$\mathbf{x}(x) = x - O = s(x, O). \tag{6.1}$$

This map is invertible, and one obtains a global coordinate system on $A^3$ by this means. Hence, as a differentiable manifold, $A^3$ is diffeomorphic to $\mathbb{R}^3$, and one says that the affine space $A^3$ *is modeled on the vector space* $\mathbb{R}^3$. If one chooses a frame in any tangent space $T_x A^3$, and thus a linear isomorphism of $\mathbb{R}^3$ with $T_x A^3$, then one can also say that $A^3$ is modeled on any of its tangent spaces.

One can define a line $[l]$ through a point $x \in A^3$ by the set of a points of the form $x + \alpha l$, where $\alpha$ is a real number and $l$ is a vector in $\mathbb{R}^3$ that defines the direction of the line. Two lines $[l]$ and $[l']$ in $A^3$ are said to be *parallel* iff the points of $[l']$ can be obtained from the points of $[l]$ by translating any point $x$ in $[l]$ to a point $x + \mathbf{s}$ in $[l']$ by means of some displacement vector $\mathbf{s}$ that is the same for all points of $[l]$. One can also say that the differential map $d\tau(\mathbf{s})$ to the map $\tau(\mathbf{s}): A^3 \to A^3$, $x \mapsto x + \mathbf{s}$ parallel-translates vectors in $T_x A^3$ to vectors in $T_{x+s} A^3$, since it is a linear isomorphism, due to the invertibility of the translation of points. The map $\tau(\mathbf{s})$ is then referred to as the *translation map* associated with $\mathbf{s}$; since the addition of vectors is commutative, it unnecessary to specify whether it is right or left translation.

Since the only frames in an affine space are in its tangent spaces, in order to define an *affine frame* $(x, \mathbf{e}_i)$ one must specify the point of application $x$ along with the linear frame $\mathbf{e}_i$ in $T_x A^3$. Thus, an affine frame is really a triple of points in the tangent bundle $T(A^3)$ that project to the same point $x$ and have linearly independent vector parts.

However, since there are no linear combinations that are defined in $A^3$, an affine frame does not give one coordinates for the points of $T(A^3)$ directly. In order to get coordinates for a point $(x, \mathbf{v}) \in T(A^3)$ from a choice of affine frame $(O, \mathbf{e}_i)$, one first parallel-translates the frame $\mathbf{e}_i$ from $O$ to all of the other tangent spaces and thus defines a global parallel frame field $\mathbf{e}_i(x)$.

This construction of $\mathbf{e}_i(x)$ allows one to give the coordinates $x^i$ of any point $x$ and the components $v^i$ of any tangent vector $\mathbf{v} \in T_x(A^3)$ by using the same global frame field for both. Namely, $x = O + x^i \mathbf{e}_i(O)$ and $\mathbf{v} = v^i \mathbf{e}_i(x)$. Thus, a point $(x, \mathbf{v}) \in T(A^3)$ gets mapped to a point $(x^i, v^i) \in \mathbb{R}^3 \times \mathbb{R}^3$ and an affine frame $(x, \mathbf{f}_i)$ gets mapped to a point $(x^i, L_j^i) \in \mathbb{R}^3 \times GL(3; \mathbb{R})$, where:

$$\mathbf{f}_i = \mathbf{e}_j(x) L_i^j. \tag{6.2}$$

An *affine transformation* $\tau: A^3 \to A^3$, $x \mapsto \tau(x)$ is defined to be a map that preserves parallelism. That is, if the lines $[l]$ and $[l']$ are parallel then the lines $\tau([l])$ and $\tau([l'])$ will also be parallel, as well. This condition includes the condition that an affine transformation takes lines to lines, which means that is a *collineation*.

If we choose a reference point $O$, and thus define a position vector field $\mathbf{x}(x)$ relative to $O$, then we can use the invertible map $\mathbf{x}: A^3 \to \mathbb{R}^3$ to define a map $\bar{\tau}: \mathbb{R}^3 \to \mathbb{R}^3$ that has the same effect on the position vectors $\mathbf{x}(x)$ of points in $\mathbb{R}^3$ that $\tau$ has on points in $A^3$. The definition is:

$$\bar{\tau}(\mathbf{x}) = \mathbf{x}(\tau(x)). \tag{6.3}$$

Since $\tau$ is a collineation of $A^3$, $\bar{\tau}$ will be a collineation of $\mathbb{R}^3$ (because a line in $A^3$ will become a line in $\mathbb{R}^3$), and thus, an invertible affine transformation of $\mathbb{R}^3$. The group $A(3; \mathbb{R})$ of invertible affine transformations on $\mathbb{R}^3$ can characterized by the semi-direct product $\mathbb{R}^3 \times_s GL(3; \mathbb{R})$, so every such transformation is described by a pair $(\mathbf{a}, L)$ that consists of a translation $\mathbf{a}$ and an invertible linear map $L$. The composition of two transformations takes the form:

$$(\mathbf{a}, L)(\mathbf{b}, M) = (\mathbf{a} + L\mathbf{b}, LM), \tag{6.4}$$

and the inverse of the element $(\mathbf{a}, L)$ is:

$$(\mathbf{a}, L)^{-1} = (-L\mathbf{a}, L^{-1}). \tag{6.5}$$

The action of $(\mathbf{a}, L)$ on $\mathbb{R}^3$ is simply:

$$(\mathbf{a}, L)\mathbf{v} = \mathbf{a} + L\mathbf{v}. \tag{6.6}$$

It is important to remember that since the association of $\tau$ with $\overline{\tau}$ depended upon the choice of $O$, so will the association of $\tau$ with some pair $(\mathbf{s}, L)$. If one chooses another point $O' = O + \mathbf{d}$ then the new position vector field will be:

$$\mathbf{x}'(x) = x - O' = x - O - \mathbf{d} = \mathbf{x}(x) - \mathbf{d},$$

and one will also have a new definition of $\overline{\tau}$ :

$$\overline{\tau}'(\mathbf{x}'(x)) = \mathbf{x}'(\tau(x)) = \mathbf{x}(\tau(x)) - \mathbf{d} = \overline{\tau}(\mathbf{x}(x)) - \mathbf{d} = \mathbf{s} - \mathbf{d} + L\mathbf{x}(x) = \mathbf{s} - \mathbf{d} + L(\mathbf{x}'(x) + \mathbf{d});$$

i.e.:

$$\overline{\tau}'(\mathbf{x}'(x)) = (\mathbf{s} + (L - I)\mathbf{d}) + L\,\mathbf{x}'(x). \tag{6.7}$$

Thus, if $\overline{\tau}$ gets associated with $(\mathbf{s}, L)$ then $\overline{\tau}'$ will get associated with $(\mathbf{s} + (L - I)\mathbf{d}, L)$. Although the linear part remains unchanged, the translational part changes by more than just $-\mathbf{d}$, namely:

$$\mathbf{s}' = \mathbf{s} + (L - I)\mathbf{d}. \tag{6.8}$$

This fact will be at the basis for our later discussion of Chasles's theorem about rigid motion.

A convenient way of representing $A(3; \mathbb{R})$ by matrices is to regard $\mathbb{R}^3$ as the space of inhomogeneous coordinates for a chart in $\mathbb{R}P^3$ and then embed $\mathbb{R}^3$ in the space $\mathbb{R}^4$ of homogeneous coordinates as the affine hyperplane $(1, x^i)$. Relative to the canonical basis for $\mathbb{R}^4$, the matrix of any affine transformation $(\mathbf{a}, L)$ then takes the form:

$$[(\mathbf{a}, L)] = \begin{bmatrix} 1 & \vdots & 0 \\ \cdots & \vdots & \cdots \\ a^i & \vdots & L^i_j \end{bmatrix}. \tag{6.9}$$

An interesting aspect of projective geometry is the fact that if one regards the linear hyperplane $(0, x^i)$ as the hyperplane at infinity then the translation subgroup of $A(3; \mathbb{R})$ acts trivially on the points at infinity, which is easy to see in this matrix representation when one sets $L^i_j = \delta^i_j$. This fact is related to the fact that parallel lines in projective spaces intersect in a point at infinity, so one can no longer define the parallel translation of lines in projective geometry.

If we introduce the Euclidian scalar product $<.,.>$ into the tangent spaces of $A^3$ then we can further reduce the group of affine transformations to the ones whose differential maps preserve the scalar product:

$$<d\tau_x(\mathbf{v}), d\tau_x(\mathbf{w})> = <\mathbf{v}, \mathbf{w}>. \qquad (6.10)$$

When we look at the image of this in $\mathbb{R}^3$ using some position vector field $\mathbf{x}(x)$, we see that if $\bar{\tau}(\mathbf{v}) = \mathbf{a} + L\mathbf{v}$ then:

$$d\bar{\tau} = L. \qquad (6.11)$$

Since this differential map is independent of the choice of point in $\mathbb{R}^3$, it becomes unnecessary to deal with the differential map and we see that we are simply restricting the linear part $L$ of the affine map $\bar{\tau}$ by:

$$<L\mathbf{v}, L\mathbf{w}> = <\mathbf{v}, \mathbf{w}>; \qquad (6.12)$$

i.e., $L$ must be an orthogonal transformation, or rotation.

In order to deal with physical motions, one must further restrict oneself to only proper rotations, and one sees that one has reduced $A(3; \mathbb{R})$ to the group $ISO(3; \mathbb{R}) = \mathbb{R}^3 \times SO(3; \mathbb{R})$ of *Euclidian spatial rigid motions*. Thus, since it is a subgroup of the semi-direct product $\mathbb{R}^3 \times SO(3; \mathbb{R})$, the products and inverses behave the same way as before.

As a real Lie group, $ISO(3; \mathbb{R})$ is non-Abelian, six-dimensional, and connected, but not simply connected, and the presence of $\mathbb{R}^3$ as a factor makes it non-compact. Since $\mathbb{R}^3$ is a normal subgroup, it also not simple. Its two-to-one simply connected covering group is $\mathbb{R}^3 \times SU(2)$.

A convenient way to represent $ISO(3; \mathbb{R})$ by matrices is to restrict the representation (6.9) accordingly, so the matrix of any rigid motion $(\mathbf{a}, R)$ takes the form:

$$[(\mathbf{a}, R)] = \begin{bmatrix} 1 & 0 \\ a^i & R^i_j \end{bmatrix}. \qquad (6.13)$$

The Lie algebra $\mathfrak{iso}(3; \mathbb{R})$ of $ISO(3; \mathbb{R})$ can be obtained by differentiation of a curve through the identity transformation and is seen to consist of the semi-direct sum $\mathbb{R}^3 \oplus_s$ $\mathfrak{so}(3; \mathbb{R})$. If $\mathbf{v} + \omega$ and $\mathbf{v}' + \omega'$ are two elements of $\mathfrak{iso}(3; \mathbb{R})$ then their Lie bracket is obtained by using bilinearity and the fact that the Lie algebra of $\mathbb{R}^3$ is Abelian:

$$[\mathbf{v} + \omega, \mathbf{v}' + \omega'] = [\mathbf{v}, \omega'] + [\omega, \mathbf{v}'] + [\omega, \omega'], \tag{6.14}$$

and if one represents $\omega$ and $\omega'$ as vectors $\boldsymbol{\omega}$ and $\boldsymbol{\omega}'$, respectively, in $(\mathbb{R}^3, \times)$ by setting $\omega = \text{ad}(\boldsymbol{\omega})$ and $\omega' = \text{ad}(\boldsymbol{\omega}')$ then one can also say that:

$$[\mathbf{v} + \boldsymbol{\omega}, \mathbf{v}' + \boldsymbol{\omega}'] = \mathbf{v} \times \boldsymbol{\omega}' + \boldsymbol{\omega} \times \mathbf{v}' + \boldsymbol{\omega} \times \boldsymbol{\omega}'. \tag{6.15}$$

The matrix representation of $\mathbf{v} + \omega$ that corresponds to (6.13) can also be obtained by differentiation at the identity:

$$[(\mathbf{v}, \omega)] = \begin{bmatrix} 0 & 0 \\ v^i & \omega^i_j \end{bmatrix}, \tag{6.16}$$

in which $\omega^i_j$ is a real, anti-symmetric 3×3 matrix.

We now extend our previous discussion of axes for $SO(3; \mathbb{R})$ to $ISO(3; \mathbb{R})$ by first pointing out that although translations are not actually linear transformations, they can still have eigenvectors. Namely, if one considers the basic definition of an eigenvector for a translation $\mathbf{a}$:

$$\mathbf{v} + \mathbf{a} = \lambda \mathbf{v} \qquad \text{or} \qquad (\lambda - 1)\mathbf{v} = \mathbf{a}$$

then one sees that all that this requires is that $\mathbf{v}$ be collinear with $\mathbf{a}$, while $\lambda$ can take on any real value except 1, unless $\mathbf{a} = 0$. Recall that in order to have an eigenvalue of 1, one must have a fixed point, and non-zero translations act without fixed points.

Now, an *axis* for a rigid motion is still an invariant line in $A^3$, so if $\mathbf{v}$ points in the direction of the line $[l]$ then its image under the rigid motion $(\mathbf{a}, R)$ will also point in that direction; i.e.:

$$\mathbf{a} + R\mathbf{v} = \lambda\mathbf{v}. \tag{6.17}$$

for some real scalar $\lambda$.

If we change to a different reference point $O'$ in $A^3$ then the translation vector $\mathbf{a}$ will become:

$$\mathbf{a}' = \mathbf{a} + (R - I)\mathbf{d}, \tag{6.18}$$

in which $\mathbf{d} = O' - O$.

If $\mathbf{d}$ is a real eigenvector of $R$ then the change of reference point has no effect on the translational part of the rigid motion. The line $[\mathbf{d}]$ would then be a rotational axis for $R$, so already we see that we have singled out a class of lines in $A^3$ that get associated with a rigid motion by way of its rotational part. However, if we consider the translational part then we can also single out a unique line among them. One does this by looking for a translation $\mathbf{d}$ that will make $\mathbf{a}'$ collinear with $\mathbf{d}$:

$$\mathbf{a} + (R - I)\mathbf{d} = \lambda \mathbf{d}.$$

Let us now denote the rotational axis of $R$ by $[l]$ and decompose both $\mathbf{a}$ and $\mathbf{d}$ into sums $\mathbf{a}_\parallel + \mathbf{a}_\perp$ and $\mathbf{d}_\parallel + \mathbf{d}_\perp$ whose component vectors are parallel to $[l]$ and perpendicular to it, respectively. This last equation then becomes two equations:

$$\mathbf{a}_\parallel = \lambda \mathbf{d}_\parallel, \qquad \mathbf{a}_\perp = [(\lambda + 1)I - R]\,\mathbf{d}_\perp\,.$$

The fact that these equations still involve the unspecified parameter $\lambda$ is due to the fact that the solution to our problem is a line, not a point. As long as one chooses a non-zero value for $\lambda$, the first equation is soluble for $\mathbf{d}_\parallel$. Since the second equation actually only involves the planar part of the rotation in the plane perpendicular to the rotational axis, the matrix $[(\lambda + 1)I - R]$ will be invertible and one can also solve for $\mathbf{d}_\perp$. Choosing $\lambda = 1$ gives the simplest solution:

$$\mathbf{d}_\parallel = \mathbf{a}_\parallel, \qquad \mathbf{d}_\perp = [2I - R]^{-1}\,\mathbf{a}_\perp\,. \qquad (6.19)$$

That is:

**Chasles's theorem [1, 2]**:  *Given a choice of reference point $O$ in $A^3$ for which a rigid motion gets associated with the pair* ($\mathbf{a}$, $R$), *one translate to another reference point $O'$ such that the rigid motion takes the form* ($\mathbf{a}_\parallel$, $R(\theta)$), *where the translation $\mathbf{a}_\parallel$ is in the direction of the rotational axis $[l]$ of $R$ that passes through $O'$ and the rotation $R(\theta)$ is about the axis $[l]$.*

A rigid motion is then canonically associated with a *central axis* $[l]$ such that the motion consists of a rotation around the axis and a translation along it. For that reason, Sir Robert Ball referred to the canonical form of a rigid motion as a *screw*. What Plücker, Klein, and Study were calling a "dyname" was then a wrench, while the French used the term "torseur," which got Anglicized to the modern term "torsor," which refers to an element of the dual space to a Lie algebra of infinitesimal motions.

Actually, the concept of a central axis went back further in time to Poinsot, who showed that an analogous (in fact, the same) axis is defined for a finite spatial distribution of force vectors that act on a rigid body. In that case, the central axis had the property that the collective effect of the forces was equivalent to a single force that acted in the direction of the axis and a force moment that acted in a plane perpendicular to the axis. Ball then referred to this configuration of force and moment as a "wrench."

The set of all oriented, orthonormal, affine frames in ($E^3$, $<.,.>$) is also the bundle $SO(E^3)$ of oriented, orthonormal *linear* frames in the tangent spaces. The group $ISO(3;\ \mathbb{R})$ of rigid motions acts on $SO(E^3)$ on the right, *but not as a structure group*. For one thing, the structure group of $SO(E^3)$ is $SO(3;\ \mathbb{R})$, not $ISO(3;\ \mathbb{R})$, and for another, the action of a structure group takes frame at a given point to other frames at that same point. The action is:

$$SO(E^3) \times ISO(3;\ \mathbb{R}) \to SO(E^3),\ ((x,\ \mathbf{e}_i),\ (s^i,\ R^i_j)) \mapsto (x',\ \mathbf{e}'_i),$$

with:

$$x' = x + s^i \mathbf{e}_i, \qquad\qquad \mathbf{e}'_i = \mathbf{e}_j R_i^j. \qquad\qquad (6.20)$$

One sees that the action of $ISO(3; \mathbb{R})$ on $SO(E^3)$ is similar to the action of $ISO(3; \mathbb{R})$ on itself by multiplication. Indeed, one recalls that $SO(E^3)$ is diffeomorphic to $ISO(3; \mathbb{R})$ as a manifold by choosing any orthonormal affine frame $(O, \mathbf{e}_i)$, mapping it to $(0, \delta_j^i)$ and mapping any other orthonormal affine frame $(x, \mathbf{f}_i)$ to the element $(s^i, R_j^i) \in ISO(3; \mathbb{R})$ that makes $x = O + s^i \mathbf{e}_i$ and $\mathbf{f}_i = \mathbf{e}_j R_i^j$. Thus, the right action of $ISO(3; \mathbb{R})$ on both $ISO(3; \mathbb{R})$ and $SO(E^3)$ commutes with that diffeomorphism. One then says that the two actions are *equivariant*. In particular, the diffeomorphism takes orbits of one action to orbits of the other.

**2. The algebra of dual numbers [3-5].** The algebra $\mathbb{D}$ of dual numbers can be regarded as an $\mathbb{R}$-algebra over $\mathbb{R}^2$, in which one notates an order pair $(a, b) \in \mathbb{R}^2$ by $a + \varepsilon b$. The number $a$ in this case is referred to as the *real* part of the dual number $a + \varepsilon b$, while $\varepsilon b$ is the *pure dual* part. Thus, one can regard the set $\{1, \varepsilon\}$ as a basis for $\mathbb{R}^2$.

The Abelian group structure on $\mathbb{D}$ is defined by vector addition of order pairs:

$$(a + \varepsilon b) + (c + \varepsilon d) = (a + c) + \varepsilon(b + d),$$

while the scalar multiplication by a real number $\lambda$ is also defined component-wise:

$$\lambda(a + \varepsilon b) = \lambda a + \varepsilon \lambda b.$$

The multiplication of dual numbers is defined by polynomial multiplication, modulo the condition on $\varepsilon$ that:

$$\varepsilon^2 = 0.$$

Hence, one can give the multiplication table for the basis elements:

$$11 = 1, \qquad 1\varepsilon = \varepsilon 1 = \varepsilon, \qquad \varepsilon\varepsilon = 0.$$

Thus, if $\alpha = a + \varepsilon b$ and $\beta = c + \varepsilon d$ then their product is defined to be:

$$\alpha\beta = ac + \varepsilon(ad + bc),$$

which is easily verified to be associative, commutative, and to have:

$$1 = 1 + \varepsilon(0)$$

for a multiplicative unity.

The fact that the multiplication distributes over addition follows from the definition of that operation and is part of the proof of the bilinearity of the product. The other part is that for any real scalar $\lambda$ and any two dual numbers $\alpha$, $\beta$, one must have:

$$(\lambda\alpha)\beta = \alpha(\lambda\beta) = \lambda(\alpha\beta).$$

Since $\varepsilon\varepsilon = 0$, one already sees that non-zero divisors of zero exist, and, in fact, any product of pure dual numbers will vanish, as well, from bilinearity. Thus, the ring is not an integral domain; i.e., the algebra is not a division algebra.

The two-real-dimensional algebra $\mathbb{D}$ can be represented quite simply by a sub-algebra of the algebra of real 3×3 matrices in a manner that is suggestive of projective geometry. Namely, one takes the dual number $a + \varepsilon b$ to the matrix:

$$[a + \varepsilon b] = \begin{bmatrix} 1 & 0 & 0 \\ 0 & a & 0 \\ b & 0 & 1 \end{bmatrix}.$$

Although $\mathbb{D}$ is not a division algebra, some elements of $\mathbb{D}$ do have multiplicative inverses. Namely, if one defines the *conjugate* to any dual number $\alpha = a + \varepsilon b$ to be:

$$\bar{\alpha} = a - \varepsilon b$$

then one sees that:

$$\alpha\,\bar{\alpha} = a^2.$$

We can then define the modulus-squared of any dual number $\alpha$ to be:

$$|\alpha|^2 = \alpha\,\bar{\alpha} = a^2,$$

and see that $|\alpha| = 0$ iff $a = 0$.

Therefore, if $\alpha$ is not a pure dual number then one can define a multiplicative inverse to $\alpha$ by way of:

$$\alpha^{-1} = \frac{\bar{\alpha}}{|\alpha|^2}.$$

One can then define the multiplicative group $D^*$ of invertible dual numbers when it is given multiplication. It is a non-compact, two-dimensional, real, Abelian, Lie group with two connected components – viz., the dual numbers with positive and negative real parts, respectively. It includes the subgroup $D_1$ of elements with unit modulus, which then all have form:

$$\alpha = \pm 1 + \varepsilon b.$$

Since any invertible dual number $\alpha$ has a non-vanishing modulus that is equal to $| a |$, one can factor the modulus out and express every element of $D^*$ in the form of a positive real number times a dual number of unit modulus:

$$\alpha = | a | \left( \pm 1 + \varepsilon \frac{b}{| a |} \right).$$

Thus, the group $D^*$ is isomorphic to the direct product $\mathbb{R}^* \times D_1$.

**3. Functions of dual numbers [3-5].** When one starts with the simplest functions $f$ : $\mathbb{D} \to \mathbb{D}$, namely, polynomial functions, one sees that if $\underline{x} = x + \varepsilon s$ is any dual number then one sees by induction that:

$$\underline{x}^n = x^n + \varepsilon\, n\, x^{n-1} s = x^n + \varepsilon\, \frac{dx^n}{dx} s. \tag{8.1}$$

Since a polynomial $P[\underline{x}]$ is a (real) linear combination of powers of $\underline{x}$, and differentiation is linear, one can say that for any polynomial function $P[\underline{x}]$ of the dual variable $\underline{x}$, one will have:

$$P[\underline{x}] = P[x] + \varepsilon\, P'[x]\, s. \tag{8.2}$$

By generalization, if one assumes that $f(x)$ is a differentiable function of $x$ then one can simply *define* the function:

$$f(\underline{x}) = f(x) + \varepsilon f'(x)\, s. \tag{8.3}$$

In particular, we will need to know that:

$$\cos(\underline{\theta}) = \cos\theta - \varepsilon\, s\, \sin\theta, \tag{8.4}$$

$$\sin(\underline{\theta}) = \sin\theta + \varepsilon\, s\, \cos\theta. \tag{8.5}$$

This also makes:

$$\cos^2(\underline{\theta}) + \sin^2(\underline{\theta}) = 1, \tag{8.6}$$

$$\cos(2\underline{\theta}) = \cos^2(\underline{\theta}) - \sin^2(\underline{\theta}), \tag{8.7}$$

$$\sin(2\underline{\theta}) = 2\sin(\underline{\theta})\cos(\underline{\theta}), \tag{8.8}$$

as in conventional trigonometry.

**4. Dual linear algebra.** The Cartesian product $\mathbb{D}^n$, whose elements all look like $(\underline{x}^1, \ldots, \underline{x}^n)$ can be treated as either an $\mathbb{R}$-vector space or a $\mathbb{D}$-module, depending upon whether one is considering scalars that come from the field $\mathbb{R}$ or the ring $\mathbb{D}$. The elements of $\mathbb{D}^n$ will be referred to as *dual vectors*.

A *module* over a ring ($^6$) (which is $\mathbb{D}$, in the present case) behaves in many ways like a vector space, except that one must take into account the possible existence of zero divisors and non-invertible elements in the ring. That is, one still has an Abelian group under addition – $\mathbb{D}^n$, in the present case – and the addition of elements is interpreted as (dual) vector addition. Furthermore, the ring $\mathbb{D}$ acts on $\mathbb{D}^n$ in the manner of scalar multiplication by way of the obvious definition:

$$\underline{\alpha}\,(\,\underline{x}^1, \,\ldots, \,\underline{x}^n\,) = (\,\underline{\alpha}\,\underline{x}^1, \,\ldots, \,\underline{\alpha}\,\underline{x}^n\,).$$

Since $\mathbb{D}$ is commutative under multiplication, it is unnecessary to distinguish left from right multiplication. Scalar multiplication has the same properties as it does for a vector space; namely, if $\underline{\alpha}$ and $\underline{\beta}$ are dual numbers, while $\underline{v}$ and $\underline{w}$ are dual vectors, then one always has:

$$1\,\underline{v} = \underline{v}, \qquad 0\,\underline{v} = 0, \qquad \underline{\alpha}\,(\,\underline{\beta}\,\underline{v}\,) = (\underline{\alpha}\,\underline{\beta})\,\underline{v}\,,$$
$$(\underline{\alpha} + \underline{\beta})\,\underline{v} = \underline{\alpha}\,\underline{v} + \underline{\beta}\,\underline{v}, \qquad \underline{\alpha}\,(\,\underline{v} + \underline{w}\,) = \underline{\alpha}\,\underline{v} + \underline{\alpha}\,\underline{w}\,.$$

However, since there are divisors of zero in $\mathbb{D}$, one will also have pairs $\underline{\alpha}\,\underline{v}$ of non-zero scalars and vectors whose product is zero, such as $\underline{\alpha} = \varepsilon\alpha$, $\underline{v} = \varepsilon\mathbf{v}$, where $\alpha$ and $\mathbf{v}$ are real.

Any dual vector $\underline{v}$ can be put into real-plus-pure-dual form $\mathbf{v} + \varepsilon\mathbf{a}$, where $\mathbf{v}$ and $\mathbf{a}$ are real vectors, as can any dual number $\underline{\alpha} = \alpha + \varepsilon\beta$, so we can also express the scalar multiplication in that way:

$$\underline{\alpha}\,\underline{v} = \alpha\mathbf{v} + \varepsilon(\alpha\mathbf{a} + \beta\mathbf{v}).$$

In particular:

$$\alpha\,\underline{v} = \alpha\mathbf{v} + \varepsilon\alpha\mathbf{a}, \qquad (1 + \varepsilon)\,\underline{v} = \mathbf{v} + \varepsilon(\mathbf{a} + \mathbf{v}).$$

Since the Abelian group $\mathbb{D}^n$ has no torsion factors – i.e., there is no non-zero integer $n$ and non-zero $\underline{v}$ such that $n\,\underline{v} = \underline{v} + \ldots + \underline{v}$ ($n$ summands) $= 0$ – the $\mathbb{D}$-module $\mathbb{D}^n$ is *free*. Hence, one can find a *basis* for it, which is a set $\{\,\underline{e}_1, \,\ldots, \,\underline{e}_n\,\}$ of $n$ dual vectors such that any dual vector $\underline{v}$ can be expressed as a linear combination with dual number coefficients:

---

($^6$) See Jacobson [**6**] or MacLane and Birkhoff [**7**] for the general theory of modules, among other references..

$$\underline{\mathbf{v}} = \sum_{i=1}^{n} \underline{v}^i \underline{\mathbf{e}}_i \ .$$

For instance, one has the usual canonical basis, where $\underline{\mathbf{e}}_i = (0, \ldots, 0, 1, 0, \ldots, 0)$, with the 1 in the $i^{\text{th}}$ place.

One must note that the existence of divisors of zero makes the usual definition of linear independence less useful, since there might be linear combinations $\sum_{i=1}^{n} \underline{x}^i \underline{\mathbf{e}}_i$ that go to zero even though not all of the $\underline{x}^i$ are zero. For instance, one might have $(\varepsilon\lambda)(\varepsilon\mathbf{e}_i)$, where $\lambda$ is real and non-zero and $\mathbf{e}_i$ is one of the canonical basis vectors. This does not imply that one cannot have a basis for a dual vector space, only that not as many sets of $n$ dual vectors can be used for that purpose.

A map $\underline{L}: V \to W$, where $V$ is an $n$-dimensional dual vector space and $W$ is an $m$-dimensional one, is said to be $\mathbb{D}$-*linear* if it takes linear combinations to linear combinations:

$$\underline{L}(\underline{\alpha}\,\underline{\mathbf{v}} + \underline{\beta}\,\underline{\mathbf{w}}) = \underline{\alpha}\underline{L}(\underline{\mathbf{v}}) + \underline{\beta}\underline{L}(\underline{\mathbf{w}}) \ .$$

When a basis $\{ \underline{\mathbf{e}}_1, \ldots, \underline{\mathbf{e}}_n \}$ has been chosen for $V$, and another basis $\{ \underline{\mathbf{e}}_1, \ldots, \underline{\mathbf{e}}_n \}$ has been chosen for $W$, any $\mathbb{D}$-linear map $\underline{L}$ can be associated with a *dual matrix* $\underline{L}_i^a$ by the usual process:

$$\underline{L}(\underline{\mathbf{e}}_i) = \underline{\mathbf{f}}_a \underline{L}_i^a \ . \tag{9.1}$$

A dual matrix can also be put into real-plus-pure-dual form:

$$\underline{L}_i^a = L_i^a + \varepsilon A_i^a \ ,$$

where $L_i^a$ and $A_i^a$ are real matrices.

The action of a dual linear map $\underline{L}$ on a dual vector $\underline{\mathbf{v}}$ can also be regarded as an action of the dual matrix $\underline{L}_j^a$ on $\underline{v}^i$ when a basis has been chosen or as an action of that matrix on the basis itself:

$$\underline{L}(\underline{\mathbf{v}}) = (\underline{L}_i^a \underline{v}^i)\underline{\mathbf{f}}_a = \underline{v}^i(\underline{L}_i^a \underline{\mathbf{f}}_a) \ .$$

Hence, one can relate the action of dual linear maps to the multiplication $\underline{L}_i^a \underline{v}^i$ of a dual matrix times a dual column vector of components. In real-plus-pure-dual form this is:

$$\underline{L}_i^a \underline{v}^i = L_i^a v^i + \varepsilon(L_i^a a^i + A_i^a v^i) \ . \tag{9.2}$$

For the sake of clarity, we shall usually omit the matrix indices, since they behave as they do in conventional linear algebra.

The product of two dual matrices $\underline{L}$ and $\underline{M} = M + \varepsilon B$ takes on the real-plus-pure-dual form:

$$\underline{L}\,\underline{M} = LM + \varepsilon(LB + AM). \qquad . \qquad (9.3)$$

The identity transformation $I: : \mathbb{D}^n \to \mathbb{D}^n$, $\underline{\mathbf{v}} \mapsto \underline{\mathbf{v}}$ still has the usual matrix $\delta^i_j$ for any basis, and a dual matrix $\underline{L}$ is invertible iff there is some dual matrix $\underline{\tilde{L}} = \tilde{L} + \varepsilon \tilde{A}$ such that:

$$\underline{L}\,\underline{\tilde{L}} = \underline{\tilde{L}}\,\underline{L} = I.$$

From (9.3), we see that if $\underline{\tilde{L}}$ exists then one must have that $\tilde{L}$ is the inverse of the real matrix $L$ (which must then be invertible) and that:

$$\tilde{A} = -\tilde{L}\,A\,\tilde{L},$$

which will exist as long as $\tilde{L}$ exists, regardless of whether $A$ is invertible.

The invertible $\mathbb{D}$-linear maps form a group, as do the invertible dual matrices, and we refer to either as $GL(n; \mathbb{D})$. That group contains $GL(n; \mathbb{R})$ as a subgroup, by way of the invertible pure real matrices. It also contains a subgroup that consists of the matrices of the form:

$$\underline{L} = I + \varepsilon A.$$

Since the product of two such matrices $\underline{L}$ and $\underline{M} = I + \varepsilon B$ is:

$$\underline{L}\,\underline{M} = I + \varepsilon(A + B),$$

the inverse of such a matrix is of the form:

$$\underline{\tilde{L}} = I - \varepsilon A,$$

which is the dual conjugate of the matrix $\underline{L}$. One then sees that this subgroup is isomorphic to the translation group of $\mathbb{R}^{n^2}$.

Note that the action of pure real matrices is linear, while the action of the latter class of matrices gives an affine transformation of the dual part of any vector:

$$(I + \varepsilon A)(\mathbf{v} + \varepsilon \mathbf{a}) = \mathbf{v} + \varepsilon(\mathbf{a} + A\mathbf{v}).$$

One can define a scalar product $<.,.>$ on $\mathbb{D}^n$ to be a symmetric, $\mathbb{D}$-bilinear functional on $\mathbb{D}^n$ that is non-degenerate in the sense that the map $\mathbb{D}^n \to \mathbb{D}^{n*}$, $\underline{\mathbf{v}} \mapsto <\underline{\mathbf{v}}, .>$ is a $\mathbb{D}$- linear isomorphism. Here, one must remember that $<.,.>$ takes its values in $\mathbb{D}$, and we are defining $\mathbb{D}^{n*}$ to be the dual space to $\mathbb{D}^n$, namely the $\mathbb{D}$-module of all $\mathbb{D}$-linear functionals on $\mathbb{D}^n$. The dual Euclidian scalar product is the one that makes the canonical basis orthonormal:

$$<\mathbf{e}_i, \mathbf{e}_j> = \delta_{ij} .$$

Thus, if two dual vectors $\underline{\mathbf{v}}$ and $\underline{\mathbf{w}}$ are expressed with respect to that basis then their scalar product takes the form:

$$< \underline{\mathbf{v}}, \underline{\mathbf{w}} > = \delta_{ij} \underline{v}^i \underline{w}^j ,$$

and if both sets of components are expressed in real-plus-pure-dual form as $v^i + \varepsilon a^i$ and $w^i + \varepsilon b^i$ then their scalar product takes the forms:

$$< \underline{\mathbf{v}}, \underline{\mathbf{w}} > = \delta_{ij} v^i w^j + \varepsilon \, \delta_{ij} \, (v^i b^j + a^i w^j) \qquad (9.4)$$

or:

$$< \underline{\mathbf{v}}, \underline{\mathbf{w}} > = <\mathbf{v}, \mathbf{w}> + \varepsilon(<\mathbf{v}, \mathbf{b}> + <\mathbf{w}, \mathbf{a}>). \qquad (9.5)$$

A $\mathbb{D}$-linear map $\underline{L}$ is said to be *dual orthogonal* iff it preserves the dual Euclidian scalar product; i.e., for all dual vectors $\underline{\mathbf{v}}$ and $\underline{\mathbf{w}}$ , one must have:

$$< \underline{L}\underline{\mathbf{v}}, \underline{L}\underline{\mathbf{w}} > = < \underline{\mathbf{v}}, \underline{\mathbf{w}} > .$$

If one expresses this in terms of components then the condition on the dual matrix $\underline{L}_j^i$ is that:

$$\delta_{kl}\underline{L}_i^k\underline{L}_j^l = \delta_{ij} \qquad\qquad \text{or} \qquad\qquad \underline{L}^{\mathrm{T}}\underline{L} = I.$$

Once again, this implies that $\underline{L}$ , as well as $\underline{L}_j^i$ , must be invertible and that the inverse of $\underline{L}_j^i$ is its transpose, in the usual sense of the word. Thus, the dual orthogonal transformations or matrices form a group $O(n; \mathbb{D})$, which contains $O(n; \mathbb{R})$ as a subgroup by way of the pure real orthogonal matrices. However, one notes that in order for a translation $\delta_j^i + \varepsilon A_j^i$ to be orthogonal its inverse, namely, $\delta_j^i - \varepsilon A_j^i$ must be its transpose, namely, $\delta_j^i + \varepsilon A_i^j$ . Thus, the matrix $A_j^i$ must be anti-symmetric. For the case of $n = 3$, which is the most interesting one to us at the moment, this means that the translation subgroup is isomorphic to $\mathbb{R}^3$ and the matrix $A_j^i$ takes the form of ad($\mathbf{a}$) for some real 3-vector $\mathbf{a}$:

$$A^i_j = \begin{bmatrix} 0 & -a^3 & a^2 \\ a^3 & 0 & -a^1 \\ -a^2 & a^1 & 0 \end{bmatrix}.$$

Thus, one sees that $O(3; \mathbb{D})$ has much in common with $IO(3; \mathbb{R})$. For one thing, any dual orthogonal 3×3 matrix $\underline{R} = R + \varepsilon A$ can be expressed as a product of a rotation and a translation:

$$\underline{R} = R(I + \varepsilon \tilde{R}A),$$

and the two subgroups $O(3; \mathbb{R})$ and $\mathbb{R}^3$ intersect only at the identity matrix. Thus, both $O(3; \mathbb{D})$ and $IO(3)$ have $\mathbb{R}^3 \times O(3; \mathbb{R})$ as their underlying group manifold. However, when one compares the group multiplications, one sees that although the association of rotations is straightforward, the association of the translations with the dual matrices of the form $I + \varepsilon A$ is not. The product $(\mathbf{a}, R)(\mathbf{a}', R')$ has a translational part $\mathbf{a} + R\mathbf{a}'$, while the product of two dual orthogonal matrices $(R + \varepsilon A)(R' + \varepsilon A')$ has a pure dual part of $AR' + RA'$, so a direct association of $\mathbf{a}$ with $A$ and $\mathbf{a}'$ and $A'$ with $\mathbf{a}'$ does not appear to be consistent.

However, $O(3; \mathbb{D})$ is, fact, isomorphic to $IO(3; \mathbb{R})$, by a different association of the translations with pure dual matrices, but the isomorphism is easier to exhibit at the infinitesimal level of Lie algebras, since the adjoint map that one uses is more germane to the Lie algebras than the Lie groups themselves. If $\mathbf{v} + \omega \in \mathfrak{iso}(3; \mathbb{R})$ then if one associates that element with the anti-symmetric dual matrix $\omega + \varepsilon \operatorname{ad}(\mathbf{v}) \in \mathfrak{so}(3; \mathbb{D})$ then one finds that not only does this give an $\mathbb{R}$-linear isomorphism of the two vector spaces, but the Lie brackets are consistent, as well:

$$[\mathbf{v} + \omega, \mathbf{v}' + \omega'] \qquad = [\mathbf{v}, \omega'] + [\omega, \mathbf{v}'] + [\omega, \omega']$$
$$= -\omega'\mathbf{v} + \omega\mathbf{v}' + [\omega, \omega'],$$

$$[\omega + \varepsilon \operatorname{ad}(\mathbf{v}), \omega' + \varepsilon \operatorname{ad}(\mathbf{v}')] \qquad = [\omega, \omega'] + \varepsilon([\operatorname{ad}(\mathbf{v}), \omega'] + [\omega, \operatorname{ad}(\mathbf{v}')])$$
$$= [\omega, \omega'] + \varepsilon([\mathbf{v}, \omega'] + [\omega, \mathbf{v}']),$$

since $[\mathbf{v}, \omega'] + [\omega, \mathbf{v}']$ is the translational part of the former expression.

In order to get some idea of why the association of group elements is more complicated, one can examine what happens to the exponential of an anti-symmetric dual matrix. First, one notes that if $\underline{\omega} = \omega + \varepsilon V$ then:

$$\underline{\omega}^n = \omega^n + \varepsilon \sum_{k=0}^{n-1} \omega^{n-k-1} V \omega^k. \tag{9.6}$$

If the matrices $\omega$ and $V$ commuted then the summation would give simply:

$$n\omega^{n-1}V = \frac{d\omega^n}{d\omega}V \,,$$

which is analogous to what happened for functions of dual variables, but, of course, matrix multiplication is not generally commutative.

When one forms the exponential sum:

$$\exp \underline{\omega}^n = \sum_{n=0}^{\infty} \frac{1}{n!}\underline{\omega}^n = \sum_{n=0}^{\infty} \frac{1}{n!}\omega^n + \varepsilon\left[\sum_{n=0}^{\infty}\frac{1}{n!}\sum_{k=0}^{n-1}\omega^{n-k-1}V\omega^k\right], \tag{9.7}$$

one sees that it has $\exp \omega$ for its real part, but a more complicated expression than one might prefer for the pure dual part. If $\omega$ and $V$ commuted then it would simplify to $(\exp \omega)V$, but that would hardly be typical.

The reduction from $O(3;\mathbb{D})$ to $SO(n;\mathbb{D})$ comes about by defining a volume element on $\mathbb{D}^n$. For a given basis $\{\underline{\mathbf{e}}_1, \ldots, \underline{\mathbf{e}}_n\}$ one can first define its reciprocal basis $\{\underline{\theta}^1, \ldots, \underline{\theta}^n\}$ on $\mathbb{D}^{n*}$ in the usual way:

$$\underline{\theta}^i(\underline{\mathbf{e}}_j) = \delta_j^i \,,$$

and then form the dual $n$-form $\underline{V} = \underline{\theta}^1 \wedge \ldots \wedge \underline{\theta}^n$, which is now a completely anti-symmetric $\mathbb{D}$-multilinear functional on $\mathbb{D}^n$. When each $\underline{\theta}^i$ is expressed in real-plus-pure-dual form as $\theta^i + \varepsilon\eta^i$, one sees that the only mixed products of real and pure dual 1-forms that survive the multiplication must include only one pure dual 1-form $\varepsilon\eta^i$, so $\underline{V}$ takes the form:

$$\underline{V} = \theta^1 \wedge \ldots \wedge \theta^n + \varepsilon(\eta^1 \wedge \theta^2 \wedge \ldots \wedge \theta^n + \ldots + \theta^1 \wedge \ldots \wedge \theta^{n-1} \wedge \eta^n). \tag{9.8}$$

which we can also write in the more concise form:

$$\underline{V} = V + e\ \varepsilon(\eta^i \wedge \#\mathbf{e}_i), \tag{9.9}$$

in which we have defined:

$$V = \theta^1 \wedge \ldots \wedge \theta^n, \qquad \#\mathbf{e}_i = i_{\mathbf{e}_i}V \,.$$

When one changes to another basis $\underline{L}(\underline{\mathbf{e}}_i) = \underline{\mathbf{e}}_j\underline{L}_i^j$ by way of an invertible $\mathbb{D}$-linear map $\underline{L}$, the effect on $\underline{V}$ is to multiply it by $\det(\underline{L}_j^i)$, which is a dual number, now, although as an algebraic expression in the components of $\underline{L}_j^i$ it is the same as in the real case.  Thus, $\underline{L}$ preserves the volume element $\underline{V}$ iff $\det(\underline{L}^i_j) = 1$, relative to the chosen basis.  The invertible $\mathbb{D}$-linear transformations that preserve the volume element or the invertible dual matrices with determinant 1 then form a subgroup $SL(n; \mathbb{D})$ of $GL(n; \mathbb{D})$, and one can also restrict $O(n; \mathbb{D})$ to $SO(n; \mathbb{D})$.

One can define dual eigenvectors and dual eigenvalues in the predictable way: A dual vector $\underline{\mathbf{v}}$ is an eigenvector of a dual linear map $\underline{L}$ with eigenvalue $\underline{\lambda} \in \mathbb{D}$ iff:

$$\underline{L}\,\underline{\mathbf{v}} = \underline{\lambda}\,\underline{\mathbf{v}}.$$

When one puts everything into real-plus-pure-dual form $- \underline{L} = L + \varepsilon A$, $\underline{\mathbf{v}} = \mathbf{v} + \varepsilon\mathbf{a}$, $\underline{\lambda} = \lambda + \varepsilon\alpha$ – one gets:

$$L\mathbf{v} + \varepsilon(L\mathbf{a} + A\mathbf{v}) = \lambda\mathbf{v} + \varepsilon(\lambda\mathbf{a} + \alpha\mathbf{v}).$$

Thus, one must necessarily have that $\mathbf{v}$ is an eigenvector of $L$ with an eigenvalue $\lambda$, but when one equates the pure dual parts of the equation, one gets:

$$L\mathbf{a} + A\mathbf{v} = \lambda\mathbf{a} + \alpha\mathbf{v},$$

which is not as strong a condition on $\mathbf{a}$ as the condition on $\mathbf{v}$.

We can rewrite this latter condition as:

$$(L - \lambda I)\mathbf{a} = - (A - \alpha I)\mathbf{v},$$

and this shows us that if $\mathbf{a}$ is not an eigenvector of $L$ with eigenvalue $\lambda$ then the matrix $(L - \lambda I)$ is invertible, and the condition on $\mathbf{a}$ is that:

$$\mathbf{a} = - (L - \lambda I)^{-1}(A - \alpha I)\mathbf{v},$$

and $\alpha$ can be arbitrary.  This is reminiscent of the fact that the eigenvalues of a given translation $\mathbf{a}$ can take on any value, depending upon what vector one chooses to be collinear to $\mathbf{a}$.

If $\mathbf{a}$ is an eigenvector of $L$ with eigenvalue $\lambda$ then if $\lambda$ is non-degenerate, $\mathbf{a}$ must be collinear with $\mathbf{v}$, and $\mathbf{v}$ must be an eigenvector of $A$ with eigenvalue $\alpha$.  This would imply that $LA\mathbf{v} = AL\mathbf{v}$, which is not as strong as saying that $A$ and $L$ must commute, unless the same condition is true for all their eigenvectors.  If $\lambda$ is degenerate then $\mathbf{a}$ can be simply contained in the same eigenspace as $\mathbf{v}$, without being collinear, but $\mathbf{v}$ would still have to be an eigenvector of $A$ with eigenvalue $\alpha$.

**5. The algebra of dual quaternions [3-5].** One can regard the real vector space $\mathbb{H}$ $\otimes$ $\mathbb{D}$ as a $\mathbb{D}$-module, which we then denote by $\mathbb{H}_{\mathbb{D}}$, by forming linear combinations of the canonical basis vectors $\mathbf{e}_{\mu}$ in $\mathbb{R}^4$ with coefficients in the ring $\mathbb{D}$ to give *dual quaternions:*

$$\underline{q} = (q^{\mu} + \varepsilon r^{\mu})\,\mathbf{e}_{\mu} = q^{\mu}\,\mathbf{e}_{\mu} + r^{\mu}\,\varepsilon\mathbf{e}_{\mu} = q + \varepsilon r. \tag{10.1}$$

Thus, one can also regard a dual quaternion as a pair of elements $q$, $r$ in $\mathbb{R}^4$; i.e., an element of $\mathbb{R}^8$. One refers to $q$ as the *quaternion* part $Q(\underline{q})$ of $\underline{q}$ and $\varepsilon r$ as the *pure dual quaternion* part $DQ(\underline{q})$. This gives two complementary projections of $\mathbb{H}_{\mathbb{D}}$ onto the two direct summands of $\mathbb{H} \oplus \mathbb{H}$, so one can say that $I = Q + DQ$. One defines them more specifically by polarizing the automorphism of $\mathbb{H}_{\mathbb{D}}$ that takes any $\underline{q} = q + \varepsilon r$ to:

$$\underline{q}^{\circ} = q - \varepsilon r, \tag{10.2}$$

which makes:

$$Q(\underline{q}) = q = \tfrac{1}{2}(\underline{q} + \underline{q}^{\circ}), \qquad DQ(\underline{q}) = \varepsilon r = \tfrac{1}{2}(\underline{q} - \underline{q}^{\circ}). \tag{10.3}$$

One can also decompose $\mathbb{H}_{\mathbb{D}}$ into a direct sum of the form $\mathbb{D} \oplus \mathbb{D}\mathbb{V}$, where $\mathbb{D}$ is a two-real-dimensional subalgebra that is isomorphic to $\mathbb{D}$ and $\mathbb{D}\mathbb{V}$ is a six-real-dimensional subspace of dual quaternions of the form:

$$\mathbf{q} + \varepsilon\mathbf{r} = (q^i + \varepsilon r^i)\,\mathbf{e}_i, \tag{10.4}$$

that one refers to as dual quaternions of vector type or, more concisely, as *dual vectors.* A typical dual quaternion is then expressed in "scalar-plus-vector" form as:

$$\underline{q} = \underline{q}^0 + \mathbf{\underline{q}}. \tag{10.5}$$

This decomposition of $\mathbb{H}_{\mathbb{D}}$ defines a decomposition $I = DS + DV$ of the identity operator into a sum of projections onto the relevant subspace One can define them more specifically by:

$$DS(\underline{q}) = \underline{q}^0 = \tfrac{1}{2}(\underline{q} + \overline{q}), \qquad DV(\underline{q}) = \mathbf{\underline{q}} = \tfrac{1}{2}(\underline{q} - \overline{q}), \tag{10.6}$$

in which the conjugate of a dual quaternion is defined by:

$$\overline{\underline{q}} = \underline{q}^0 - \mathbf{\underline{q}} = \overline{q} + \varepsilon\overline{r}. \tag{10.7}$$

The algebra of $\mathbb{H}_\mathbb{D}$ is simply the one that it inherits from $\mathbb{H}$ by $\mathbb{D}$-bilinearity, modulo the relation $\varepsilon^2 = 0$. One can first extend the basis $\{\mathbf{e}_\mu, \mu = 0, \ldots, 3\}$ for $\mathbb{R}^4$ to the basis $\{\mathbf{e}_\mu, \varepsilon\mathbf{e}_\mu, \mu = 0, \ldots, 3\}$ for $\mathbb{R}^8$ and then define the missing products:

$$\mathbf{e}_\mu(\varepsilon\mathbf{e}_\nu) = (\varepsilon\mathbf{e}_\mu)\mathbf{e}_\nu = \varepsilon\,(\mathbf{e}_\mu\mathbf{e}_\nu), \qquad (\varepsilon\mathbf{e}_\mu)(\varepsilon\mathbf{e}_\nu) = 0\,. \tag{10.8}$$

One can also define the product of any two dual quaternions $q + \varepsilon r$ and $q' + \varepsilon r'$ by $\mathbb{D}$-bilinearity:

$$\underline{q}\,\underline{q}' = (q + \varepsilon r)(q' + \varepsilon r') = qq' + \varepsilon(rq' + qr')\,. \tag{10.9}$$

Thus, $\mathbb{H}_\mathbb{D}$ is still an associative, but not commutative, ring and has a unity element 1 that is still defined by $\mathbf{e}_0$, with a center that is defined by all of the dual scalars, which have the form $(q^0 + \varepsilon r^0)\,\mathbf{e}_0$.

In particular, the square of a dual quaternion takes the form:

$$\underline{q}^2 = q^2 + \varepsilon(qr + rq)\,. \tag{10.10}$$

One finds that the product of dual quaternions admits an expansion that is analogous to the one for real quaternions:

$$\underline{q}\,\underline{q}' = (\underline{q}, \underline{q}') + \underline{q}^0\underline{\mathbf{q}}' + \underline{q}'^0\underline{\mathbf{q}} + \underline{\mathbf{q}}\times\underline{\mathbf{q}}'\,, \tag{10.11}$$

in which

$$(\underline{q}, \underline{q}') = DS(\underline{q}\,\underline{q}') = \underline{q}^0\underline{q}'^0 - \sum_{i=1}^{3}\delta_{ij}\underline{q}^i\underline{q}'^{\,j}\,, \tag{10.12}$$

and one finds that if $\underline{\mathbf{q}} = \mathbf{q} + \varepsilon\mathbf{r}$ and $\underline{\mathbf{q}}' = \mathbf{q}' + \varepsilon\mathbf{r}'$ then:

$$\underline{\mathbf{q}}\times\underline{\mathbf{q}}' = \tfrac{1}{2}[\underline{\mathbf{q}}, \underline{\mathbf{q}}'] = \mathbf{q}\times\mathbf{q}' + \varepsilon(\mathbf{q}\times\mathbf{r}' + \mathbf{r}\times\mathbf{q}')\,. \tag{10.13}$$

For dual quaternions of dual vector type, one then has:

$$\underline{\mathbf{q}}\,\underline{\mathbf{q}}' = -\left(\sum_{i=1}^{3}\delta_{ij}\underline{q}^i\underline{q}'^{\,j}\right) + \underline{\mathbf{q}}\times\underline{\mathbf{q}}'\,. \tag{10.14}$$

It is also useful to know that:

$$(\underline{\mathbf{a}}\times\underline{\mathbf{b}})\times\underline{\mathbf{c}} = <\underline{\mathbf{a}}, \underline{\mathbf{c}}>\underline{\mathbf{b}} - <\underline{\mathbf{a}}, \underline{\mathbf{b}}>\underline{\mathbf{c}}\,, \tag{10.15}$$

in which:

$$<\underline{\mathbf{a}}, \underline{\mathbf{b}}> = -DS(\underline{\mathbf{a}}\,\underline{\mathbf{b}}) = \sum_{i=1}^{3}\delta_{ij}\underline{a}^i\,\underline{b}^j\,. \tag{10.16}$$

Like the algebra $\mathbb{D}$ itself, the algebra $\mathbb{H}_{\mathbb{D}}$ has divisors of zero; in particular, from the last set of equations in (10.8), one sees that the product of any two pure dual quaternions is zero. They are then nilpotents of degree two; in fact they are the only ones. In order to verify this, one goes back to the expression (10.10) for $\underline{q}^2$ and sets it equal to zero. This gives:

$$q^2 = 0, \qquad qr + rq = 0.$$

Hence, either $q = 0$ or $q$ is a non-trivial nilpotent real quaternion, which we have seen is impossible.

As for idempotents, one sets $\underline{q}^2 = \underline{q}$ in (10.10) and deduces the conditions:

$$q^2 = q, \qquad qr + rq = r.$$

Thus $q$ must be an idempotent in $\mathbb{H}$, which can only mean 0 or 1. Either of these cases give the same result that $r$, and therefore $q$, vanishes. Hence, there are no non-trivial idempotents in $\mathbb{H}_{\mathbb{D}}$, either.

In order to find the invertible elements, one can look at (10.9) when:

$$qq' = 1, \qquad rq' + qr' = 0.$$

If $q$ is invertible then this can be solved uniquely by setting:

$$q' = q^{-1} = \frac{\overline{q}}{\|q\|^2}, \qquad r' = -q^{-1} r \, q^{-1}.$$

Thus, $\underline{q} = q + \varepsilon r$ is invertible iff $q$ is invertible iff $q \neq 0$, and:

$$\underline{q}^{-1} = q^{-1} - \varepsilon \, q^{-1} r \, q^{-1}. \tag{10.17}$$

Hence, $\mathbb{H}_{\mathbb{D}}$, unlike $\mathbb{H}$, is not a division algebra, although it contains a subset $\underline{Q}^*$ that defines a multiplicative group, namely, the set complement of the vector subspace of pure dual quaternions. As a set, it takes the form of the product $Q^* \times \mathbb{R}^4$, since the only restriction on the quaternion part $q$ is that it be non-zero, while there is no restriction placed on the pure dual quaternion $r$. The group structure is somewhat more involved, since although the non-zero quaternions still have the same group structure − namely, $Q^*$ − as before, the pure dual quaternions by themselves have no group structure (since their product is always 0), and the extension of $Q^*$ to the pure dual quaternions does not have an obvious interpretation, at the moment.

Nonetheless, one finds that the dual quaternions of the form $1 + \varepsilon q$ do form a subgroup of $\underline{Q}^*$, and the product of two of them takes the form:

$$(1 + \varepsilon q)(1 + \varepsilon q') = 1 + \varepsilon(q + q'). \tag{10.18}$$

Thus, the subgroup that they define is isomorphic to the four-dimensional translation group. Therefore, since the subgroup of non-zero pure quaternions and the subgroup of translations have only the element 1 in common and both are four-dimensional, we see that the group $\underline{Q}^*$ is eight dimensional.

In order to go further into the structure of the group $\underline{Q}^*$, we now examine the nature of the most natural scalar product that we can define on $\mathbb{H}_\mathbb{D}$, which is analogous to the one that we defined on $\mathbb{H}$, except that it take its values in $\mathbb{D}$, not $\mathbb{R}$:

$$<\underline{q}, \underline{q}'> \ = DS(\underline{q}\overline{\underline{q}}') = <q, q'> + \varepsilon(<r, q'> + <q, r'>), \tag{10.19}$$

In particular:

$$<\underline{q}, \underline{q}> \ = \ \| \, q \, \|^2 + 2\varepsilon <q, r> \equiv \|\underline{q}\|^2, \tag{10.20}$$

and a dual quaternion $\underline{q}$ is said to be a *unit dual quaternion* iff $\|\underline{q}\| = 1$, which is true iff:

$$\| \, q \, \| = 1, \qquad <q, r> = 0. \tag{10.21}$$

The first condition says that the quaternion part $q$ must lie on the unit sphere in $\mathbb{H}$. The second condition defines a homogeneous quadratic hypersurface in $\mathbb{R}^8$, and therefore a quadric in $\mathbb{R}\mathbb{P}^7$ that is called the *Study quadric*. When written out in terms of scalar and vector parts, it reads:

$$q^0 r^0 + <\mathbf{q}, \mathbf{r}> \ = \sum_{\mu=0}^{3} q^\mu r_\mu = 0. \tag{10.22}$$

When one passes from a non-zero dual quaternion $\underline{q}$ to the line $[\underline{q}]$ through the origin that it defines, the first condition in (10.21) becomes superfluous.

We now see that for unit dual quaternions the expression (10.17) for the inverse becomes:

$$\underline{q}^{-1} = \overline{q} - \varepsilon \, \overline{q} \, r \overline{q} \, . \tag{10.23}$$

One can now regard the set $\underline{Q}_1$ of all unit dual quaternions as a group, since if $\underline{q}$ and $\underline{q}'$ are unit dual quaternions then the product $\underline{q}\,\underline{q}' = qq' + \varepsilon(qr' + rq')$ is also a unit dual quaternion. This follows from the fact that:

$$<\underline{q}\,\underline{q}', \ \underline{q}\,\underline{q}'> \ = DS(\underline{q}\,\underline{q}' \ \overline{\underline{q}}'\,\overline{\underline{q}} \,) = 1.$$

The group $\underline{Q}_1$ includes $SU(2)$ as a subgroup in the form of the unit quaternions, as well as the translation group $\mathbb{R}^3$, in the form of all elements of the form $1 + \varepsilon \mathbf{s}$. Indeed, for a general element $q + \varepsilon \mathbf{s}$, the conditions that $\| q \| = 1$ and $<q, \mathbf{s}> = <\mathbf{q}, \mathbf{s}> = 0$ amount to the statement that $q$ describes a point on the unit 3-sphere in $\mathbb{H}$, while $\mathbf{s}$ lies in the plane tangent to it. Thus, as a manifold, we can think of $\underline{Q}_1$ as the manifold $TS^3 = S^3 \times \mathbb{R}^3$.

One should note that, from (10.20), $\mathbb{H}_{\mathbb{D}}$ admits *null* elements, for which $\| \underline{q} \| = 0$. In fact, $\underline{q}$ is a null dual quaternion iff it is a pure dual quaternion.

When the scalar product that we just defined is restricted to dual quaternions of vector type – i.e., dual vectors – one gets:

$$< \underline{\mathbf{q}}, \underline{\mathbf{q}}' > \; = <\mathbf{q}, \mathbf{q}'> + \varepsilon (<\mathbf{r}, \mathbf{q}'> + <\mathbf{q}, \mathbf{r}'>). \qquad (10.24)$$

The norm-squared is then:

$$\| \underline{\mathbf{q}} \|^2 = < \underline{\mathbf{q}}, \underline{\mathbf{q}} > \; = \| \mathbf{q} \|^2 + 2\varepsilon <\mathbf{q}, \mathbf{r}>, \qquad (10.25)$$

and a unit dual quaternion $\underline{\mathbf{q}}$ of dual vector type must satisfy:

$$\| \mathbf{q} \|^2 = 1, \qquad <\mathbf{q}, \mathbf{r}> = 0, \qquad (10.26)$$

which represents a unit vector $\mathbf{q}$ in $\mathbb{R}^3$ and a vector $\mathbf{r}$ of unspecified length that is perpendicular to it. One then refers to such a $\underline{\mathbf{q}}$ as a *dual unit vector*. The set of all dual unit vectors is then $TS^2$, which only locally looks like $S^2 \times \mathbb{R}^2$, since $S^2$ is not parallelizable. Thus, that set cannot be a group manifold, since every Lie group is parallelizable. Indeed, the product of two dual unit vectors $\mathbf{q} + \varepsilon \mathbf{r}$ and $\mathbf{q}' + \varepsilon \mathbf{r}'$ is $\mathbf{q}\mathbf{q}' + \varepsilon (\mathbf{r}\mathbf{q}' + \mathbf{q}\mathbf{r}')$, which does not have to be a dual vector, since it has a dual scalar part equal to $- <\mathbf{q}, \mathbf{q}'> - \varepsilon (<\mathbf{r}, \mathbf{q}'> + <\mathbf{q}, \mathbf{r}'>) = - < \underline{\mathbf{q}}, \underline{\mathbf{q}}' >$.

We can express any dual quaternion in a polar form that is analogous to the one obtained for real quaternions, namely:

$$\underline{q} = \| \underline{q} \| \, (\cos \tfrac{1}{2} \underline{\theta} + \sin \tfrac{1}{2} \underline{\theta} \, \underline{\mathbf{u}}). \qquad (10.27)$$

The factor in parentheses then represents a typical unit dual quaternion where $\underline{\theta} = \theta + \varepsilon s$, is a dual angle and $\underline{\mathbf{u}} = \mathbf{u} + \varepsilon \mathbf{m}$ is a dual unit vector.

From (10.27), the polar form of a unit dual quaternion is then:

$$\underline{u} = \cos \tfrac{1}{2} \underline{\theta} + \sin \tfrac{1}{2} \underline{\theta} \, \underline{\mathbf{u}}$$

$$= (\cos \tfrac{1}{2}\,\theta + \sin \tfrac{1}{2}\,\theta\ \mathbf{u}) + \varepsilon\,[-\tfrac{s}{2}\,(\sin \tfrac{1}{2}\,\theta - \cos \tfrac{1}{2}\,\theta\mathbf{u}) + \sin \tfrac{1}{2}\,\theta\mathbf{m}], \qquad (10.28)$$

which is analogous to the form that we introduced for real quaternions. This time, the rigid motion involves a dual axis $\underline{\mathbf{u}} = \mathbf{u} + \varepsilon\mathbf{m}$, $(\mathbf{m} = \mathbf{x} \times \mathbf{u})$ that represents a dual unit vector and a dual angle $\underline{\theta} = \theta + s$ that combines a rotation around $\mathbf{u}$ by the angle $\theta$ with a translation along $\mathbf{m}$ through a distance $s$. This differs from the previous case of real quaternions also by the fact that the rotational axis $\underline{\mathbf{u}}$ no longer has to go through the origin.

There is a simple isomorphic representation of the Lie algebra $\mathfrak{iso}(3)$ in the Lie algebra $\underline{\mathbf{q}}_0$ of dual vectors given the commutator bracket. If one associates the infinitesimal rigid motion $\boldsymbol{\omega} + \mathbf{v}$ with the dual vector $\tfrac{1}{2}\,(\boldsymbol{\omega} + \varepsilon\mathbf{v})$ then one sees that, from the bilinearity of the commutator bracket, the Lie bracket of two such infinitesimal motions:

$$[\boldsymbol{\omega} + \mathbf{v}, \boldsymbol{\omega}' + \mathbf{v}'] = [\boldsymbol{\omega}, \boldsymbol{\omega}'] + [\boldsymbol{\omega}, \mathbf{v}'] + [\mathbf{v}, \boldsymbol{\omega}'] = \boldsymbol{\omega} \times \boldsymbol{\omega}' + \boldsymbol{\omega} \times \mathbf{v}' + \mathbf{v} \times \boldsymbol{\omega}'$$

(since $[\mathbf{v}, \mathbf{v}'] = 0$ for the translations) is consistent with the commutator of the corresponding dual vectors:

$$\tfrac{1}{2}\,[\boldsymbol{\omega} + \varepsilon\mathbf{v}, \boldsymbol{\omega}' + \varepsilon\mathbf{v}'] = \tfrac{1}{2}\,\{[\boldsymbol{\omega}, \boldsymbol{\omega}'] + \varepsilon\{[\boldsymbol{\omega}', \mathbf{v}] + [\boldsymbol{\omega}, \mathbf{v}']\}\} = \boldsymbol{\omega} \times \boldsymbol{\omega}' + \varepsilon(\boldsymbol{\omega} \times \mathbf{v}' + \mathbf{v} \times \boldsymbol{\omega}'),$$

in which we have used the fact that $[\mathbf{q}, \mathbf{q}'] = 2\,\mathbf{q} \times \mathbf{q}'$ for quaternions of vector type.

**6. The action of rigid motions on dual quaternions**. The group of unit dual quaternions can be used to represent rigid motions of affine $E^3$ and its bundle $SO(E^3)$ of orthonormal affine frames.

The action of unit dual quaternions on dual quaternions is simply the predictable extension of the action $q' = u\,q\,\bar{u}$ of a unit quaternion $u$ on a quaternion $q$ to an action of a unit dual quaternion $\underline{u}$ on a dual quaternion $\underline{q}$, namely:

$$\underline{Q}_1 \times \mathbb{H}_{\mathbb{D}} \to \mathbb{H}_{\mathbb{D}}, \ \underline{q} \mapsto \underline{q}' = \underline{u}\,\underline{q}\,\overline{\underline{u}}. \qquad (11.1)$$

The proof that the transformation of $\mathbb{H}_{\mathbb{D}}$ is a $\mathbb{D}$-linear, surjective, and two-to-one follows from the same proofs as in the case of $\mathbb{H}$, since we are still assuming that $\underline{u}$ is invertible.

The proof that the action is by isometries of $<.,.>$ is straightforward, since:

$$<\underline{p}', \underline{q}'> = DS(\underline{u}\,\underline{p}\,\overline{\underline{u}}\,\underline{u}\,\overline{\underline{q}}\,\overline{\underline{u}}) = DS(\underline{u}\,\underline{p}\,\overline{\underline{q}}\,\overline{\underline{u}}) = DS(\underline{p}\,\overline{\underline{q}}) = <\underline{p}, \underline{q}>,$$

in which we have implicitly used the fact that the action (11.1) has $DS\mathbb{H}_\mathbb{D}$ and $DV\mathbb{H}_\mathbb{D}$ as invariant subspaces. Therefore, the action thus defined takes dual scalars to other dual scalars and dual vectors to dual vectors.

Since the action (11.1) is linear in $\underline{q}$ for each $\underline{u}$ and has $DV\mathbb{H}_\mathbb{D}$ for an invariant subspace, if one chooses a basis $\{\underline{\mathbf{e}}_i, i = 1, 2, 3\}$ for $DV\mathbb{H}_\mathbb{D}$ then one can associate the action of $\underline{u}$ with an invertible dual 3×3 matrix $\underline{R}_j^i$ by way of:

$$\underline{u}\,\underline{\mathbf{e}}_i\,\overline{\underline{u}} = \underline{\mathbf{e}}_j\underline{R}_i^j. \tag{11.2}$$

The fact that the action is by isometries then implies that $\underline{R}_j^i \in SO(3; \mathbb{D})$ and the resulting map $\underline{Q}_1 \to SO(3; \mathbb{D})$, $\underline{u} \mapsto \underline{R}_j^i$ is then a two-to-one homomorphism of the group of unit dual quaternions with the group of three-dimensional rigid motions, as represented by the volume-preserving, dual orthogonal transformations. The fact that the association preserves the product follows from the fact that the product $\underline{\mathbf{u}}\,\underline{\mathbf{u}}'$ of two unit dual quaternions acts on a dual orthonormal frame $\underline{\mathbf{e}}_i$ to give:

$$\underline{\mathbf{u}}\,\underline{\mathbf{u}}'\underline{\mathbf{e}}_i\overline{\underline{\mathbf{u}}\,\underline{\mathbf{u}}'} = \underline{\mathbf{u}}(\underline{\mathbf{u}}'\underline{\mathbf{e}}_i\overline{\underline{\mathbf{u}}}')\overline{\underline{\mathbf{u}}} = (\underline{\mathbf{u}}\,\underline{\mathbf{e}}_j\overline{\underline{\mathbf{u}}})\underline{R}_i'^j = \underline{\mathbf{e}}_j\underline{R}_k'^j R_i^k. $$

Thus, the association takes the quaternion product $\underline{\mathbf{u}}\,\underline{\mathbf{u}}'$ to the matrix product $\underline{R}_k'^j R_i^k$, so it is an order-reversing two-to-one isomorphism from $\underline{Q}_1$ to $SO(3; \mathbb{D})$.

The transition from an action of $\underline{Q}_1$ on dual vectors in $DV\mathbb{H}_\mathbb{D}$ to an action on orthonormal affine frames $(x, \mathbf{f}_i)$ at its various points is also straightforward, since one has the vector $\mathbf{x}$ from $O$ to $x$, while the unit vectors $\mathbf{f}_i$ define moments $\mathbf{m}_i = \mathbf{x} \times \mathbf{f}_i$ about $O$. The orthonormal affine frame $(x, \mathbf{f}_i)$ at the point $x \in E^3$ then becomes the three dual unit vectors:

$$\underline{\mathbf{f}}_i = \mathbf{f}_i + \varepsilon\,\mathbf{m}_i. \tag{11.3}$$

One finds that:

$$<\underline{\mathbf{f}}_i, \underline{\mathbf{f}}_j> = \delta_{ij} + \varepsilon(<\mathbf{f}_i, \mathbf{m}_j> + <\mathbf{m}_i, \mathbf{f}_j>). \tag{11.4}$$

One then observes that:

$$<\mathbf{f}_i, \mathbf{m}_j> + <\mathbf{m}_i, \mathbf{f}_j> = \det[\mathbf{f}_i \mid \mathbf{x} \mid \mathbf{f}_j] + \det[\mathbf{x} \mid \mathbf{f}_i \mid \mathbf{f}_j] = 0, \tag{11.5}$$

since switching any two columns in a matrix changes the sign of the determinant. We can therefore assert that if $(x, \mathbf{e}_i)$ is an orthonormal affine frame at a point $x$ in affine $E^3$ then the dual frame $\underline{\mathbf{f}}_i$ that was defined in (11.3) is an orthonormal dual frame in $DV\mathbb{H}_\mathbb{D}$.

Since the action (11.1) is by $\mathbb{D}$-linear isometries, and $(x, \mathbf{f}_i) = (O, \mathbf{e}_i)(x^i, R_i^j)$, one sees that:

$$\underline{\mathbf{f}}_i = \mathbf{e}_j R_i^j + \varepsilon \, \mathbf{x} \times \mathbf{e}_i = \mathbf{e}_j R_i^j + \varepsilon \, \mathbf{e}_i \, \mathrm{ad}(\mathbf{x}) = \mathbf{e}_j \, ( \, R_i^j + \varepsilon \, [\mathrm{ad}(\mathbf{x})]_i^j \, ) = \mathbf{e}_j \, \underline{R}_i^j \, ,$$

in which we have introduced the right-adjoint operator for the vector $\mathbf{x}$:

$$\mathbf{y} \, \mathrm{ad}(\mathbf{x}) = \mathbf{x} \times \mathbf{y} = -\, \varepsilon_{ijk} \, x^k \, y^j \, \mathbf{e}_i \, , \tag{11.6}$$

and the dual matrix:

$$\underline{R}_i^j = R_i^j + \varepsilon \, x^k [\mathrm{ad}(\mathbf{e}_k)]_i^j = R_i^j - \varepsilon \, \varepsilon_{ijk} x^k \, . \tag{11.7}$$

One finds that there is an analogue of Rodrigues's formula that applies to dual vectors:

$$\underline{\mathbf{u}} \, \mathbf{x} \, \overline{\underline{\mathbf{u}}} = \cos \underline{\theta} \, \underline{\mathbf{x}} + (1 - \cos \underline{\theta}) < \underline{\mathbf{u}}, \underline{\mathbf{x}} > \underline{\mathbf{u}} + \sin \underline{\theta} \, \underline{\mathbf{u}} \times \underline{\mathbf{x}} \, . \tag{11.8}$$

**7. Some line geometry.** One use for the set of dual unit vectors is that they give a faithful representation of the manifold of all lines in $E^3$, which we regard as an affine space, not a vector space, now. First, one represents a line $[l]$ in $E^3$ by means of two (tangent) vectors $\mathbf{x}$ and $\mathbf{u}$, where $\mathbf{x}$ is the displacement vector $x - O$ that takes one from a chosen reference point $O$ to a point $x$ on $[l]$ and $\mathbf{u}$ is a unit vector tangent to $x$ that defines the direction of $[l]$. Since $-\mathbf{u}$ would also define the direction of $[l]$, unless one orients the line (which would then make it a "spear," in the language of Study), the set of all lines through $x$ is a manifold that is diffeomorphic to $\mathbb{R}P^2$, which is doubly covered by $S^2$.

Thus, one can associate $[l]$ with a line through the origin in $\mathbb{H}$ that lies in the subspace of pure quaternions by taking $\mathbf{u}$ to the corresponding pure quaternion, and thus with two points of intersection with the unit sphere in the subspace. As for the vector $\mathbf{x}$, one finds it more convenient to define the *moment* $\mathbf{m}$ of $\mathbf{u}$ about $O$ by way of:

$$\mathbf{m} = \mathbf{x} \times \mathbf{u}, \tag{12.1}$$

which is then independent of the choice of $x$, but dependent on the choice of $O$, since any other point $x'$ on $[l]$ could be expressed by a position vector of the form $\mathbf{x} + \alpha \mathbf{u}$, for some scalar $\alpha$, and the moment of $\mathbf{u}$ about $O$ would then go to:

$$(\mathbf{x} + \alpha \mathbf{u}) \times \mathbf{u} = \mathbf{m} + \alpha \, \mathbf{u} \times \mathbf{u} = \mathbf{m}.$$

One can define a canonical choice of $x$ by specifying that $\mathbf{x}$ be perpendicular to $\mathbf{u}$.

A different choice of $O$ – say, $O' = O + \mathbf{s}$ – would then make the new moment of $\mathbf{u}$ take the form:

$$\mathbf{m}' = (\mathbf{x} + \mathbf{s}) \times \mathbf{u} = \mathbf{m} + \mathbf{s} \times \mathbf{u},$$

and this would be unchanged only if $\mathbf{s}$ were parallel to $\mathbf{u}$. Since this defines a line $[\mathbf{a}]$ through $O$ that is parallel to $[l]$, one can also think of $\mathbf{m}$ as the moment of $[l]$ about $[\mathbf{a}]$.

Another advantage of using $\mathbf{m}$, in place of $\mathbf{x}$ is that $\mathbf{m}$ will automatically be perpendicular to $\mathbf{u}$. Thus, if one associates the line $[l]$ with the pair of vectors $(\mathbf{u}, \mathbf{m})$ and then with the dual vector $\underline{l} = \mathbf{u} + \varepsilon \mathbf{m}$, one finds that $\underline{l}$ is also a dual unit vector. Conversely, every dual unit vector $\mathbf{u} + \varepsilon \mathbf{m}$ (and its negative) defines a line in $E^3$ by way of the line through the origin of $\mathbb{H}$ that contains $\mathbf{u}$ and the vector $\mathbf{m}$ that is tangent to the unit 2-sphere, which then gives the moment of the line about $O$, and thus, the canonical normal $\mathbf{x}$ from $O$ to that line.

If one has two lines $[l]$ and $[l']$ in $E^3$ that have a distance of closest approach $s$ that lies along a common perpendicular to both lines, which is described by the points $x$ and $x'$, and define an angle $\theta$ when they are parallel-translated along that line until they intersect then one finds that:

$$<\mathbf{u}, \mathbf{u}'> = \cos \theta, \tag{12.2}$$

$$<\mathbf{x}, \mathbf{u}'> + <\mathbf{x}', \mathbf{u}> = -\det[\mathbf{x}' - \mathbf{x} \mid \mathbf{u} \mid \mathbf{u}'] = -s \sin \theta. \tag{12.3}$$

When the two lines are represented by dual unit vectors $\underline{l}$ and $\underline{l}'$, one finds, from (8.4), that the latter set of equations consolidate into:

$$<\underline{l}, \underline{l}'> = \cos \underline{\theta}, \qquad \underline{\theta} = \theta + \varepsilon s. \tag{12.4}$$

Thus, the orthogonality of the lines, in the dual sense, is equivalent to the statement that the lines intersect at a right angle, since the real part of $\underline{\theta}$ describes the angle between them and dual part describes the distance of closest approach.

One also finds that there is an analogue of the usual formula for the cross product:

$$\underline{l} \times \underline{l}' = \sin \underline{\theta} \; \underline{\mathbf{n}}, \tag{12.5}$$

in which $\underline{\mathbf{n}}$ is the dual unit vector that is orthogonal to the plane of $\underline{l}$ and $\underline{l}'$ in the right-hand sense.

The key to making the association of a rigid motion of $E^3$ with $\underline{\mathbf{u}}$ and $\underline{\theta}$ is given by Chasles's theorem, which we discussed above. The dual unit vector $\underline{\mathbf{u}} = \mathbf{u} + \varepsilon \mathbf{m}$ then defines central axis $[l]$ of the rigid motion with respect to some chosen point $O$ in $E^3$.

## 8. The kinematics of translating rigid bodies.

When a rigid body is allowed to translate, as well as rotate, the rotating orthonormal frame $\mathbf{f}_i(t)$ at the fixed point $O$ becomes the orthonormal frame $(x(t), \mathbf{f}_i(t))$ at a moving point $x(t)$. One can then represent $(x(t), \mathbf{f}_i(t))$ in the form:

$$(x(t), \mathbf{f}_i(t)) = (x_0, \mathbf{f}_{0j})(s^j(t), R_i^j(t)) = (x_0 + \mathbf{s}(t), \mathbf{f}_{0j} R_i^j(t)), \tag{13.1}$$

in which $x_0 = x(0)$, $\mathbf{f}_{0j} = \mathbf{f}_i(0)$, and $\mathbf{s}(t) = s^j(t)\mathbf{f}_{0j}$. One can also express this relationship by the pair of equations:

$$x(t) = x_0 + \mathbf{s}(t), \qquad \mathbf{f}_i(t) = \mathbf{f}_{0j} R_i^j(t). \tag{13.2}$$

If one assumes that all functions of time are sufficiently differentiable then a first differentiation gives:

$$\mathbf{v} = \frac{dx}{dt} = \dot{\mathbf{s}}, \qquad\qquad \dot{\mathbf{f}}_i = \frac{d\mathbf{f}_i}{dt} = \mathbf{f}_{0j}\,\dot{R}_i^{\ j}\,. \qquad\qquad (13.3)$$

One solves for $(x_0, \mathbf{f}_{0i})$ in terms of $(x, \mathbf{f}_i)$:

$$x_0 = x - \mathbf{s}, \qquad \mathbf{f}_{0i} = \mathbf{f}_i\,\tilde{R}_i^{\ j}\,, \qquad\qquad (13.4)$$

and substitutes this in (13.3) to get:

$$\mathbf{v} = \dot{\mathbf{s}}, \qquad\qquad \dot{\mathbf{f}}_i = \mathbf{f}_i\,\omega_i^{\ j}\,, \qquad\qquad (13.5)$$

in which we have introduce the *angular velocity* of the moving frame with respect to the initial one:

$$\omega_i^{\ j} = \dot{R}_k^{\ j}\,\tilde{R}_i^{\ k}\,. \qquad\qquad (13.6)$$

A second differentiation of (13.3) gives the acceleration of the moving frame relative to the initial one:

$$\mathbf{a} = \frac{d\mathbf{v}}{dt} = \ddot{\mathbf{s}}, \qquad\qquad \ddot{\mathbf{f}}_i = \frac{d\dot{\mathbf{f}}_i}{dt} = \mathbf{f}_{0j}\,\ddot{R}_i^{\ j}\,. \qquad\qquad (13.7)$$

When one substitutes for $\mathbf{f}_{0j}$, one gets:

$$\ddot{\mathbf{f}}_i = \mathbf{f}_j\,\alpha_i^{\ j}\,, \qquad\qquad (13.8)$$

in which we have introduced the angular acceleration of the moving frame:

$$\alpha_i^{\ j} = \tilde{R}_k^{\ j}\,\ddot{R}_i^{\ k}\,. \qquad\qquad (13.9)$$

One can also differentiate (13.5) to get:

$$\ddot{\mathbf{f}}_i = \dot{\mathbf{f}}_j\,\omega_i^{\ j} + \mathbf{f}_j\,\dot{\omega}_i^{\ j} = \mathbf{f}_j(\omega_k^{\ j}\omega_i^{\ k} + \dot{\omega}_i^{\ j})\,, \qquad\qquad (13.10)$$

which also makes:

$$\alpha_i^{\ j} = \omega_k^{\ j}\omega_i^{\ k} + \dot{\omega}_i^{\ j}\,, \qquad\qquad (13.11)$$

since $\mathbf{f}_i$ is a frame.

If one now represents the orthonormal frame $(x_0, \mathbf{f}_{0i})$ by the dual unit vector $\underline{\mathbf{f}}_{0i} = \mathbf{f}_{0i} + \varepsilon\mathbf{m}_{0i}$, with the predictable definition for $\mathbf{m}_{0i}$, then if the rigid motion $(s^i(t), R_i^{\ j}(t))$ is represented by the unit dual quaternion $\underline{q}(t) = q(r) + \varepsilon s(t)$, the time evolution of the initial frame $(x_0, \mathbf{f}_{0i})$ under the action of the one-parameter family of rigid motions can be expressed in a form that follows from (11.1):

$$\underline{\mathbf{f}}_i(t) = \underline{q}(t)\,\underline{\mathbf{f}}_{0i}\,\overline{\underline{q}}(t)\,. \qquad\qquad (13.12)$$

If we assume that the curve $q(t)$ in $Q_1$ is sufficiently differentiable then one can express the velocity of the moving frame $\underline{\mathbf{f}}_i(t)$ relative to the initial frame in the form:

$$\underline{\mathbf{v}}_i = \frac{d\underline{\mathbf{f}}_i}{dt} = \dot{\underline{q}}\,\underline{\mathbf{f}}_{0i}\,\overline{\underline{q}} + \underline{q}\,\underline{\mathbf{f}}_{0i}\,\dot{\overline{\underline{q}}}\,. \qquad (13.13)$$

When we solve (13.12) for $\underline{\mathbf{f}}_{0i} = \overline{\underline{q}}\,\underline{\mathbf{f}}_i\,\underline{q}$ and substitute in (13.13), we get the velocity relative to the moving frame:

$$\underline{\mathbf{v}}_i = \underline{\boldsymbol{\omega}}\underline{\mathbf{f}}_i + \underline{\mathbf{f}}_i\,\overline{\underline{\boldsymbol{\omega}}} = [\underline{\boldsymbol{\omega}},\underline{\mathbf{f}}_i]\,, \qquad (13.14)$$

in which we have introduced the *absolute velocity*:

$$\underline{\boldsymbol{\omega}} = \dot{\underline{q}}\,\overline{\underline{q}}\,, \qquad (13.15)$$

and used the fact that $\underline{\boldsymbol{\omega}}$ is a dual vector, so $\overline{\underline{\boldsymbol{\omega}}} = -\,\underline{\boldsymbol{\omega}}$.

The absolute velocity $\underline{\omega} = \boldsymbol{\omega} + \varepsilon \mathbf{v}$ corresponds to an element $\boldsymbol{\omega} + \mathbf{v}$ of the Lie algebra $\mathfrak{iso}(3)$ of the group of rigid motions in $E^3$, so it consists of a rotational part $\boldsymbol{\omega}$ and a translational part $\mathbf{v}$.

Another differentiation of (13.13) gives the *acceleration* relative to the initial frame:

$$\underline{\mathbf{a}}_i = \frac{d\underline{\mathbf{v}}_i}{dt} = \ddot{\underline{q}}\,\underline{\mathbf{f}}_{0i}\,\overline{\underline{q}} + 2\dot{\underline{q}}\,\underline{\mathbf{f}}_{0i}\,\dot{\overline{\underline{q}}} + \underline{q}\,\underline{\mathbf{f}}_{0i}\,\ddot{\overline{\underline{q}}}\,, \qquad (13.16)$$

and upon substituting for $\underline{\mathbf{f}}_{0i}$, one gets the *absolute acceleration*:

$$\underline{\mathbf{a}}_i = [\underline{\boldsymbol{\alpha}},\underline{\mathbf{f}}_i] - 2\underline{\boldsymbol{\omega}}\underline{\mathbf{f}}_i\,\underline{\boldsymbol{\omega}}, \qquad (13.17)$$

into which we have introduced the generalized angular acceleration:

$$\underline{\boldsymbol{\alpha}} = \ddot{\underline{q}}\,\overline{\underline{q}}\,. \qquad (13.18)$$

One can also differentiate (13.14) to get:

$$\underline{\mathbf{a}}_i = [\dot{\underline{\boldsymbol{\omega}}},\underline{\mathbf{f}}_i] + [\underline{\boldsymbol{\omega}},\dot{\underline{\mathbf{f}}}_i] = [\dot{\underline{\boldsymbol{\omega}}},\underline{\mathbf{f}}_i] + [\underline{\boldsymbol{\omega}},[\underline{\boldsymbol{\omega}},\underline{\mathbf{f}}_i]]\,, \qquad (13.19)$$

which makes:

$$[\underline{\boldsymbol{\alpha}},\underline{\mathbf{f}}_i] = [\dot{\underline{\boldsymbol{\omega}}},\underline{\mathbf{f}}_i] + [\underline{\boldsymbol{\omega}},[\underline{\boldsymbol{\omega}},\underline{\mathbf{f}}_i]] + 2\underline{\boldsymbol{\omega}}\underline{\mathbf{f}}_i\,\underline{\boldsymbol{\omega}}\,. \qquad (13.20)$$

# CHAPTER IV

## COMPLEX QUATERNIONS

Since the algebra of complex quaternions $\mathbb{H}_\mathbb{C}$ comes about by taking the real tensor product $\mathbb{H} \otimes_\mathbb{R} \mathbb{C}$, in a manner that is analogous to the way that dual quaternions came about from the tensor product $\mathbb{H} \otimes_\mathbb{R} \mathbb{D}$, we will proceed in a manner that is analogous to what we did in the last chapter. The main difference is in the fact that the algebra $\mathbb{C}$ is now a division algebra, as well as a field, since $i^2 = -1$, as opposed to the way that $\varepsilon^2 = 0$. In fact, one can continuously deform the linear automorphism $i$ into the nilpotent $\varepsilon$ by representing both of them as 2×2 real matrices. If one defines the one-parameter family of matrices:

$$\sigma(\lambda) = \begin{bmatrix} 0 & 1 \\ -\lambda & 0 \end{bmatrix} \tag{13.21}$$

then one sees that for $\lambda = 0$ the matrix represents $\varepsilon$, while for $\lambda = +1$ it represents $i$.

**1. Functions of complex variables**. Since the theory of functions of complex variables is quite vast and the elements are commonly taught, we shall simply summarize some of the formulas that are relevant to our immediate purposes.

If:

$$z = x + iy = r(\cos\theta + i\sin\theta)$$

is a complex number then some of the basic functions that one encounters in terms of real variables take the complex form:

$$z^n = r^n (\cos n\theta + i\sin n\theta), \tag{14.1}$$

$$e^z = e^x(\cos y + i\sin y), \qquad \text{so} \qquad e^{iy} = \cos y + i\sin y, \tag{14.2}$$

$$\begin{aligned} \sin z &= \sin(x+iy) = \sin x\cos iy + \cos x\sin iy \\ &= \sin x\cosh y + i\cos x\sinh y, \end{aligned} \tag{14.3}$$

$$\begin{aligned} \cos z &= \cos(x+iy) = \cos x\cos iy - \sin x\sin iy \\ &= \cos x\cosh y - i\sin x\sinh y, \end{aligned} \tag{14.4}$$

since:

$$\sin ix = i\sinh x, \qquad \cos ix = \cosh x. \tag{14.5}$$

Some other useful formulas that we will need are the complex analogues of the usual real formulas:

$$\cos^2 z + \sin^2 z = 1, \tag{14.6}$$

$$\cos (z + z') = \cos z \cos z' - \sin z \sin z', \tag{14.7}$$

$$\sin (z + z') = \sin z \cos z' + \cos z \sin z', \tag{14.8}$$

so, in particular:

$$\sin 2z = 2 \sin z \cos z, \tag{14.9}$$

$$\cos 2z = \cos^2 z - \sin^2 z. \tag{14.10}$$

One also finds that:

$$(\sin z)^* = \sin \alpha^*, \qquad (\cos z)^* = \cos z^*. \tag{14.11}$$

**2. The group of complex rotations.** Although our ultimate objective will be the representation of the proper, orthochronous Lorentz group, nevertheless, for the purposes of quaternions, its representations by means of $SO(3; \mathbb{C})$ and $SL(2; \mathbb{C})$ will appear most directly. Thus, we shall begin by examining what happens when one complexifies Euclidian space $E^3$ to $E^3_{\mathbb{C}}$ and then show how that relates to the more physically familiar Lorentz group.

We define $E^3_{\mathbb{C}}$ to be $(\mathbb{C}^3, <.,.>)$, where $<.,.>$ is the Euclidian scalar product. That is, a complex frame $\{\mathbf{e}_i, i = 1, 2, 3\}$ allows one to express every complex vector $\mathbf{v}$ by a linear combination $v^i \mathbf{e}_i$ with complex components $v^i$, and that frame is said to be *orthonormal* iff:

$$<\mathbf{e}_i, \mathbf{e}_i> = \delta_{ij}, \tag{15.1}$$

which makes:

$$<\mathbf{v}, \mathbf{w}> = \delta_{ij} \, v^i \, w^j = \sum_{i=1}^{3} v^i w^i, \tag{15.2}$$

and this differs from the corresponding real expression only by the fact that the resulting number is complex. Thus, the definitions of orthogonality and normality do not change.

However, when one defines the norm-squared of any complex vector:

$$\| \mathbf{v} \|^2 = \sum_{i=1}^{3} (v^i)^2, \tag{15.3}$$

one finds that, unlike the real analogue, it does not have to be positive-definite. That is, non-zero solution of the quadratic equation $\| \mathbf{v} \|^2 = 0$ can exist, and one calls such vectors *null vectors*. If one puts the vector $\mathbf{v}$ into real-plus-imaginary form $\mathbf{a} + i\mathbf{b}$ then one sees that:

$$\| \mathbf{v} \|^2 = \| \mathbf{a} \|^2 - \| \mathbf{b} \|^2 + 2i <\mathbf{a}, \mathbf{b}>. \tag{15.4}$$

In order for this to vanish, the real and imaginary parts must satisfy the conditions:

$$\| \mathbf{a} \| = \| \mathbf{b} \|, \qquad <\mathbf{a}, \mathbf{b}> = 0. \qquad (15.5)$$

That is, the real vectors $\mathbf{a}$ and $\mathbf{b}$ must have the same length and they must be orthogonal.

This situation is quite fundamental to the theory of electromagnetism, in which the vector $\mathbf{a}$ becomes the electric field strength $\mathbf{E}$ and the vector $\mathbf{b}$ becomes the magnetic field strength $\mathbf{B}$. The two vectors can be combined into a complex vector $\mathbf{E} + i\mathbf{B}$, which is a concept that goes back at least as far as Riemann [1], and was expanded upon by Conway [2], Silberstein [3], Majorana [4], and Oppenheimer [5] in various contexts. They can also be combined into a 2-form $d\tau \wedge E + \#\mathbf{B}$, where $d\tau$ is the proper time 1-form that allows one to decompose four-dimensional Minkowski spacetime $\mathfrak{M}^4$ into a one-dimensional proper time axis $T$ and a complementary spatial subspace $\Sigma$, and $\# : \Lambda_1\Sigma \to \Lambda^2\Sigma$, $\mathbf{v} \mapsto i_v\mathcal{V}_s$ is the Poincaré isomorphism that one gets from a choice of spatial volume element $\mathcal{V}_s$. The null vectors or 2-forms include the fields of electromagnetic waves, but not exclusively.

However, since our immediate interest in this monograph is kinematics, we shall not go further into such matters at the moment. We shall, however, revisit them in the context of complex line geometry later in this chapter.

A *complex orthogonal* transformation is still defined to be a $\mathbb{C}$-linear isomorphism $L$: $E_{\mathbb{C}}^3 \to E_{\mathbb{C}}^3$ with the property that:

$$<L\mathbf{v}, L\mathbf{w}> = <\mathbf{v}, \mathbf{w}> \qquad \text{for all } \mathbf{v}, \mathbf{w}, \qquad (15.6)$$

and this still implies the basic property of the matrix $[L]$ of any complex orthogonal transformation that:

$$[L]^{-1} = [L]^{\mathrm{T}}. \qquad (15.7)$$

This still implies that:

$$\det(L) = \pm 1, \qquad (15.8)$$

although the group $O(3; \mathbb{C})$ of all complex orthogonal transformations of $E_{\mathbb{C}}^3$ does not split into two connected components, because $+ 1$ and $- 1$ can be connected to each other in the complex plane by a continuous path that does not go through 0. The set of transformations with the positive sign on their determinant includes the identity, and is therefore a subgroup, which we denote $SO(3; \mathbb{C})$ and call the *proper complex orthogonal group in three-dimensions*. If one introduces an orientation on $\mathbb{C}^3$ then one can think of its elements a orientation-preserving complex orthogonal transformations, since they all have $\det(L) = 1$.

However, just as we now have null vectors in $E_{\mathbb{C}}^3$, we also find that although the typical element $L \in SO(3; \mathbb{C})$ can be expressed in real-plus-imaginary form, that representation is not as physically illuminating as when one uses polar decomposition to express $L$ as a product $RB$ of a real rotation $R$ and a matrix $B$ that will be seen to correspond to a Lorentz boost. Once again, the presence of null vectors complicates the Gram-Schmidt process by which one orthonormalizes $L$ into $R$, but we shall see that the expression of matrices in real-plus-imaginary form makes this decomposition quite elementary in the context of the Lie algebra $\mathfrak{so}(3; \mathbb{C})$ of infinitesimal generators of one-parameter subgroups of orientation-preserving complex orthogonal transformations.

Meanwhile, one can still use the complexified elementary rotation matrices $R(\theta, 0, 0)$, $R(0, \phi, 0)$, $R(0, 0, \psi)$ as the generators of all complex rotations, except that now the Euler angles are complex. However, using the identities that we established in the previous section, one can factor an elementary complex rotation into the product of a real rotation and an imaginary one. We illustrate this for matrices in $SO(2; \mathbb{C})$, but the principle is the same for the elementary rotations in $SO(3; \mathbb{C})$. If $\alpha = \theta + i\beta$ is a complex angle then:

$$\begin{bmatrix} \cos\alpha & -\sin\alpha \\ \sin\alpha & \cos\alpha \end{bmatrix} = \begin{bmatrix} \cos\theta\cosh\beta - i\sin\theta\sinh\beta & -\sin\theta\cosh\beta - i\cos\theta\cosh\beta \\ \sin\theta\cosh\beta + i\cos\theta\cosh\beta & \cos\theta\cosh\beta - i\sin\theta\sinh\beta \end{bmatrix}$$

$$= \begin{bmatrix} \cos\theta & -\sin\theta \\ \sin\theta & \cos\theta \end{bmatrix} \begin{bmatrix} \cosh\beta & -i\sinh\beta \\ i\sinh\beta & \cosh\beta \end{bmatrix}.$$

The right-hand matrix can also be written in the form:

$$\begin{bmatrix} \cos i\beta & -\sin i\beta \\ \sin i\beta & \cos i\beta \end{bmatrix},$$

which makes it clear that one is dealing with a planar rotation through an imaginary angle.

One finds that in this two-complex-dimensional case the imaginary rotations do form a group under multiplication, and it is of real dimension one, but since $\cosh\beta$ and $\sinh\beta$ are asymptotic to $e^{\beta}$, the Lie group $SO(2; i\mathbb{R})$ that is generated by all such matrices is not compact. Since it connected, it must therefore be diffeomorphic to a line, while the real rotations form a one-real-dimensional Lie group that is diffeomorphic to a circle. In fact, the group $SO(2; \mathbb{C})$ then becomes isomorphic to the group $(\mathbb{C}^*, \times)$ of non-zero complex numbers under multiplication.

One notes that the eigenvalues of matrices in $SO(2; \mathbb{C})$ immediately take the form:

$$\lambda = \cos\alpha + i\sin\alpha,$$

but since $\alpha$ is complex, when one expands it into its real components $\theta + i\beta$ the ultimate result is:

$$\lambda = e^{\beta}(\cos\theta + i\sin\theta),$$

which again includes all non-zero complex numbers.

However, although one can commute the two matrices in this case, that is only because they act about the same complex axis. In three complex dimensions, when two rotations act about different axes they do not commute, in general. Hence, if one factors $R(\alpha, \beta, \gamma)$ into the product $R(\alpha)\, R(\beta)\, R(\gamma)$ of three complex elementary rotations, and then factors the complex rotations into the products of real and imaginary rotations, then the resulting sequence:

$$R(\theta)\, R(\beta_x)\, R(\phi)\, R(\beta_y)\, R(\psi)\, R(\beta_x)$$

cannot generally be rearranged into a product of three real rotations and a product of three imaginary ones.

By differentiating a curve through the identity transformation, one can see that a typical element of $\mathfrak{so}(3; \mathbb{C})$ is again an anti-symmetric matrix $\varpi$, except that now it consists of complex entries. One still has the condition $\mathrm{Tr}(\varpi) = 0$ that corresponds to the condition $\det(L) = 1$ on the finite transformations.

When one uses the bilinearity of the Lie bracket, which is now assumed to be $\mathbb{C}$-bilinearity, one sees that if one expresses $\varpi$ as $\omega + i\zeta$ then one gets:

$$[\varpi, \varpi'] = [\omega, \omega'] - [\zeta, \zeta'] + i([\omega, \zeta'] + [\zeta, \omega']). \tag{15.9}$$

In particular, $[\omega, \omega']$ and $[i\zeta, i\zeta']$ both belong to $\mathfrak{so}(3; \mathbb{R})$, while $[\omega, i\zeta]$ will belong to its imaginary complement $i\,\mathfrak{so}(3; \mathbb{R})$. Hence, although $\mathfrak{so}(3; \mathbb{R})$ is a subalgebra of $\mathfrak{so}(3; \mathbb{C})$, its complement $i\,\mathfrak{so}(3; \mathbb{R})$ is not. Therefore, the decomposition $\mathfrak{so}(3; \mathbb{C}) = \mathfrak{so}(3; \mathbb{R}) \oplus i\,\mathfrak{so}(3; \mathbb{R})$ of vector spaces does not correspond to a direct sum of subalgebras.

If one takes a real basis for $\mathfrak{so}(3; \mathbb{C})$ in the form of $\{\mathbf{e}_i, i\mathbf{e}_i, i = 1, 2, 3\}$ then one finds the commutation relations for the Lie algebra $\mathfrak{so}(3; \mathbb{C})$ immediately from those of $\mathfrak{so}(3; \mathbb{R})$ and bilinearity:

$$[\mathbf{e}_i, \mathbf{e}_j] = \varepsilon_{ijk}\, \mathbf{e}_k, \qquad [\mathbf{e}_i, i\mathbf{e}_j] = \varepsilon_{ijk}\, i\mathbf{e}_k, \qquad [i\mathbf{e}_i, i\mathbf{e}_j] = -\,\varepsilon_{ijk}\, \mathbf{e}_k\ . \tag{15.10}$$

Thus, although the vector space of real vectors defines a sub-Lie-algebra, from the last set of relations, the vector space of imaginary ones does not.

When one looks at the situation involving the eigenvalues of complex orthogonal transformations, one sees that there are two immediate differences from the real case: Firstly, polynomials with complex coefficients do not have to always admit conjugate pairs of roots, and secondly, they are always factorizable into the product of powers of linear factors; i.e., they are always *reducible*. The general form for a characteristic polynomial will then be $(\lambda - \lambda_1)(\lambda - \lambda_1)(\lambda - \lambda_3)$, naively.

Thus, when one considers that for an orthogonal $L$ the equation $L\mathbf{v} = \lambda\mathbf{v}$ is equivalent to the equation $L^{\mathrm{T}}\mathbf{v} = (1/\lambda)\,\mathbf{v}$, and that both $L$ and $L^{\mathrm{T}}$ have the same eigenvalues, one sees that the only symmetry we can generally find in the roots of the characteristic polynomial for a given $L$ is that if $\lambda$ is a root then so is $1/\lambda$. Hence, we can further specify the form as $(\lambda \pm 1)(\lambda - \lambda_1)(\lambda - 1/\lambda_2)$, since the only numbers that equal their reciprocals are $\pm 1$. However, $\lambda = -1$ does not correspond to a proper rotation.

The fact that any proper complex orthogonal transformation $L$ must have 1 as an eigenvalue still implies the existence of an axis for any complex rotation. However, it is now a complex line, which is then a real plane in $\mathbb{C}^3$. Hence, it consists of a real axis for the real rotational part, as well as another one for the imaginary rotation.

### 3. The algebra of complex quaternions [6-10].

If $\{\mathbf{e}_\mu,\ \mu = 0, \ldots, 3\}$ is the canonical basis for $\mathbb{C}^4$ then a typical complex quaternion can be represented in the form:

$$q = q^0 + q^i \mathbf{e}_i\,, \tag{16.1}$$

in which the components $q^\mu = p^\mu + ir^\mu$ are now assumed to be complex numbers. Therefore, analogous to what we did with dual quaternions, we can also express a typical complex quaternion as the sum of a real quaternion and an imaginary one:

$$q = p + ir = (p^0 + p^i \mathbf{e}_i) + i(r^0 + r^i \mathbf{e}_i). \tag{16.2}$$

Thus in addition to the quaternion conjugation automorphism, one can also introduce complex conjugation and the adjunction operator:

$$\overline{q} = \overline{p} + i\overline{r}\,, \qquad\qquad q^* = p - ir, \qquad\qquad q^\dagger = \overline{q}^* = \overline{p} - i\overline{r}\,. \tag{16.3}$$

These automorphisms define projections of $\mathbb{H}_{\mathbb{C}}$ onto direct summands in decomposition of $\mathbb{H}_{\mathbb{C}}$ by polarizing the identity operator:

$$I = CS + CV = \mathrm{Re} + \mathrm{Im} = H^+ + H^-\,, \tag{16.4}$$

in which one then has:

$$CS(q) = \tfrac{1}{2}(q + \overline{q})\,, \qquad CV(q) = \tfrac{1}{2}(q - \overline{q})\,, \tag{16.5}$$

$$\mathrm{Re}(q) = \tfrac{1}{2}(q + q^*)\,, \qquad \mathrm{Im}(q) = \tfrac{1}{2}(q - q^*)\,, \tag{16.6}$$

$$H^+(q) = \tfrac{1}{2}(q + q^\dagger)\,, \qquad H^-(q) = \tfrac{1}{2}(q - q^\dagger)\,. \tag{16.7}$$

Note that a self-adjoint quaternion will have the form:

$$q = q^0 + i\,q^i\,\mathbf{e}_i \qquad\qquad (q^\mu \text{ all real}), \tag{16.8}$$

while an anti-self-adjoint quaternion will have the form:

$$q = i\, q^0 + q^i\, \mathbf{e}_i \qquad (q^\mu \text{ all real}). \qquad (16.9)$$

The real dimensions of the spaces of complex scalars, complex vectors, real quaternions, imaginary quaternions, self-adjoint complex quaternions, and anti-self-adjoint complex quaternions are then 2, 6, 4, 4, 4, 4, respectively.

The multiplication of two complex quaternions $q$, $q'$ follows from the assumption of bilinearity, if one preserves the same multiplication table for the basis elements as in the real case and uses the basic property $i^2 = -1$:

$$q\, q' = (pp' - rr') + i(pr' + rp'). \qquad (16.10)$$

which can also be expressed in complex scalar plus complex vector form:

$$q\, q' = (q^0\, q'^0 - <\mathbf{q}, \mathbf{q}'>) + (q^0\mathbf{q}' + q'^0\mathbf{q} + \mathbf{q} \times \mathbf{q}'), \qquad (16.11)$$

in which all of the scalars and vectors are complex, now.

This multiplication is still associative, but not commutative, and has a unity element 1, just like the real quaternions, but it is no longer a division algebra since it has divisors of zero. In fact, one can show that the only complex division algebra, up to isomorphism is $\mathbb{C}$ itself (see Dickson [**11**], pp. 126).

In order to give an example of a pair of divisors of zero, one again looks at $q\bar{q}$ :

$$q\bar{q} = (q^0)^2 + <\mathbf{q}, \mathbf{q}> \equiv \|\, q\,\|^2. \qquad (16.12)$$

Whereas, in the real case this would have to vanish for any non-zero $q$, in the complex case, this is no longer true. One then refers to the complex quaternions for which $\|\, q\,\|$ vanishes as the *null quaternions.* Hence, any null quaternion and its conjugate represent divisors of zero.

One can introduce the following complex scalar products by using complex components for the quaternions, this time:

$$(q, q') = CS(qq') = q^0 q'^0 - <\mathbf{q}, \mathbf{q}'> = \eta_{\mu\nu} q^\mu q'^\nu, \qquad (16.13)$$
$$<q, q'> = CS(q\bar{q}') = q^0 q'^0 + <\mathbf{q}, \mathbf{q}'> = \delta_{\mu\nu} q^\mu q'^\nu. \qquad (16.14)$$

Thus, the first one makes $\mathbb{C}^4$ into a complex Minkowski space, while the second one makes it into a complex Euclidian one. From the definition of the (Euclidian) norm in (16.12), one can also say that:

$$\|\, q\,\|^2 = <q, q>. \qquad (16.15)$$

If one puts $q$ and $q'$ into real-plus-imaginary form $p + ir$, $p' + ir'$ then their scalar product takes the form:

$$<q, q'> = <p, p'> - <r, r'> + i(<p, r'> + <p', r>). \qquad (16.16)$$

Thus:

$$\|\, q\,\|^2 = \|\, p\,\|^2 - \|\, r\,\|^2 + 2i <p, r>. \qquad (16.17)$$

One can then characterize a null quaternion by the pair of quadratic conditions on the real quaternions $p$, $r$:

$$\| p \|^2 = \| r \|^2, \qquad <p, r> = 0. \qquad (16.18)$$

Thus, the $p$ and $r$ must both lie on a real 3-sphere of radius $\| p \|^2$, or rather, they must each lie separately on a disjoint pair of real 3-spheres in $\mathbb{C}^4 = \mathbb{R}^4 \times \mathbb{R}^4$, while the second condition once more singles out the Study quadric in $\mathbb{R}^4 \times \mathbb{R}^4$. It is interesting that the change of coefficient ring from $\mathbb{D}$ to $\mathbb{C}$ has not affected the condition for the vanishing of the pure dual or imaginary part of the scalar product, even though one has defined two different real algebras over $\mathbb{R}^4 \times \mathbb{R}^4$. This is related to the fact that in a sense the translations of $E^3$ are a non-relativistic ($c \rightarrow \infty$) version of the Lorentz boosts.

We can examine the possible existence of nilpotents of degree two and idempotents by specializing the general expression (16.11) for the product of complex quaternions to an expression for the square of one:

$$q^2 = (q^0)^2 - <\mathbf{q}, \mathbf{q}> + 2q^0\mathbf{q} . \qquad (16.19)$$

Of course, the only essential difference between this expression and the corresponding one for real quaternions is in the fact that now everything is complex. However, that still implies some new consequences.

First we set $q^2 = 0$ and get the same conditions as in the real case, which we repeat for the sake of logical continuity:

$$(q^0)^2 = <\mathbf{q}, \mathbf{q}>, \qquad q^0\mathbf{q} = 0.$$

As before, setting $\mathbf{q} = 0$ still makes $q^0 = 0$, which is still trivial, and setting $q^0 = 0$ still makes $<\mathbf{q}, \mathbf{q}> = 0$, but now that we are dealing with complex vectors this equation can admit non-trivial solutions. If we express $\mathbf{q}$ in real-plus-imaginary form as $\mathbf{p} + i\mathbf{r}$ then this condition expands into:

$$<\mathbf{p}, \mathbf{p}> = <\mathbf{r}, \mathbf{r}>, \qquad <\mathbf{p}, \mathbf{r}> = 0. \qquad (16.20)$$

As we shall see when we discuss complex line geometry, the quadric that is defined by the last condition in the three-complex-dimensional vector space of pure quaternions relates to something else that has been well-studied ([7]), namely, the Klein quadric. In fact, in electromagnetism, if one replaces $\mathbf{p}$ and $\mathbf{r}$ with $\mathbf{E}$ and $\mathbf{B}$, respectively, then one finds that the two conditions on nilpotents of degree two amount to the same necessary (but not sufficient) conditions on electromagnetic field strengths in order for them to represent the fields of electromagnetic waves.

We thus conclude that $\mathbb{H}_\mathbb{C}$ does, in fact, admit nilpotent elements of degree two, which are moreover, physically significant.

---

[7]  No pun intended!

As for idempotents, if we set $q^2 = q$ in (16.19) then this gives the necessary conditions:

$$q^0 = (q^0)^2 - <\mathbf{q}, \mathbf{q}>, \quad \mathbf{q} = 2q^0\mathbf{q}. \tag{16.21}$$

These are also the same as for the real case, except that now everything is complex. If one sets $\mathbf{q} = 0$ in the latter equation then this would imply $q^0 = 0$ in the former one, as before. Otherwise, $q^0 = 1/2$, as before, which still implies that:

$$<\mathbf{q}, \mathbf{q}> = -\tfrac{1}{4}, \tag{16.22}$$

except that now it admits non-trivial solutions. In real-plus-imaginary form, it gives:

$$<\mathbf{p}, \mathbf{p}> - <\mathbf{q}, \mathbf{q}> = -\tfrac{1}{4}, \qquad <\mathbf{p}, \mathbf{q}> = 0, \tag{16.23}$$

which differs from the nilpotent case in the first equation, but not the second one.

These necessary conditions are clearly sufficient.

If we replace $\mathbf{q}$ with $\|\mathbf{q}\|\mathbf{u}$, where $\|\mathbf{u}\| = 1$, then one finds that $\|\mathbf{q}\| = i/2$. Thus, we can express any idempotent in $\mathbb{H}_{\mathbb{C}}$ in the form:

$$q = \tfrac{1}{2}(1 + i\mathbf{u}). \tag{16.24}$$

A complex quaternion is a *unit quaternion* iff $\|q\|^2 = 1$, which then leads to the pair of real quadrics in $\mathbb{R}^4 \times \mathbb{R}^4$:

$$\|p\|^2 - \|r\|^2 = 1, \qquad <p, r> = 0. \tag{16.25}$$

Thus, one is still dealing with the Study quadric, although the quadric that is defined by the unity constraint is no longer homogeneous. However, if one reverts to the complex form of the norm-squared:

$$\|q\|^2 = (q^0)^2 + (q^1)^2 + (q^2)^2 + (q^3)^2 \tag{16.26}$$

then one sees that the unit complex quaternions simply define a complex 3-sphere of unit radius, just as the unit real quaternions defined a real 3-sphere of unit radius.

One now sees how to define the inverse element to any invertible element from this. If $q$ is not a null quaternion then one sets:

$$q^{-1} = \frac{\overline{q}}{\|q\|^2}, \tag{16.27}$$

and one sees that $q^{-1}$ is, in fact, the multiplicative inverse of $q$. Of course, this, too, is simply the complex analogue of the result for real quaternions.

Thus, the multiplicative group $\mathbb{C}Q^*$ of all invertible (i.e., non-null) complex quaternions can be factored into a product $\mathbb{C}^* \times \mathbb{C}Q_1$ of the multiplicative group of non-zero complex numbers and the multiplicative group of unit complex quaternions. The group $\mathbb{C}Q_1$ then consists of all points on a complex-Euclidian 3-sphere in $\mathbb{C}^4$ of unit radius; in fact, as we shall demonstrate in the next subsection, it is a complex Lie group that is isomorphic to $SL(2; \mathbb{C})$, which is a two-to-one simply-connected covering group of the proper, orthochronous Lorentz group.

In analogy to what we de did for real and dual quaternions, we find that there is also a polar form for any non-null complex quaternion:

$$q = \| q \| (\cos \tfrac{1}{2} \alpha + \sin \tfrac{1}{2} \alpha \, \hat{\mathbf{u}}), \tag{16.28}$$

in which the complex angle $\alpha = \theta + i\beta$ represents both an angle of rotation $\theta$ around a spatial axis, which is generated by the real vector $\mathbf{a}$ and a boost with a rapidity parameter $\beta$ in a direction that is defined by a real vector $\mathbf{b}$. One can then obtain the angle $\alpha$ from any non-null $q$ by means of:

$$\cos \tfrac{1}{2} \alpha = \frac{q^0}{\| q \|}. \tag{16.29}$$

The complex vector $\hat{\mathbf{u}} = \mathbf{a} + i\mathbf{b}$ is then assumed to be a complex unit vector, so:

$$1 = \| \hat{\mathbf{u}} \|^2 = - \hat{\mathbf{u}} \, \hat{\mathbf{u}} = - \mathbf{aa} + \mathbf{bb} - i(\mathbf{ab} + \mathbf{ba}) = \| \mathbf{a} \|^2 - \| \mathbf{b} \|^2 + 2i\!<\!\mathbf{a}, \mathbf{b}\!>, \tag{16.30}$$

which imposes the conditions on $\mathbf{a}$ and $\mathbf{b}$ that follow from the restriction of (16.25) to complex quaternions of vector type:

$$1 = \| \mathbf{a} \|^2 - \| \mathbf{b} \|^2, \qquad <\!\mathbf{a}, \mathbf{b}\!> = 0. \tag{16.31}$$

One can obtain $\hat{\mathbf{u}}$ from:

$$\mathbf{q} = \| \mathbf{q} \| \hat{\mathbf{u}}, \qquad \| \mathbf{q} \| = \| q \| \sin \tfrac{1}{2} \alpha, \tag{16.32}$$

as long as $\| \mathbf{q} \|$ is non-null. If $\mathbf{q} = \mathbf{p} + i\mathbf{r}$ then one gets the individual component vectors $\mathbf{a}$ and $\mathbf{b}$ of $\hat{\mathbf{u}}$ from:

$$\| \mathbf{q} \| \, \hat{\mathbf{u}} = \| \mathbf{q} \| (\mathbf{a} + i\mathbf{b}) = \mathbf{p} + i\mathbf{r},$$

which makes:

$$\mathbf{a} = \frac{\mathbf{p}}{\| \mathbf{q} \|}, \qquad \mathbf{b} = \frac{\mathbf{r}}{\| \mathbf{q} \|}. \tag{16.33}$$

When one takes the commutator bracket of two complex quaternions, one gets:

$$[q, q'] = 2 \, \mathbf{q} \times \mathbf{q}'. \tag{16.34}$$

This tells us that the complex vector quaternions define a complex three-dimensional Lie algebra that is isomorphic to $\mathfrak{so}(3; \mathbb{C})$, and that the center of the quaternion algebra consists of all complex quaternions of complex scalar type. One finds that $\mathfrak{so}(3; \mathbb{C})$ is also isomorphic to $\mathfrak{sl}(2; \mathbb{C})$, as well as $\mathfrak{so}(1, 3)$. In fact, when one expresses an element of $\mathfrak{so}(3; \mathbb{C})$ in real-plus-imaginary form:

$$\boldsymbol{\Omega} = \boldsymbol{\omega} + i\boldsymbol{\beta}, \tag{16.35}$$

one finds that $\boldsymbol{\omega}$ represents an infinitesimal Euclidian rotation (i.e., an element of $\mathfrak{so}(3; \mathbb{R})$), while $\boldsymbol{\beta}$ is effectively an infinitesimal Lorentz boost.

It is amusing that although one is often told in elementary special relativity that the vector cross product is no longer useful in four real dimensions, nonetheless, its extension to three *complex* dimensions has a fundamental special-relativistic significance, after all.

For the sake of completeness, we include the anti-commutator bracket of two complex quaternions:

$$\{q, q'\} = 2((q, q') + q^0 \mathbf{q}' + q'^0 \mathbf{q}). \tag{16.36}$$

When $q$ and $q'$ are complex vectors, one gets:

$$\{\mathbf{q}, \mathbf{q}'\} = -2{<}\mathbf{q}, \mathbf{q}'{>}, \tag{16.37}$$

and one has:

$$\mathbf{q}\,\mathbf{q}' = -{<}\mathbf{q}, \mathbf{q}'{>} + \mathbf{q} \times \mathbf{q}', \tag{16.38}$$

as in the real case.

Equation (16.37) is suggestive of the Clifford algebra of real $E^3$ − which is also of real dimension eight − although one finds that $\mathbb{H}_{\mathbb{C}}$ is not the completion of the even subalgebra of $\mathcal{C}(3; \delta_{ij})$ by associating the imaginary quaternions with the odd subspace, but the complexification of that even real subalgebra to the even complex subalgebra of the Clifford algebra over $E_{\mathbb{C}}^3$.

It will be useful to note the following facts, which are to be contrasted with the corresponding dual results. If $\mathbf{q} = \mathbf{p} + i\mathbf{r}$, $\mathbf{q}' = \mathbf{p}' + i\mathbf{r}'$ then:

$$\mathbf{q} \times \mathbf{q}' = \mathbf{p} \times \mathbf{p}' - \mathbf{r} \times \mathbf{r}' + i(\mathbf{r} \times \mathbf{p}' + \mathbf{p} \times \mathbf{r}'), \tag{16.39}$$

and the triple vector product rule still holds for complex vectors:

$$(\mathbf{a} \times \mathbf{b}) \times \mathbf{c} = {<}\mathbf{a}, \mathbf{c}{>}\,\mathbf{b} - {<}\mathbf{a}, \mathbf{b}{>}\,\mathbf{c}. \tag{16.40}$$

When we get to the representation of spinors by complex quaternions, we shall need to know a bit about the left and right ideals of the algebra $\mathbb{H}_{\mathbb{C}}$. A left ideal $\mathcal{I}$ of $\mathbb{H}_{\mathbb{C}}$ is, by definition, a linear subspace such that $\mathbb{H}_\mathbb{C}\,\mathcal{I} \le \mathcal{I}$, and therefore a sub-algebra. If one has an idempotent $\varepsilon$ then it will generate a left ideal $\mathcal{I}(\varepsilon) = \mathbb{H}_\mathbb{C}\,\varepsilon$. As mentioned above, the element $\varepsilon^\mathfrak{L} = 1 - \varepsilon$ is also an idempotent that is orthogonal to $\varepsilon$ so one has a decomposition of unity:

$$1 = \varepsilon + \varepsilon^\mathfrak{L}.$$

The left ideal generated by $\varepsilon^\mathfrak{L}$ will then be $\mathcal{I}(\varepsilon^\mathfrak{L}) = \mathbb{H}_\mathbb{C}\,\varepsilon^\mathfrak{L}$, and one has a decomposition:

$$\mathbb{H}_\mathbb{C} = \mathcal{I}(\varepsilon) \oplus \mathcal{I}(\varepsilon^\mathfrak{L}).$$

Since we have four complex dimensions to start with, naively, the only distinct non-trivial possibilities for direct sum decompositions are into vector subspaces of 1+3 dimensions and 2+2 dimensions. However, one immediately sees that a one-dimensional ideal in $\mathbb{H}_\mathbb{C}$ − i.e., a complex line through the origin − cannot exist, since the line through the origin of any non-zero complex quaternion will be rotated by some quaternion, so the only possibility is a pair of complementary two-complex dimensional sub-algebras.

We next examine the nature of an idempotent $\varepsilon$ in $\mathbb{H}_\mathbb{C}$ more closely in its scalar-plus-vector representation:

$$\varepsilon = \varepsilon^0 + \boldsymbol{\varepsilon}.$$

From the definition of an idempotent, one has:

$$\varepsilon^2 = (\varepsilon, \varepsilon) + 2\varepsilon^0\boldsymbol{\varepsilon} \;=\; \varepsilon^0 + \boldsymbol{\varepsilon}\,.$$

This makes:

$$\varepsilon^0 = (\varepsilon, \varepsilon) = \tfrac{1}{2}, \tag{16.41}$$

so:

$$\varepsilon^\mathfrak{L} = (1 - \varepsilon^0) - \boldsymbol{\varepsilon} = \bar{\varepsilon}\,. \tag{16.42}$$

Hence, any idempotent $\varepsilon$ is a null quaternion:

$$\| \varepsilon \|^2 = \varepsilon\bar{\varepsilon} \;=\; \varepsilon\varepsilon^\mathrm{c} = 0. \tag{16.43}$$

Furthermore, any other element $q\varepsilon$ in $\mathcal{I}(\varepsilon)$ will have to be a null quaternion, since:

$$\| q\varepsilon \|^2 = q\varepsilon\overline{q\varepsilon} = q\varepsilon\bar{\varepsilon}\,\bar{q} = 0.$$

Any linear combination of such elements can be written in the form:

$$\sum \lambda_a (q^a \varepsilon) = \left(\sum \lambda_a q^a\right)\varepsilon = q\varepsilon,$$

so every element of $\mathcal{I}(\varepsilon)$ is a null quaternion.

Therefore, any left ideal in $\mathbb{H}_{\mathbb{C}}$ defines a two-dimensional vector subspace in the three-dimensional quadric hypersurface that is defined by the null quaternions.

As we observed above, there are only two (complex) degrees of freedom in the choice of idempotents in $\mathbb{H}_{\mathbb{C}}$, since their scalar part is always 1/2, while their vector part must lie on the complex unit sphere $\mathbb{C}S^2$ in $\mathbb{C}^3$. Therefore, since the null quaternions lie on a three-complex-dimensional hypersurface, not all null quaternions are going to be idempotents. In particular, the nilpotents of degree two are also null quaternions.

One example of an idempotent can be given by assuming that $\mathbf{u}$ is in the $z$ direction:

$$\varepsilon = \tfrac{1}{2}(1 \pm i\mathbf{e}_3). \tag{16.44}$$

The complementary idempotent $\varepsilon^c$ is then the conjugate quaternion, which simply inverts the choice of sign.

One can adapt the basis $\{\mathbf{e}_\mu\}$ to the direct sum by defining:

$$\boldsymbol{l}_0 = \tfrac{1}{2}[(\mathbf{e}_2 - i\mathbf{e}_1) + \chi(\mathbf{e}_0 + i\mathbf{e}_3)], \qquad \boldsymbol{l}_1 = \tfrac{1}{2}[-\varphi(\mathbf{e}_2 + i\mathbf{e}_1) + (\mathbf{e}_0 - i\mathbf{e}_3)], \tag{16.45}$$

$$\boldsymbol{l}_2 = \tfrac{1}{2}[(\mathbf{e}_2 - i\mathbf{e}_1) + \varphi(\mathbf{e}_0 + i\mathbf{e}_3)], \qquad \boldsymbol{l}_3 = \tfrac{1}{2}[-\chi(\mathbf{e}_2 + i\mathbf{e}_1) + (\mathbf{e}_0 - i\mathbf{e}_3)], \tag{16.46}$$

in which:

$$\varphi = \frac{\varepsilon^0 - i\varepsilon^3}{\varepsilon^2 + i\varepsilon^1}, \qquad \chi = \frac{\varepsilon^0 + i\varepsilon^3}{\varepsilon^2 + i\varepsilon^1}. \tag{16.47}$$

From these two equations, and the conditions on $\varepsilon$ that are imposed by the fact that it is an idempotent, one can solve for the components of $\varepsilon$ in terms of $\varphi, \chi$:

$$\varepsilon^0 = \tfrac{1}{2}, \qquad \varepsilon^1 = -\frac{i}{2}\frac{1 - \varphi\chi}{\varphi - \chi}, \qquad \varepsilon^2 = \frac{1}{2}\frac{1 + \varphi\chi}{\varphi - \chi}, \qquad \varepsilon^3 = -\frac{i}{2}\frac{\varphi + \chi}{\varphi - \chi}. \tag{16.48}$$

Thus, we have one way of parameterizing the two-dimensional imaginary sphere that the idempotents represent.

Analogous remarks apply to the case of right ideals. If confusion might arise, we could distinguish left ideals from right ideals by means of appropriate subscripts, but we shall generally treat them separately in the sequel.

**4. The action of the Lorentz group on complex quaternions.** We shall first observe that the extension of the representation of real quaternions by 2×2 complex matrices to a representation of complex quaternions still comes from the association of the basis elements $\mathbf{e}_\mu$ with the 2×2 complex matrices $\tau_\mu$. Thus the matrix $[q]$ that gets associated with a complex quaternion $q$ is still of the form:

$$[q] = q^\mu \, \tau_\mu \; = \begin{bmatrix} q^0 + iq^1 & q^2 - iq^3 \\ q^2 + iq^3 & q^0 - iq^1 \end{bmatrix}. \tag{17.1}$$

except that now the components $q^\mu$ are generally complex.

Since the complex vector spaces $\mathbb{H}_\mathbb{C}$ and $M(2; \mathbb{C})$ both have complex dimension four, this association is a $\mathbb{C}$-linear isomorphism of those vector spaces. Thus, it is by going from real numbers to complex ones that one completes the association of quaternions with 2×2 complex matrices, as the complex quaternions of real type still define a real subspace of $M(2; \mathbb{C})$ of real dimension four.

Furthermore, since the product of complex quaternions still goes to the corresponding product of matrices:

$$[qq'] = [q][q'],$$

the association is also an isomorphism of complex algebras. One also still has:

$$\det[q] = \| \, q \, \|^2,$$

except that the determinant and norm involved are complex, now. Thus, the null quaternions go to matrices of zero determinant, which are then the non-invertible ones, and the unit quaternions go to elements of $SL(2; \mathbb{C})$. In fact, that association is also an isomorphism of complex Lie groups of complex dimension three. This association of unit complex quaternions with matrices in $SL(2; \mathbb{C})$ also shows quite clearly that the latter complex Lie group is diffeomorphic, as a complex manifold, to the complex 3-sphere.

Now that one also has an adjunction automorphism defined, one finds that, in fact:

$$[q]^\dagger = [q^\dagger]. \tag{17.2}$$

Thus, self-adjoint complex quaternions go to Hermitian matrices, while the anti-self-adjoint quaternions go to anti-Hermitian matrices. One finds that when one polarizes the matrices of $M(2; \mathbb{C})$ that have zero trace – i.e., the elements of $\mathfrak{sl}(2; \mathbb{C})$ – with respect to the Hermitian conjugate operation, the effect is to express any element of that Lie algebra as a sum of a zero-trace anti-Hermitian matrix, which then belongs to the Lie algebra $\mathfrak{su}(2)$, and a zero-trace Hermitian one. Since the Lie bracket of two Hermitian matrices is anti-Hermitian:

$$[H, H']^\dagger = (HH' - H'H)^\dagger = H'^\dagger H^\dagger - H^\dagger H'^\dagger = H'H - HH' = - [H, H'],$$

those matrices do not define a complex Lie subalgebra of $\mathfrak{sl}(2; \mathbb{C})$, but only a three-real-dimensional subspace.

One can easily define the isomorphism of Lie algebras $\mathfrak{sl}(2; \mathbb{C})$ with $\mathfrak{so}(3; \mathbb{C})$, since both have complex dimension three. First, one defines their $\mathbb{C}$-linear isomorphism as vector spaces by the association of the basis vectors $\tau_i$ with the elementary three-dimensional infinitesimal rotation matrices $J_i$, $i = 1, 2, 3$. One then observes that since the basis elements satisfy the same commutation relations the linear isomorphism is a Lie algebra isomorphism. Under this association, one sees that anti-Hermitian matrices go to real infinitesimal rotation matrices, while the Hermitian ones go to imaginary rotation matrices. As we saw above, this means that the anti-Hermitian matrices represent infinitesimal generators of real Euclidian rotations, while the Hermitian ones represent the infinitesimal generators of real Minkowski space boosts; i.e., pure Lorentz transformations.

Thus, we see that the Lie group of unit complex quaternions is isomorphic to $SL(2; \mathbb{C})$, which doubly covers the proper, orthochronous Lorentz group $SO_+(3, 1)$. Therefore, if one can represent Minkowski space as a vector subspace in $\mathbb{H}_\mathbb{C}$ then one can represent the action of $SO_+(3, 1)$ on Minkowski space by the action of unit complex quaternions on other quaternions.

In fact, we saw above that the basic automorphisms of $\mathbb{H}_\mathbb{C}$ define several subspaces that have real dimension four. In particular, the spaces of real, imaginary, self-adjoint, and anti-self-adjoint quaternions all have real dimension four. Furthermore, one can define a scalar product on each of them that makes them isometric to real Minkowski space. In the case of the real and imaginary quaternions, the scalar product is $(q, q') = CS(qq')$, while in the case of self-adjoint and anti-self-adjoint quaternions, the scalar product is $<q, q'> = CS(q\overline{q'})$.

The problem at hand is to find linear actions of $\mathbb{C}Q_1$ on $\mathbb{H}_\mathbb{C}$ that leave these subspaces invariant and act by isometries on them. There are five basic actions of the group $\mathbb{C}Q_1$ of unit complex quaternions on $\mathbb{H}_\mathbb{C}$. We introduce the following terminology:

1. Left-multiplication:

$$\mathbb{C}Q_1 \times \mathbb{H}_\mathbb{C} \to \mathbb{H}_\mathbb{C}, \qquad (u, q) \mapsto uq,$$

2. Right-multiplication:

$$\mathbb{H}_\mathbb{C} \times \mathbb{C}Q_1 \to \mathbb{H}_\mathbb{C}, \qquad (u, p) \mapsto qu,$$

3. Complex congruence:

$$\mathbb{C}Q_1 \times \mathbb{H}_\mathbb{C} \to \mathbb{H}_\mathbb{C}, \qquad (u, q) \mapsto uqu^*,$$

4. Conjugate congruence:

$$\mathbb{C}Q_1 \times \mathbb{H}_\mathbb{C} \to \mathbb{H}_\mathbb{C}, \qquad (u, q) \mapsto uq\overline{u},$$

5.  Adjoint congruence:

$$\mathbb{C}Q_1 \times \mathbb{H}_{\mathbb{C}} \to \mathbb{H}_{\mathbb{C}}, \qquad (u, q) \mapsto uqu^{\dagger}.$$

There is also a "chiral" action of the product group $\mathbb{C}Q_1 \times \mathbb{C}Q_1$:

$$\mathbb{C}Q_1 \times \mathbb{C}Q_1 \times \mathbb{H}_{\mathbb{C}} \to \mathbb{H}_{\mathbb{C}}, \qquad (u, u', q) \mapsto uqu'.$$

One can also consider the conjugate actions of all of the above, when the element $u$ gets applied to $q$ by way of $\bar{u}$. As it turns out, going from an action to its conjugate action amounts to the difference between vectors and covectors; i.e., duality in the real linear spaces that are being represented by subspaces of quaternions. However, since any unit quaternion $p$ has a conjugate $\bar{u}$ that is also a unit quaternion the conjugate actions of $\mathbb{C}Q_1$ will always have the same invariant subspaces as the direct action.

The actual representation of the conventional vector spaces of physical tensors, such as scalars, vectors, covectors, bivectors, 2-forms, 3-vectors, and 3-forms, and spinors then comes down to identifying invariant subspaces of the various actions that were defined above that have the same dimensions as the corresponding spaces of real or complex tensor or spinor objects with those invariant subspaces.

The action that we are calling complex congruence above is certainly a linear action and has $\mathbb{H}$ and $i\mathbb{H}$ as invariant subspaces. However, when one takes the scalar product $(p', q')$ of two real or imaginary quaternions $p' = upu^*$, $q' = uqu^*$ one gets:

$$(p', q') = CS(p', q') = CS(upu^* uqu^*). \tag{17.3}$$

One sees that since generally $u^* \neq \bar{u}$ the product in parentheses does not reduce to $CS(upqu^*)$, which would then reduce to $(p, q)$. Thus, the action in question does not generally act by isometries on its invariant subspaces.

A more physically convenient action is what we are calling adjoint congruence, which takes all self-adjoint quaternions to other self-adjoint quaternions, and similarly for the anti-self-adoint ones. This action then corresponds to the action of matrices in $SL(2; \mathbb{C})$ on matrices in $M(2; \mathbb{C})$ by an analogous conjugation that involves the Hermitian adjoint. The fact that this action leaves the subspaces $H_+$ and $H_-$ invariant follows from the general discussion in Chap. I, sec. 3.

One finds that the linear action just defined is also by isometries of the restriction of the complex quaternion scalar product to self-adjoint and anti-self-adjoint complex quaternions. Since these latter quaternions take the forms (16.8) and (16.9), respectively, the restrictions in question take the form:

$$<q, q'>_+ = \eta_{\mu\nu} q^\mu q^\nu, \qquad\qquad <q, q'>_- = - <q, q'>_+ , \tag{17.4}$$

respectively. Thus, the restrictions of the complex quaternion scalar product to those subspaces make them isometric to Minkowski space, as well as linearly isomorphic. The fact that linear action in question preserves this scalar product follows from direct calculation: If $x' = uxu^\dagger$, $y' = uyu^\dagger$ then one has:

$$<x', y'>_+ = CS(x'\overline{y'}) = CS(uxu^\dagger \overline{u}^\dagger \overline{y}\, \overline{u}) = CS(ux\overline{y}\,\overline{u}) = CS(u\overline{u})CS(x\overline{y}) = <x, y>_+,$$

and similarly for $<x', y'>_-$.

Since Minkowski space vectors can be represented as self-adjoint or anti-self-adjoint complex quaternions, so can linear frames; in particular, one can represent a Lorentzian frame $\{\mathbf{f}_\mu, \mu = 0, \ldots, 3\}$ by a corresponding Lorentzian frame in $H_+$ or $H_-$, which we shall denote by the same symbols. If one then defines the transformation of this frame by a unit complex quaternion $u$ by:

$$\mathbf{f}'_\mu = u\, \mathbf{f}_\mu\, u^\dagger \tag{17.5}$$

then one sees that since $\mathbf{f}_\mu$ is a frame on $H_\pm$ the new frame $\mathbf{f}'_\mu$ can be expressed in terms of it by way of a coefficient matrix $L^\nu_\mu$:

$$\mathbf{f}'_\mu = \mathbf{f}_\nu L^\nu_\mu. \tag{17.6}$$

Since the transformation that is defined by $u$ is an isometry of $H_\pm$, one infers that the matrix $L^\nu_\mu$ must be Lorentz-orthogonal. One can also see that the matrix $L^\nu_\mu$ must be real, since the fact that the transformation takes self-adjoint elements to other self-adjoint elements implies that $[\mathbf{f}'_\mu]^\dagger = \mathbf{f}'_\mu$, and when one expands this using (17.6), one gets:

$$[\mathbf{f}_\nu L^\nu_\mu]^\dagger = [\mathbf{f}_\nu]^\dagger [L^\nu_\mu]^* = \mathbf{f}_\nu [L^\nu_\mu]^* = \mathbf{f}_\nu L^\nu_\mu,$$

which is possible only if:

$$[L^\nu_\mu]^* = L^\nu_\mu.$$

Furthermore, one sees that since $u$ and $-u$ both produce the same Lorentz transformation, the association of $\pm u$ to $L^\nu_\mu$ becomes the 2-1 covering map $SL(2; \mathbb{C}) \rightarrow SO_+(3, 1)$.

One also sees that since the antipodal pair $\{u, -u\}$ on the complex 3-sphere is associated with a complex line through the origin of $\mathbb{C}^4$ – namely, all points of the form $\lambda u$, with $\lambda$ complex – it then defines a point in $\mathbb{C}P^3$. This shows one a direct path to proving that the identity component of the Lorentz group is diffeomorphic to $\mathbb{C}P^3$ as a manifold.

The action of $\mathbb{C}Q_1$ on $\mathbb{H}_\mathbb{C}$ that takes the form of conjugate congruence has the spaces of complex scalar and complex vectors as invariant subspaces, from the general considerations regarding automorphisms of algebras above. It also acts by isometries, since $q' = uq\overline{u}$ then:

$$q'\,\overline{q'} = uq\overline{u}u\overline{q}\,\overline{u} = q\,\overline{q}\,.$$

If one reverts to the polar form for the unit complex quaternion $u$ then one finds for their action on complex vectors $\mathbf{v}$ in the manner that is currently at issue that:

$$
\begin{aligned}
u\mathbf{v}\overline{u} &= (\cos\tfrac{1}{2}\alpha + \sin\tfrac{1}{2}\alpha\ \mathbf{u})\ \mathbf{v}\ (\cos\tfrac{1}{2}\alpha - \sin\tfrac{1}{2}\alpha\ \mathbf{u}) \\
&= \cos\alpha\,\mathbf{v} + (1 - \cos\alpha)\ <\mathbf{u},\mathbf{v}>\ \mathbf{u} + \sin\alpha\ \mathbf{u}\times\mathbf{v},
\end{aligned}
\tag{17.7}
$$

which still has the form of Rodrigues's formula, except that now the angle $\alpha$ and unit vector $\mathbf{u}$ are both complex. Thus, one is now dealing with a rotation along one axis and a boost along another.

Since the complex vector space of complex quaternions of vector type is $\mathbb{C}$-linearly isomorphic to $\mathbb{C}^3$ by a choice of complex 3-frame – such as $\mathbf{e}_i$ – one can also represent the action of a unit complex quaternion $u$ on complex 3-frames by a 3×3 complex matrix $L_i^j$:

$$u\,\mathbf{e}_i\,\overline{u} = \mathbf{e}_j L_i^j\,. \tag{17.8}$$

Because this action is also an isometry of the complex Euclidian scalar product, one sees that the matrix $L_i^j$ must be complex orthogonal. Thus, since $\pm\,u$ once more produce the same $L_i^j$, the association of $\pm\,u$ with $L_i^j$ amounts to the two-to-one covering map $SL(2; \mathbb{C}) \to SO(3; \mathbb{C})$, and one also finds that $SO(3; \mathbb{C})$ is isomorphic to $SO_+(3, 1)$. One also notes that the covering $SL(2; \mathbb{C}) \to SO(3, \mathbb{C})$ is simply the complexification of the covering $SU(2) \to SO(3; \mathbb{R})$, while the covering $\mathbb{C}S^3 \to \mathbb{C}\mathrm{P}^3$ is the complexification of the covering $S^3 \to \mathbb{R}\mathrm{P}^3$. Thus, one can regard the transition from non-relativistic physics to relativistic physics as having as much to do with the transition from real to complex numbers as it does with the transition from three dimensions to four.

Because the complex quaternions of vector type represent a real vector space of dimension six, it seems, on the surface of things, that they would be less well-adapted to the motions of points in a four-dimensional real vector space, such as Minkowski space. However, this is only partially true, since they are eminently adapted to the problem of describing the motions of lines and 2-planes in $\mathbb{R}^4$, as well bivectors and 2-forms, which are at the root of modern electromagnetism. We shall discuss this shortly, but first we want to discuss how the action of the Lorentz group on spinors can be represented by a linear action of the unit complex quaternions on the complex quaternions.

One can represent $SL(2; \mathbb{C})$ spinors by means of complex quaternions, as long one has chosen an idempotent $\varepsilon$. As discussed above, such an element generates both a left ideal $\mathcal{I}_L(\varepsilon) = \mathbb{H}_\mathbb{C}\varepsilon$ and a right ideal $\mathcal{I}_R(\varepsilon) = \varepsilon\,\mathbb{H}_\mathbb{C}$, both of which are two-dimensional sub-algebras that are composed of nothing but null quaternions. We also note that any left ideal is, by definition, an invariant subspace of the action of $SL(2; \mathbb{C})$ on $\mathbb{H}_\mathbb{C}$ by left-multiplication. Analogously, any right ideal is an invariant subspace of the action of $SL(2; \mathbb{C})$ on $\mathbb{H}_\mathbb{C}$ by right-multiplication.

The representation of a unit quaternion $u$ by an invertible 4×4 complex matrix $[L(u)]^\mu_\nu$ that one gets from left-multiplication is the one that we discussed in Chapter I that one gets by choosing a basis for $\mathbb{H}_\mathbb{C}$ and using the structure constants $a^\mu_{\kappa\nu}$ and the components $u^\kappa$ to define:

$$[L(u)]^\mu_\nu = a^\mu_{\kappa\nu}u^\kappa. \qquad (17.9)$$

In the case of the structure constants for the quaternions (which are the same regardless of the coefficient ring), one can simply write out the components of the product $uq$ and identify the matrix that takes $q^\mu$ to $(uq)^\mu$. This makes:

$$[L(u)]^\mu_\nu = \begin{bmatrix} u^0 & -u^1 & -u^2 & -u^3 \\ u^1 & u^0 & -u^3 & u^2 \\ u^2 & u^3 & u^0 & -u^1 \\ u^3 & -u^2 & u^1 & u^0 \end{bmatrix} = a^\mu_{0\nu}u^0 + a^\mu_{1\nu}u^1 + a^\mu_{2\nu}u^2 + a^\mu_{3\nu}u^3, \qquad (17.10)$$

with

$$a^\mu_{0\nu} = I, \quad a^\mu_{1\nu} = \begin{bmatrix} 0 & -1 & 0 & 0 \\ 1 & 0 & 0 & 0 \\ 0 & 0 & 0 & -1 \\ 0 & 0 & 1 & 0 \end{bmatrix}, \quad a^\mu_{2\nu} = \begin{bmatrix} 0 & 0 & -1 & 0 \\ 0 & 0 & 0 & 1 \\ 1 & 0 & 0 & 0 \\ 0 & -1 & 0 & 0 \end{bmatrix}, \quad a^\mu_{3\nu} = \begin{bmatrix} 0 & 0 & 0 & -1 \\ 0 & 0 & -1 & 0 \\ 0 & 1 & 0 & 0 \\ 1 & 0 & 0 & 0 \end{bmatrix}. \qquad (17.11)$$

With the notation of Chapter II, we can express these matrices in the form:

$$a^\mu_{0\nu} = \begin{bmatrix} \tau_0 & 0 \\ 0 & \tau_0 \end{bmatrix}, \quad a^\mu_{1\nu} = -\begin{bmatrix} \tau_2 & 0 \\ 0 & \tau_2 \end{bmatrix}, \quad a^\mu_{2\nu} = i\begin{bmatrix} 0 & \tau_1 \\ -\tau_1 & 0 \end{bmatrix}, \quad a^\mu_{3\nu} = i\begin{bmatrix} 0 & \tau_3 \\ -\tau_3 & 0 \end{bmatrix}. \qquad (17.12)$$

Similarly, for right multiplication one finds that the matrix $[R(u)]^\mu_\nu$ comes from:

$$[R(u)]^\mu_\nu = a^\mu_{\nu\kappa}u^\kappa, \qquad (17.13)$$

which makes:

$$[R(u)]^\mu_\nu = \begin{bmatrix} u^0 & -u^1 & -u^2 & -u^3 \\ u^1 & u^0 & u^3 & -u^2 \\ u^2 & -u^3 & u^0 & u^1 \\ u^3 & u^2 & -u^1 & u^0 \end{bmatrix} = a^\mu_{\nu 0}u^0 + a^\mu_{\nu 1}u^1 + a^\mu_{\nu 2}u^2 + a^\mu_{\nu 3}u^3 , \qquad (17.14)$$

with:

$$a^\mu_{\nu 0} = I, \quad a^\mu_{\nu 1} = \left[\begin{array}{cc:cc} 0 & -1 & 0 & 0 \\ 1 & 0 & 0 & 0 \\ \hdashline 0 & 0 & 0 & 1 \\ 0 & 0 & -1 & 0 \end{array}\right], \quad a^\mu_{\nu 2} = \left[\begin{array}{cc:cc} 0 & 0 & -1 & 0 \\ 0 & 0 & 0 & -1 \\ \hdashline 1 & 0 & 0 & 0 \\ 0 & 1 & 0 & 0 \end{array}\right], \quad a^\mu_{\nu 3} = \left[\begin{array}{cc:cc} 0 & 0 & 0 & -1 \\ 0 & 0 & 1 & 0 \\ \hdashline 0 & -1 & 0 & 0 \\ 1 & 0 & 0 & 0 \end{array}\right], \qquad (17.15)$$

or

$$a^\mu_{\nu 0} = \left[\begin{array}{c:c} \tau_0 & 0 \\ \hdashline 0 & \tau_0 \end{array}\right], \quad a^\mu_{\nu 1} = \left[\begin{array}{c:c} -\tau_2 & 0 \\ \hdashline 0 & \tau_2 \end{array}\right], \quad a^\mu_{\nu 2} = \left[\begin{array}{c:c} 0 & -\tau_0 \\ \hdashline \tau_0 & 0 \end{array}\right], \quad a^\mu_{\nu 3} = \left[\begin{array}{c:c} 0 & \tau_2 \\ \hdashline -\tau_2 & 0 \end{array}\right]. \qquad (17.16)$$

One can easily show these actions preserve the scalar product $<.,.>$, since if $q' = uq$ then:

$$q'\overline{q'} = uq\overline{q}\,\overline{u} = u\,\overline{u} \cdot q\overline{q} = q\overline{q} ,$$

while if $q' = qu$ then:

$$q'\overline{q'} = qu\,\overline{u}\,\overline{q} = q\overline{q} .$$

Of course, since the ideals $\mathcal{I}_L(\varepsilon)$ and $\mathcal{I}_R(\varepsilon)$ are composed exclusively of null quaternions, the scalar product is always zero.

Similarly, $\mathcal{I}_{L,R}(\varepsilon^c) = \mathcal{I}_{L,R}(\overline{\varepsilon})$ is an invariant subspace under these actions, respectively, and thus, if $\varepsilon$ is a primitive idempotent then this defines a decomposition of $\mathbb{H}_\mathbb{C}$ into irreducible representations:

$$\mathbb{H}_\mathbb{C} = \mathcal{I}_L(\varepsilon) \oplus \mathcal{I}_L(\overline{\varepsilon}) = \mathcal{I}_R(\varepsilon) \oplus \mathcal{I}_R(\overline{\varepsilon}) .$$

Since the invariant subspaces $\mathcal{I}_L(\varepsilon)$ or $\mathcal{I}_R(\varepsilon)$, as well as their complements, are two-dimensional, if one chooses a complex 2-frame $\{\mathbf{f}_a , a = 1, 2\}$ in any of them then the action of $\mathbb{C}Q_1$ by left or right multiplication will produce another complex 2-frame in the same space, regardless of what unit quaternion $u$ one chooses to act on the frame. Hence, one can associate $u$ with an invertible 2×2 matrix $\Lambda^b_a$ or $\Gamma^b_a$ by way of:

$$u\,\mathbf{f}_a = \mathbf{f}_b\Lambda^b_a , \qquad\qquad \mathbf{f}_a u^\dagger = \mathbf{f}_b\Gamma^b_a , \qquad (17.17)$$

depending upon whether one is dealing with left or right translation, respectively.

Since the action preserves the scalar product, the matrix will have determinant 1; i.e., it will belong to $SL(2; \mathbb{C})$. Thus, the action of the group of complex unit quaternions on the invariant subspaces of left and right multiplication behaves like the defining representation of $SL(2; \mathbb{C})$ on $\mathbb{C}^2$ by left or right matrix multiplication on column or row vectors, respectively.

Following Blaton [**6**], we will call quaternions, when given the action of right-multiplication by $u^\dagger$, where $u$ is a unit quaternion, *semi-quaternions of the first kind* and when they are given the action of left-multiplication by $u$, they will be called *semi-quaternions of the second kind*. In order to distinguish them, we shall also use a single underbar for the first kind and a double underbar for the second kind.

One immediately sees that products of the form $\underline{p}\underline{\underline{q}}$ also transform like vectors, since:

$$\underline{p}'\underline{\underline{q}}' = u\underline{p}\underline{\underline{q}}u^\dagger = u(\underline{p}\underline{\underline{q}})u^\dagger. \tag{17.18}$$

This is analogous to the way that one expresses vectors in Minkowski space as tensor products of spinors. Of course, the components of that tensor product will define a 2×2 complex matrix, and we know that $SL(2; \mathbb{C})$ acts on such things by matrix conjugation, since it acts on the column vectors of $\mathbb{C}^2$ by left-multiplication and the row vectors of $\mathbb{C}^{2*}$ by right-multiplication.

In general, products of the form $\underline{p}\,\overline{\underline{q}}$ are invariant under right-multiplication, while products of the form $\overline{\underline{p}}\,\underline{q}$ are invariant under left-multiplication:

$$\underline{p}'\overline{\underline{q}}' = \underline{p}u^\dagger\overline{u}^\dagger\overline{\underline{q}} = \underline{p}(\overline{u}u)^\dagger\overline{\underline{q}} = \underline{p}\,\overline{\underline{q}},$$
$$\overline{\underline{\underline{p}}}'\underline{\underline{q}}' = \overline{\underline{\underline{p}}}\,\overline{u}\,u\,\underline{\underline{q}} = \overline{\underline{\underline{p}}}\,\underline{\underline{q}}.$$

A semi-quaternion $q$ is of the first (second, resp.) kind iff $q^\dagger$ is of the second (first, resp.) kind:

$$(\underline{q}\,u^\dagger)^\dagger = u\,\underline{q}^\dagger, \qquad (u\underline{\underline{q}})^\dagger = \underline{\underline{q}}^\dagger u^\dagger.$$

If one decomposes a semi-quaternion of the first kind $\underline{q}$ into its projections $\underline{q}\varepsilon$ and $\underline{q}\overline{\varepsilon}$ in the left ideals $I_l(\varepsilon)$ and $I_l(\overline{\varepsilon})$ for a primitive idempotent $\varepsilon$ and its conjugate $\overline{\varepsilon}$ then $\underline{q}\varepsilon$ and $\underline{q}\overline{\varepsilon}$ will be referred to as *spinors of the first kind*. Similarly, the decomposition into right ideals gives *spinors of the second kind*. The general quaternion will then represent a sum of linearly independent spinors of the same kind, or a *bispinor*:

$$q = q\varepsilon + q\overline{\varepsilon} = \varepsilon q + \overline{\varepsilon}q. \tag{17.19}$$

It is important to point out that the definition of a spinor in this way clearly depends upon the choice of idempotent $\varepsilon$.

The "chiral" action of $CQ_1$ on $\mathbb{H}_{\mathbb{C}}$ that we defined above, which takes $(u_1, u_2, q)$ to $u_1 q u_2$ is also an isometry of the complex Euclidian scalar product, since if $q' = u_1 q u_2$ then one must have:

$$q' \overline{q}' = u_1 q u_2 \overline{u}_2 \overline{q} \, \overline{u}_1 = q \overline{q} \, .$$

Thus, if we express this action by means of an invertible 4×4 complex matrix $M_\nu^{\,\mu}$ by means of:

$$u_1 \mathbf{e}_\mu \, u_2 = \mathbf{e}_\nu M_\mu^{\,\nu}$$

then we can see that one must have that $M_\nu^{\,\mu} \in SO(4; \mathbb{C})$. One thus defines a homomorphism $SL(2; \mathbb{C}) \times SL(2; \mathbb{C}) \mapsto SO(4; \mathbb{C})$, $(u_1, u_2) \mapsto M_\nu^{\,\mu}$ that also defines an isomorphism at the level of Lie algebras.

The chiral action can also be written as the product of a left-translation matrix and a right translation matrix. In fact, the associativity of the quaternion multiplication implies that the two matrices must commute:

$$(u_1 \mathbf{e}_\mu) \, u_2 = \mathbf{e}_\nu L_\mu^{\,\nu}(u_1) R_\mu^{\,\kappa}(u_2) \, , \quad u_1 (\mathbf{e}_\mu \, u_2) = \mathbf{e}_\nu R_\kappa^{\,\nu}(u_2) L_\mu^{\,\kappa}(u_1) \, ,$$

so

$$L_\kappa^{\,\nu}(u_1) R_\mu^{\,\kappa}(u_2) = R_\kappa^{\,\nu}(u_2) L_\mu^{\,\kappa}(u_1) \, .$$

One can identify various subgroup actions by restricting the chiral action to preserving various invariant subspaces. For instance, if one demands that it take real quaternions to real ones and imaginary quaternions to imaginary ones then this would make $q'^{\,*} = \pm \, q'$, which would imply that:

$$(u_1 q u_2)^* = u_1^* q^* u_2^* = \pm \, u_1^* q u_2^* = \pm \, u_1 q u_2 \, .$$

If one pre-multiplies both sides of the last equality by $\tilde{u}_1$ and post-multiplies it by $\tilde{u}_2$ then one gets $\tilde{u}_1 u_1^* q^* u_2^* \tilde{u}_2 = q$ and if this is to be true for all real $q$ then one must have:

$$u_1 = u_1^* \, , \qquad u_2 = u_2^* \, ;$$

i.e., $u_1$ and $u_2$ must be real quaternions.

Similarly, if one considers the corresponding statement regarding the matrix $M_\nu^{\,\mu}$ then one must have:

$$(\mathbf{e}_\nu M_\mu^{\,\nu})^* = \mathbf{e}_\nu^* (M_\mu^{\,\nu})^* = \pm \, \mathbf{e}_\nu (M_\mu^{\,\nu})^* = \pm \, \mathbf{e}_\nu M_\mu^{\,\nu}$$

This then implies that $(M_\mu^\nu)^* = M_\nu^\mu$, which makes the matrix real. Thus, one now has a homomorphism $SU(2) \times SU(2) \to SO(4; \mathbb{R})$, $(u_1, u_2) \mapsto M_\nu^\mu$.

If the action is to take scalars to scalars and vectors to vectors then one must have:

$$\overline{u_1 \mathbf{e}_\mu u_2} = \overline{u}_2 \overline{\mathbf{e}}_\mu \overline{u}_1 = \pm \overline{u}_2 \mathbf{e}_\mu \overline{u}_1 = \pm u_1 \mathbf{e}_\mu u_2,$$

which leads to the condition:

$$u_1 = \overline{u}_2,$$

and the action reduces to that of conjugate congruence.

In order to get back to the proper, orthochronous, Lorentz group, one needs only to restrict the action in such a way that it takes (anti) self-adjoint quaternions to other (anti) self-adjoint ones:

$$q'^\dagger = u_2^\dagger q^\dagger u_1^\dagger = \pm u_2^\dagger q u_1^\dagger = \pm q' = \pm u_1 q u_2,$$

which implies that one must have:

$$u_1 = u_2^\dagger.$$

The action then reduces to adjoint congruence.

Since this also implies that the matrix $L_\mu^\nu(u_1)$ must be the complex conjugate of $R_\mu^\nu(u_2)$, we have the following theorem that goes back to Einstein and Mayer [**12**], and is also discussed by Scherrer [**13**], Blaton [**6**], and Lanczos [**7**]:

**Theorem:**

*Any proper, orthogonal Lorentz matrix $M_\nu^\mu$ can be expressed as the product $L_\kappa^\mu(u_1) R_\nu^\kappa(u_2)$ of two matrices such that*:

1. $L_\kappa^\mu(u_1) R_\nu^\kappa(u_2) = R_\kappa^\mu(u_2) L_\nu^\kappa(u_1)$.
2. $R_\mu^\nu(u_2) = (L_\mu^\nu(u_1))^*$.

We summarize the various actions of the group of complex unit quaternions in Table 1.

**5. Some complex line geometry**. In order to see how bivectors relate to lines, one needs to define the Plücker-Klein embedding of the manifold of lines in $\mathbb{RP}^3$ in the vector space $\Lambda_2 \mathbb{R}^4$ of bivectors over $\mathbb{R}^4$. First, one notes that under the projection $\mathbb{R}^4 - \{0\} \to \mathbb{RP}^3$, $\mathbf{x} \mapsto [\mathbf{x}]$, which takes any point in $\mathbb{R}^4$ that is not the origin to the line through the origin that goes through it, a line $[\mathbf{x}, \mathbf{y}]$ in $\mathbb{R}\mathrm{P}^3$ will be the projection of a 2-plane through the origin in $\mathbb{R}^4$. If one spans that plane by means of two linearly independent vectors – say $\mathbf{x}$ and $\mathbf{y}$ – then there is a bivector $\mathbf{x} \wedge \mathbf{y}$ that gets associated with that line.

Table 1. Representation of metric spaces by invariant subspaces of $\mathbb{H}_\mathbb{C}$.

| Metric space | Invariant subspace | Isometric action of unit quaternions |
|---|---|---|
| $(\mathbb{C}, |\,|^2)$ | Scalar quaternions | Conjugate congruence |
| $(\mathbb{C}^2, 0)$ | Left or right ideals | Left or right multiplication |
| $(\mathbb{R}^4, \delta_{\mu\nu})$ | Real or imaginary quaternions | Complex congruence |
| $(\mathbb{R}^4, \eta_{\mu\nu})$ | (anti-) self-dual quaternions | Adjoint congruence |
| $(\mathbb{C}^3, \delta_{ij})$ | Pure quaternions | Conjugate congruence |

However, the choice of spanning vectors is not unique, and if one chooses any other pair – say $\mathbf{x}'$ and $\mathbf{y}'$ – then they will be related to the first pair by an invertible linear transformation:

$$\begin{bmatrix} \mathbf{x}' \\ \mathbf{y}' \end{bmatrix} = \begin{bmatrix} a_1^1 & a_2^1 \\ a_1^2 & a_2^2 \end{bmatrix} \begin{bmatrix} \mathbf{x} \\ \mathbf{y} \end{bmatrix}. \tag{18.1}$$

One then finds that

$$\mathbf{x}' \wedge \mathbf{y}' = \det[a]\, \mathbf{x} \wedge \mathbf{y}. \tag{18.2}$$

Thus, any two choices for spanning vectors will produce bivectors that differ only by a non-zero scalar multiple. Hence, they all define the same point $[\mathbf{x} \wedge \mathbf{y}]$ in $\mathrm{P}\Lambda_2\mathbb{R}^4$, which is the five-dimensional real projective space of lines through the origin of $\Lambda_2\mathbb{R}^4$. If $\mathbb{R}\mathrm{P}_1^3$ is the manifold of lines in $\mathbb{R}\mathrm{P}^3$ then the map $\mathbb{R}\mathrm{P}_1^3 \to \mathrm{P}\Lambda_2\mathbb{R}^4$, $[\mathbf{x}, \mathbf{y}] \mapsto [\mathbf{x} \wedge \mathbf{y}]$ that takes a line to the equivalence class of bivectors that are associated with it is an embedding that one calls the *Plücker-Klein embedding*.

The image of the latter embedding does not consist of all bivectors, but only ones that are *decomposable*; i.e., of the form $\mathbf{x} \wedge \mathbf{y}$, instead of $\mathbf{x} \wedge \mathbf{y} + \mathbf{v} \wedge \mathbf{w}$. Decomposable bivectors $\mathbf{b}$ have the characteristic property that:

$$\mathbf{b} \wedge \mathbf{b} = 0. \tag{18.3}$$

This is a homogeneous, quadratic condition on the bivectors, which then defines a quadric hypersurface in $\Lambda_2\mathbb{R}^4$, and because of the homogeneity, in $P\Lambda_2\mathbb{R}^4$, as well. This quadric is called the *Klein quadric*, and this shows that manifold of lines in $\mathbb{R}P^3$ is four-dimensional.

The connection with complex quaternions is straightforward when one first notes that the real vector space $\Lambda_2\mathbb{R}^4$ is six-dimensional, which is also the real dimension of the subspace of complex quaternions of vector type. One first assumes that $\mathbb{R}^4$ has been given a "time-space decomposition" into a direct sum $\mathbb{R} \oplus \mathbb{R}^3$, where, for example, $\mathbf{e}_0$ might span the $\mathbb{R}$ summand and $\{\mathbf{e}_i, i = 1, 2, 3\}$ might span the $\mathbb{R}^3$ summand. One then notes that a basis for $\Lambda_2\mathbb{R}^4$ can be defined by $\{\boldsymbol{\varepsilon}_i, {}^*\boldsymbol{\varepsilon}_i, i = 1, 2, 3\}$, in which:

$$\boldsymbol{\varepsilon}_i = \mathbf{e}_0 \wedge \mathbf{e}_i, \qquad {}^*\boldsymbol{\varepsilon}_i = \tfrac{1}{2}\varepsilon_{ijk}\,\mathbf{e}_j \wedge \mathbf{e}_k. \tag{18.4}$$

This not only defines a basis for $\Lambda_2\mathbb{R}^4$ as a real vector space of dimension six, but if we define the linear isomorphism $* : \Lambda_2\mathbb{R}^4 \to \Lambda_2\mathbb{R}^4$, $\mathbf{b} \mapsto {}^*\mathbf{b}$ by its effect on the basis elements:

$$*(\boldsymbol{\varepsilon}_i) = {}^*\boldsymbol{\varepsilon}_i, \qquad *({}^*\boldsymbol{\varepsilon}_i) = -\boldsymbol{\varepsilon}_i \tag{18.5}$$

then we find that the map $*$ also allows one to define a complex structure on $\Lambda_2\mathbb{R}^4$ by simply setting:

$$i\mathbf{b} = {}^*\mathbf{b}, \qquad \text{so} \qquad (\alpha + i\beta)\mathbf{b} = \alpha\mathbf{b} + \beta{}^*\mathbf{b}. \tag{18.6}$$

One thus has a way of defining complex scalar multiplication on $\Lambda_2\mathbb{R}^4$ that makes $\{\boldsymbol{\varepsilon}_i, i = 1, 2, 3\}$ into a complex basis. Under this complex structure the bivectors of "electric" type are the ones in the real subspace spanned by $\{\boldsymbol{\varepsilon}_i, i = 1, 2, 3\}$, while the ones of "magnetic" type are in the real subspace spanned by $\{{}^*\boldsymbol{\varepsilon}_i, i = 1, 2, 3\}$. Thus, the electric bivectors correspond to the real subspace of the complex vector space, while the magnetic ones correspond to the imaginary subspace. This suggest an obvious $\mathbb{C}$-linear isomorphism of $\Lambda_2\mathbb{R}^4$ to $\mathbb{C}^3$ that takes the complex basis $\{\boldsymbol{\varepsilon}_i, i = 1, 2, 3\}$ to the canonical basis $\{\mathbf{e}_i, i = 1, 2, 3\}$ in $\mathbb{C}^3$. The association of bivectors then takes the form:

$$E^i \boldsymbol{\varepsilon}_i + B^i *\boldsymbol{\varepsilon}_i \mapsto (E^i + iB^i) \, \mathbf{e}_i \, .$$

As mentioned previously, this association of a bivector (or 2-form, for that matter) with a complex 3-vector goes back to lectures of Riemann on partial differential equations in mathematical physics, and was resurrected numerous times by many other researchers to this day. The subsequent association of a bivector with a complex quaternion of vector type then becomes obvious if one regards $\mathbf{e}_i$ as also spanning the vector subspace of $\mathbb{H}$, which we have been treating as an algebra over $\mathbb{C}^4$.

The action of $\mathbb{C}Q_1$ on bivectors over $\mathbb{R}^4$ or lines in $\mathbb{R}P^3$ that is most appropriate is then conjugate congruence, which has the complex quaternions of vector type as an invariant subspace. This gives us a different way of geometrically characterizing the unit complex quaternions of vector type. As we mentioned previously, since $\hat{\mathbf{u}} = \mathbf{a} + i\mathbf{b}$ must satisfy $\| \hat{\mathbf{u}} \|^2 = - \hat{\mathbf{u}} \, \hat{\mathbf{u}} = 1$, one must have $<\mathbf{a}, \mathbf{a}> - <\mathbf{b}, \mathbf{b}> = 1$ and $<\mathbf{a}, \mathbf{b}> = 0$. Thus, the (real) spatial vectors $\mathbf{a}$ and $\mathbf{b}$ must be orthogonal, which means that their tips generate a line, namely, $(1 -\lambda)\mathbf{a} + \lambda\mathbf{b}$, and the vectors themselves span a plane through the origin.

**6. The kinematics of Lorentzian frames**. The (proper, orthochronous) Lorentz group can be represented by either its defining representation of $SO_+(3, 1)$, which acts on real Minkowski space $\mathfrak{M}^4$, $SO(3; \mathbb{C})$, whose defining representation is on complex Euclidian space $E_{\mathbb{C}}^3$, or $SL(2; \mathbb{C})$, whose defining representation is on $\mathbb{C}^2$, which is implicitly given the trivial scalar product. In the last section, we saw how each of the representations can take the form of linear actions of the group $\mathbb{C}Q_1$ of complex unit quaternions on various invariant subspaces of $\mathbb{H}_{\mathbb{C}}$ that were associated with each action. Moreover, we saw how that action would affect the various types of frames that were appropriate to each action.

In order to go on to relativistic kinematics, we only need to start with a differentiable curve in the Lie group $\mathbb{C}Q_1$ and differentiate the action of that Lie group on the invariant subspaces in each case. Hence, the actions will all have in common some things that pertain to the differentiation of curves in $\mathbb{C}Q_1$.

If $u(t) = u^\mu(t)\mathbf{e}_\mu$ is such a sufficiently differentiable curve then its velocity vector field will be a vector field on that curve:

$$\dot{u}(t) = \frac{du}{dt} = \frac{du^{\mu}}{dt}\mathbf{e}_{\mu}, \tag{19.1}$$

as will its acceleration vector field:

$$\ddot{u}(t) = \frac{d\dot{u}}{dt} = \frac{d^2u}{dt^2} = \frac{d^2u^{\mu}}{dt^2}\mathbf{e}_{\mu}. \tag{19.2}$$

If one right-translates all points of the curve $u(t)$ to the identity element 1 by means of $u(t)^{-1}$ then the differential map to each individual right translation takes the curve $\dot{u}(t)$ in the tangent spaces $T_{u(t)}\mathbb{C}Q_1$ to a curve in $T_1\mathbb{C}Q_1$:

$$\omega(t) = \dot{u}(t)\,u(t)^{-1}. \tag{19.3}$$

Since $T_1\mathbb{C}Q_1$ can be identified with the Lie algebra of $\mathbb{C}Q_1$, which is isomorphic to $\mathfrak{sl}(2;\,\mathbb{C})$, one can think of the elements $\omega(t)$ as being infinitesimal Lorentz transformations, which makes the curve $\omega(t)$ represent a relativistic analogue of angular velocity for whatever frame $\mathbb{C}Q_1$ is acting on. However, since $\mathfrak{sl}(2;\,\mathbb{C})$ decomposes into a direct sum of vector spaces $\mathfrak{su}(2) \oplus \mathfrak{h}(2)$, the transformations of $\omega(t)$ include both infinitesimal rotations and infinitesimal boosts, and one can generally represent $\omega(t)$ in the form $\varpi(t) + \eta(t)$, where $\varpi(t)$ represents a curve in $\mathfrak{su}(2)$ and $\eta(t)$ represents a curve in $\mathfrak{h}(2)$.

Since the acceleration $\ddot{u}(t)$ is a curve in the second tangent bundle $T_{\dot{u}(t)}T_{u(t)}\mathbb{C}Q_1$, when one right-translates $u(t)$ back to 1, the effect is to right-translate $\dot{u}(t)$ to $\omega(t)$ and produce a curve in $T_{\omega(t)}T_1\mathbb{C}Q_1 = T_{\omega(t)}\mathfrak{sl}(2;\,\mathbb{C})$, namely:

$$\alpha(t) = \ddot{u}(t)\,u(t)^{-1}. \tag{19.4}$$

This then represents a relativistic analogue of angular acceleration.

If one takes the time derivative of $\omega(t)$ − namely:

$$\dot{\omega} = \ddot{u}u^{-1} + \dot{u}\dot{u}^{-1} = \ddot{u}u^{-1} - \dot{u}u^{-1}\dot{u}u^{-1} = \alpha - \omega\omega,$$

then one sees that:

$$\alpha = \dot{\omega} + \omega\omega. \tag{19.5}$$

When one decomposes $\omega$ into $\varpi + \eta$, this decomposes $\alpha$ into:

$$\alpha = (\dot{\varpi} + \varpi\varpi) + (\dot{\eta} + \eta\eta) + \{\varpi,\,\eta\}. \tag{19.6}$$

Thus, the contributions from the rotations and boosts must be augmented by a coupling term.

Let us now apply this to the action of $\mathbb{C}Q_1$ on Minkowski space, as it is represented by either invariant subspace $H^\pm$, namely, by adjoint congruence, which takes any $(u, q)$ to $uqu^\dagger$. If we regard the curve $q(t)$ as being produced by the action of $u(t)$ on some initial element $q_0 = q(0)$, namely:

$$q(t) = u(t)\, q_0\, u(t)^\dagger, \tag{19.7}$$

then its velocity vector field takes the form:

$$\dot q = \dot u q_0 u^\dagger + u q_0 \dot u^\dagger, \tag{19.8}$$

in which we have dropped the explicit reference to the curve parameter, for brevity.

This velocity amounts to the one that is observed by an "inertial" observer, whose velocity relative to $q(t)$ is then $\dot u$. If one substitutes:

$$q_0 = u^{-1}\, q\, u^{-\dagger} \tag{19.9}$$

in (19.8) then the result is:

$$\dot q = \dot u u^{-1} q u^{-\dagger} u^\dagger + u u^{-1} q u^{-\dagger} \dot u^\dagger = \omega q + q \omega^\dagger . \tag{19.10}$$

This is then the form that is taken by the velocity of $q(t)$ with respect to a co-moving – or non-inertial – observer.

Now that we are dealing with the Lie algebra $\mathfrak{sl}(2; \mathbb{C})$, we see that we cannot simply assume that $\omega$ is anti-Hermitian, since that is only true for the part of it that belongs to $\mathfrak{su}(2)$. The other part is then Hermitian, and if we express $\omega = \varpi + \eta$, such that $\varpi$ is anti-Hermitian and $\eta$ is Hermitian, then we see that this makes:

$$\dot q = [\varpi, q] + \{\eta, q\}. \tag{19.11}$$

A second differentiation of $q(t)$ give the acceleration in an inertial frame:

$$\ddot q = \ddot u q_0 u^\dagger + 2 \dot u q_0 \dot u^\dagger + u q_0 \ddot u^\dagger, \tag{19.12}$$

and in a co-moving frame:

$$\begin{aligned}
\ddot q &= \ddot u u^{-1} q u^{-\dagger} u^\dagger + 2 \dot u u^{-1} q u^{-\dagger} \dot u^\dagger + u u^{-1} q u^{-\dagger} \ddot u^\dagger \\
&= \alpha q + q \alpha^\dagger + 2 \omega q \omega^\dagger .
\end{aligned} \tag{19.13}$$

If we replace the moving point $q(t)$ with the moving frame:

$$\mathbf{f}_\mu(t) = u(t)\, \mathbf{f}_{0\mu}\, u(t)^\dagger \tag{19.14}$$

then one finds that its velocity takes the forms:

$$\dot{\mathbf{f}}_\mu \;=\; \dot{u}\,\mathbf{f}_{0\mu}u^\dagger + u\,\mathbf{f}_{0\mu}\dot{u}^\dagger \;=\; \omega\mathbf{f}_\mu + \mathbf{f}_\mu\,\omega^\dagger, \qquad (19.15)$$

while its acceleration takes the forms:

$$\ddot{\mathbf{f}}_\mu = \ddot{u}\,\mathbf{f}_{0\mu}u^\dagger + 2\dot{u}\,\mathbf{f}_{0\mu}\dot{u}^\dagger + u\,\mathbf{f}_{0\mu}\ddot{u}^\dagger = \alpha\mathbf{f}_\mu + \mathbf{f}_\mu\,\alpha^\dagger + 2\omega\mathbf{f}_\mu\,\omega^\dagger. \qquad (19.16)$$

If we wish to now examine the action of $\mathbb{C}Q_1$ on bivectors, we recall that they are modeled by the invariant subspace $\mathbf{C}V$ of complex vector quaternions under the action of conjugate congruence, which takes $(u, \mathbf{q})$ to $u\mathbf{q}\bar{u}$ . Thus, if $\mathbf{q}_0 = \mathbf{q}(0)$ represents an initial bivector then its time evolute can be defined by:

$$\mathbf{q}(t) = u(t)\,\mathbf{q}_0\,\bar{u}(t)\,. \qquad (19.17)$$

Although we are using a different automorphism in order to define the action of $\mathbb{C}Q_1$ , nonetheless, the calculations that follow are essentially the same, except for a change of symbol. Thus, velocity and acceleration in an inertial frame take the form:

$$\dot{q} \;=\; \dot{u}q_0\bar{u} + uq_0\dot{\bar{u}}\,, \qquad \ddot{q} \;=\; \ddot{u}q_0\bar{u} + 2\dot{u}q_0\dot{\bar{u}} + uq_0\ddot{\bar{u}}\,, \qquad (19.18)$$

while in a co-moving frame they become:

$$\dot{q} = \omega q + q\bar{\omega}, \qquad \ddot{q} = \alpha q + q\bar{\alpha} + 2\omega q\bar{\omega}\,. \qquad (19.19)$$

The kind of frame that is most appropriate to $CV$ is an oriented, orthonormal, complex 3-frame $\{\mathbf{f}_i, i = 1, 2, 3\}$. When one substitutes $\mathbf{f}_i(t)$ for $q(t)$ and $\mathbf{f}_{0i}$ for $q_0$, the last two sets of equations take the analogous forms:

$$\dot{\mathbf{f}}_i \;=\; \dot{u}\,\mathbf{f}_{0i}\,\bar{u} + u\,\mathbf{f}_{0i}\,\dot{\bar{u}} \qquad\qquad = \omega\mathbf{f}_i + \mathbf{f}_i\bar{\omega}\,, \qquad (19.20)$$

$$\ddot{\mathbf{f}}_i \;=\; \ddot{u}\,\mathbf{f}_{0i}\,\bar{u} + 2\dot{u}\,\mathbf{f}_{0i}\,\dot{\bar{u}} + u\,\mathbf{f}_{0i}\,\ddot{\bar{u}} \;=\; \alpha\mathbf{f}_i + \mathbf{f}_i\bar{\alpha} + 2\omega\mathbf{f}_i\,\bar{\omega}\,. \qquad (19.21)$$

The kinematics of spinors (see, e.g., Gürsey [**14**] or Proca [**15**]) involves the action of $\mathbb{C}Q_1$ on the invariant subspaces of either right or left multiplication, namely, the left or right ideals. If $\mathbb{H}_\mathbb{C} = \mathcal{I}_L(\varepsilon) \oplus \mathcal{I}_L(\bar{\varepsilon})$ is a decomposition of $\mathbb{H}_\mathbb{C}$ into left ideals relative to a primitive idempotent $\varepsilon$ then the action of $\mathbb{C}Q_1$ on $\mathcal{I}_L(\varepsilon)$ or $\mathcal{I}_L(\bar{\varepsilon})$ takes the form $(u, q) \mapsto uq$. Thus, if $u(t)$ is a sufficiently differentiable curve in $\mathbb{C}Q_1$ and $q_0$ is an element of either ideal then one defines a sufficiently differentiable curve $q(t)$ in the respective ideal by way of:

$$q(t) = u(t) \, q_0 \, . \tag{19.22}$$

Differentiation gives the velocity and acceleration in the inertial frame as:

$$\dot{q} = \dot{u} q_0 \, , \qquad \ddot{q} = \ddot{u} q_0 \, , \tag{19.23}$$

and in the co-moving frame, one substitutes $q_0 = u^{-1} q$ in order to get:

$$\dot{q} = \omega q, \qquad \ddot{q} = \alpha q. \tag{19.24}$$

The kind of frame that is most suited to this action is a complex, null 2-frame $\{\mathbf{f}_a, a = 1, 2\}$ for either $\mathcal{I}_L(\varepsilon)$ or $\mathcal{I}_L(\overline{\varepsilon})$. If one substitutes $\mathbf{f}_{0a}$ for $q_0$ and $\mathbf{f}_a$ for $q$ then the kinematical equations take the forms:

$$\dot{\mathbf{f}}_a = \dot{u} \, \mathbf{f}_{0a} = \omega \, \mathbf{f}_a \, , \qquad \qquad \ddot{\mathbf{f}}_a = \ddot{u} \, \mathbf{f}_{0a} = \alpha \, \mathbf{f}_a \, . \tag{19.25}$$

The only difference between the left action and the right action is that the right action also involves taking the adjoint of $u$ before acting to the right. Thus, the curve $q(t)$ in a right ideal comes from:

$$q(t) = q_0 \, u(t)^{\dagger} \, . \tag{19.26}$$

The kinematical equations can be obtained from (19.23) and (19.24) by inspection:

$$\dot{q} = q_0 \dot{u}^{\dagger} = q \omega^{\dagger}, \qquad \ddot{q} = q_0 \ddot{u}^{\dagger} = q \alpha^{\dagger}, \tag{19.27}$$

and the kinematical equations for moving frames become:

$$\dot{\mathbf{f}}_a = \mathbf{f}_{0a} \dot{u}^{\dagger} = \mathbf{f}_a \omega^{\dagger}, \qquad \qquad \ddot{\mathbf{f}}_a = \mathbf{f}_{0a} \ddot{u}^{\dagger} = \mathbf{f}_a \, \alpha^{\dagger}. \tag{19.28}$$

# CHAPTER V

## COMPLEX DUAL QUATERNIONS

**1. The group of complex rigid motions.** The group $ISO(3; \mathbb{C})$ of complex rigid motions is defined by complexification of the corresponding real group. That is, one starts with complex three-dimensional affine Euclidian space $E_{\mathbb{C}}^3 = (A_{\mathbb{C}}^3, \delta_{ij})$ upon which one has an action of the three-dimensional complex translation group $\mathbb{C}^3$.

Thus, we must begin with complex three-dimensional affine space $A_{\mathbb{C}}^3$. This is simple a space on which one has defined a simply transitive action of the complex translation group $\mathbb{C}^3$, which can either be written in the form $x \mapsto y = x + z^i$ or as an anti-symmetric function from $A_{\mathbb{C}}^3 \times A_{\mathbb{C}}^3$ to $\mathbb{C}^3$ that takes $(x, y)$ to $y - x = z^i$. Hence, if one chooses a point $O \in A_{\mathbb{C}}^3$ to serve as "origin" or reference point then any $x \in A_{\mathbb{C}}^3$ can be associated with an ordered triple of complex numbers $(z^1, z^2, z^3)$ that makes $x = O + z^i$. If one chooses a complex 3-frame $\{\mathbf{e}_1, \mathbf{e}_2, \mathbf{e}_3\}$ in $T_O A_{\mathbb{C}}^3$ then $z^i$ can be associated with the tangent vector $\mathbf{z} = z^i \mathbf{e}_i$. The group $A(3; \mathbb{C})$ of complex affine transformations acts on $GL(A_{\mathbb{C}}^3)$ on the right as:

$$(x, \mathbf{f}_i)(z^i, \tilde{L}_j^i) = (x + z^i \mathbf{f}_i, \mathbf{f}_j \tilde{L}_i^j), \tag{20.1}$$

so it acts on the coordinates $(x^i, f_j^i)$ of a complex affine frame $(x, \mathbf{f}_i)$ on the right by:

$$(z^i, L_j^i)(x^i, f_j^i) = (z^i + L_j^i x^j, L_k^i f_j^k). \tag{20.2}$$

One introduces a complex Euclidian structure on $T(A_{\mathbb{C}}^3)$ in the form of a complex scalar product on the tangent vectors $<\mathbf{v}_x, \mathbf{w}_x>$ at each point. A complex linear frame $\mathbf{e}_i$ in $T_z A_{\mathbb{C}}^3$ is said to be *orthonormal* iff:

$$<\mathbf{e}_i, \mathbf{e}_j> = \delta_{ij}, \tag{20.3}$$

so if $\mathbf{v}_z = v^i \mathbf{e}_i$ and $\mathbf{w}_z = w^i \mathbf{e}_i$ then:

$$<\mathbf{v}_z, \mathbf{w}_z> = \delta_{ij} v^i w^j. \tag{20.4}$$

A *complex rigid frame* $(z, \mathbf{e}_i)$ on $E_{\mathbb{C}}^3$ then consists of a point $z$ in complex three-dimensional affine space $A_{\mathbb{C}}^3$ and a complex-orthonormal frame $\{\mathbf{e}_i, i = 1, 2, 3\}$ in the tangent space $T_z A_{\mathbb{C}}^3$. A *complex rigid motion* is then a transformation $T$ of $A_{\mathbb{C}}^3$ whose differential preserves the Euclidian scalar product:

$$<dT/_z\mathbf{v}, dT/_z\mathbf{w}> = <\mathbf{v}, \mathbf{w}>,$$

where **v**, **w** $\in T_z A_{\mathbb{C}}^3$, so $dT/z\mathbf{v}$, $dT/z\mathbf{v} \in T_{T(z)}A_{\mathbb{C}}^3$, and also preserve the orientation of any frame.  One can then show that $dT/z$ is invertible at every $z$, and by the inverse function theorem, $T$ is locally invertible.

Since the differential of any uniform translation is zero, the group $\mathbb{C}^3$ will be a subgroup of $ISO(3; \mathbb{C})$, as will the subgroup $SO(3; \mathbb{C})$ of all complex, orientation-preserving rotations of any tangent space.  In fact, one can represent $ISO(3; \mathbb{C})$ as the semi-direct product $\mathbb{C}^3 \times_s SO(3; \mathbb{C})$, although the representation depends upon a choice of complex rigid frame.  Thus, the complex Lie group $ISO(3; \mathbb{C})$ has a complex dimension of six or a real dimension of twelve.  As a complex manifold, it is diffeomorphic to the product $\mathbb{C}^3 \times \mathbb{C}P^3$, so it is non-compact and connected, but not simply connected.  Its simply connected covering group is $\mathbb{C}^3 \times_s SL(2; \mathbb{C})$.

If $(z^i,\ R_j^i)$, $(w^j,\ S_j^i)\ \in \mathbb{C}^3 \times_s SO(3; \mathbb{C})$ then their product is:

$$(z^i, R_j^i)(w^j, S_j^i)\ = (z^i + R_j^i w^j,\ R_k^i S_j^k). \qquad (20.5)$$

The group $\mathbb{C}^3 \times_s SO(3; \mathbb{C})$ also acts on complex rigid frames in an analogous way:

$$(z, \mathbf{e}_i)(s^i, R_j^i)\ = (z + s^i\mathbf{e}_i,\ \mathbf{e}_j R_i^j). \qquad (20.6)$$

One can represent an element $(s^i, R_j^i)$ of $\mathbb{C}^3 \times_s SO(3; \mathbb{C})$ by an invertible 4×4 complex matrix if one treats the coordinates $z^i$ of $\mathbb{C}^3$ as inhomogeneous coordinates for a Plücker coordinate chart on $\mathbb{C}P^3$ and regards $\mathbb{C}^3$ as embedded in the space of homogeneous coordinates $\mathbb{C}^4 - \{0\}$ as the affine hyperplane $z^0 = 1$.  The matrix of $(s^i, R_j^i)$ is then the complexification of the corresponding real one:

$$\left[\begin{array}{c|c} 1 & 0 \\ \hline s^i & R_j^i \end{array}\right].$$

One notices that this construction is not as natural for the covering group $\mathbb{C}^3 \times_s SL(2; \mathbb{C})$, since $SL(2; \mathbb{C})$ is more closely related to the geometry of $\mathbb{C}P^2$.  However, if one restricts oneself to a subset of the form $\mathbb{C}^2 \times SL(2; \mathbb{C})$ then one can represent an element $(s^a,\ L_b^a)$ as an invertible complex 3×3 matrix:

$$\left[\begin{array}{c|c} 1 & 0 \\ \hline s^a & R^a_b \end{array}\right].$$

One notes, from (20.5), that, in fact, no $\mathbb{C}^2$ subspace $\Pi$ of $\mathbb{C}^3$ will define a subgroup of the form $\mathbb{C}^2 \times_s SO(3; \mathbb{C})$, since the issue is whether every complex rotation $R^i_j z^j$ of a $z^i \in \Pi$ will still be a vector in $\Pi$. That is, $\Pi$ will have to be an invariant subspace of the action of $SO(3; \mathbb{C})$ on $\mathbb{C}^3$ by left-multiplication.

Since every complex rotation has an axis $[I]$ – which is, of course, a complex line, not a real one – that is a one-dimensional invariant subspace of $\mathbb{C}^3$, its orthogonal complement $[I]^\perp$ is also an invariant plane of the rotation that is, moreover, linearly isomorphic to $\mathbb{C}^2$. However, since different rotations will generally have non-collinear axes, it is impossible to find one axis such that all rotations will have the same invariant plane.

Since $\mathbb{C}^2$ is isomorphic to $\mathbb{R}^4$ as a real Lie group and $SO(3; \mathbb{C})$ is isomorphic to $SO_+(3, 1)$, finding such invariant subspaces would be essential to representing transformations of the Poincaré group $\mathbb{R}^4 \times_s SO_+(3, 1)$ as complex rigid motions. However, we now see that the complex rigid motions are not an extension of the Poincaré group.

That does not mean that complex rigid motions have no physical relevance. Since their most natural action is by matrix multiplication on vectors and covectors in $\mathbb{C}^3$, the issue is what physical relevance that space would have. However, we have already pointed out that the spaces of 2-forms and bivectors on a four-dimensional vector space, when given a complex structure, can be modeled by that complex vector space. Hence, one would expect that the physical applications of complex rigid motions would relate to the time evolution of electromagnetic fields, such as electromagnetic waves. Therefore, we shall confine our attention to that application.

## 2. The algebra of complex dual numbers.

One can define the algebra $\mathbb{C}\mathbb{D}$ of complex dual numbers quite simply by saying that it is $\mathbb{D} \otimes_\mathbb{R} \mathbb{C}$, which means that one starts with the same basic elements 1 and $\varepsilon$ as a complex basis, as in the real case, and forms all complex linear combinations $a + \varepsilon b$, where $a$ and $b$ are complex numbers, now. We will then call $a$ the (*complex*) *scalar* part of the complex dual number and $\varepsilon b$ is the *pure (complex) dual* part.

Thus, the basic vector space on which the algebra is defined is $\mathbb{C}^2$, this time. If one prefers to regard $\mathbb{C}\mathbb{D}$ as an algebra over $\mathbb{R}^4$, instead, then one regards the four elements 1, $i$, $\varepsilon$, $i\varepsilon$, which are assumed to be linearly independent, as a basis. The multiplication rules for the basis elements are the ones that follow naturally from associativity. We summarize them in a table:

| | 1 | $i$ | $\varepsilon$ | $i\varepsilon$ |
|---|---|---|---|---|
| 1 | 1 | $i$ | $\varepsilon$ | $i\varepsilon$ |
| $i$ | $i$ | $-1$ | $i\varepsilon$ | $-\varepsilon$ |
| $\varepsilon$ | $\varepsilon$ | $i\varepsilon$ | 0 | 0 |
| $i\varepsilon$ | $i\varepsilon$ | $-\varepsilon$ | 0 | 0 |

The rules of addition and multiplication are unchanged from the real case, at least formally. Thus, the ring $\mathbb{CD}$ is a commutative ring with unity, with zero divisors defined by pairs of pure complex dual numbers; thus, it is not a division algebra. The conjugation operation that takes $\alpha$ to $\bar{\alpha}$ is defined essentially as before, as is the resulting modulus-squared. Since $|\alpha| = a$ is still definite – although not *positive*-definite – the multiplicative inverse to $\alpha$ is again defined whenever $\alpha$ is not a pure complex dual number and has formally the same expression as before:

$$\alpha^{-1} = \frac{\bar{\alpha}}{|\alpha|^2}\,.$$

Therefore, the invertible complex dual numbers define a non-compact, two-dimensional, Abelian, complex Lie group $CD^*$ whose two connected components are defined by the half-planes in $\mathbb{C}^2$ that lie on either side of the "complex $y$-axis." The subgroup $CD_1$ of complex dual numbers of unit modulus is only slightly more involved, this time, since the complex dual numbers of unit *dual* modulus must not be confused with the complex numbers of unit modulus, whose modulus-squared is defined $zz^*$. Thus, a typical element of $CD_1$ still has the form:

$$\alpha = \pm 1 + \varepsilon b,$$

except that $b$ is complex, this time, and a typical element of $CD^*$ can be expressed in the form of a product of a non-zero complex number and a complex dual number of unit modulus:

$$\alpha = a\left(1 + \frac{\varepsilon b}{a}\right).$$

This brings us to the essential differences between $\mathbb{D}$ and $\mathbb{CD}$, which are mostly concerned with the fact that since a complex dual number $\alpha = (a + ib) + \varepsilon(c + id)$ can also be expressed as the sum $\alpha = \beta + i\gamma$ of two real dual numbers $\beta$ and $\gamma$ by way of:

$$\alpha = (a + \varepsilon c) + i(b + \varepsilon d),$$

one can define another automorphism of the algebra that takes $\alpha$ to its complex conjugate:

$$\alpha^* = \beta - i\gamma = (a - ib) + \varepsilon(c - id) = (a + \varepsilon c) - i(b + \varepsilon d).$$

One can then combine the two types of conjugation that we have defined on complex dual numbers to give the *adjunction* automorphism, which takes $\alpha = a + \varepsilon b$ to:

$$\alpha^\dagger = \overline{\alpha}^* = a^* - \varepsilon b^* = \overline{\beta} - i\overline{\gamma},$$

which makes:

$$\alpha\,\alpha^\dagger = (a + \varepsilon b)(a^* - \varepsilon b^*) = |\,a\,|^2 + \varepsilon(ab^* + a^*b).$$

**3. Functions of complex dual numbers.** The only differences between the formulas that we derived for functions of dual variables in chapter III and the ones for complex dual variables are based in the fact that now they reduce to sums of functions of complex variables. Hence, one must be more careful about the differentiation of such functions, since complex differentiation is more restrictive than real differentiation in that complex functions must satisfy the Cauchy-Riemann equations in order to be continuously differentiable – or *holomorphic*. As a consequence, such functions are also complex analytic, in the sense that they can be expanded into a convergent Taylor series, at least locally.

We simply repeat the formulas of the previous section on functions of real dual variables, starting with the power formula:

$$\underline{z}^n = z^n + \varepsilon\,n\,z^{n-1}s = z^n + \varepsilon\,\frac{dz^n}{dz}\,s. \tag{22.1}$$

More generally, a polynomial $P[\underline{z}]$ is a complex linear combination of powers of $\underline{z}$, and for any polynomial function $P[\underline{z}]$ of the complex dual variable $\underline{z}$, one will have:

$$P[\underline{z}] = P[z] + \varepsilon\,P'[z]\,s. \tag{22.2}$$

This generalizes to the definition:

$$f(\underline{x}) = f(z) + \varepsilon f'(z)\,s. \tag{22.3}$$

if one assumes that $f(z)$ is a holomorphic function of $z$.

In particular, the trigonometric formulas remain intact with complex dual arguments:

$$\cos(\underline{\alpha}) = \cos\alpha - \varepsilon\,s\,\sin\alpha, \tag{22.4}$$

$$\sin(\underline{\alpha}) = \sin\alpha + \varepsilon\,s\,\cos\alpha. \tag{22.5}$$

$$\cos^2(\underline{\alpha}) + \sin^2(\underline{\alpha}) = 1, \tag{22.6}$$

$$\cos(2\underline{\alpha}) = \cos^2(\underline{\alpha}) - \sin^2(\underline{\alpha}), \tag{22.7}$$

$$\sin(2\underline{\alpha}) = 2\sin(\underline{\alpha})\cos(\underline{\alpha})\,. \tag{22.8}$$

This time, $\underline{\alpha} = \alpha + \varepsilon\,s$ is the sum of complex numbers, so we resort to the relevant formulas in the section on functions of complex variables for the verification of the present formulas.

**4. Complex-dual linear algebra.** Much of what we said previously in chapter III in the context of dual linear algebra carries over to the case of complex-dual linear algebra by complexification. In particular, one is still dealing with a module over the ring $\mathbb{CD}$, rather than a vector over any field, and the model for such a $\mathbb{CD}$-module is the Cartesian product $\mathbb{CD}^n$ of $n$ copies of $\mathbb{CD}$, whose elements then look like $\underline{\mathbf{z}} = (\underline{z}^1, \ldots, \underline{z}^n)$, where the coordinates $\underline{z}^i$ are complex-dual numbers, this time.

One forms linear combinations $\underline{\alpha}\,\underline{\mathbf{z}} + \underline{\beta}\,\underline{\mathbf{w}}$ coordinate-wise, as before, and a basis for an $n$-dimensional complex-dual vector space $V$ is still a set $\{\underline{\mathbf{e}}_1, \ldots, \underline{\mathbf{e}}_n\}$ of $n$ complex-dual vectors such that any complex-dual vector $\underline{\mathbf{z}}$ in $V$ can be expressed as a linear combination:

$$\underline{\mathbf{z}} = \sum_{i=1}^{n} \underline{z}^i \underline{\mathbf{e}}_i\,. \tag{23.1}$$

Once again, the issue of linear independence is complicated by the fact that the existence of divisors of zero in $\mathbb{CD}$ makes it possible for $\underline{z}^i \underline{\mathbf{e}}_i$ to be zero without all of the $\underline{z}^i$ being individually zero. However, this does not preclude the existence of bases, it only reduces the number of acceptable sets of $n$ complex-dual vectors that could serve as bases to essentially complex bases – i.e., ones with no non-vanishing complex dual part.

Any complex-dual vector $\underline{\mathbf{z}}$ can be decomposed into complex-plus-pure-complex-dual form, real-dual-plus-imaginary-dual form, or a sum of four real vectors:

$$\underline{\mathbf{z}} = \mathbf{z} + \varepsilon\mathbf{w} = \underline{\mathbf{x}} + i\underline{\mathbf{y}} = \mathbf{x} + \varepsilon\mathbf{a} + i\mathbf{y} + i\varepsilon\mathbf{b},$$

in which $\mathbf{z}$ and $\mathbf{w}$ are complex $n$-vectors, $\underline{\mathbf{x}}$ and $\underline{\mathbf{y}}$ are real-dual, and $\mathbf{x}$, $\mathbf{a}$, $\mathbf{y}$, $\mathbf{b}$ are all real $n$-vectors.

A complex-dual linear map $\underline{L}: V \to W$ from an $n$-dimensional $\mathbb{CD}$-linear space $V$ to an $m$-dimensional one $W$ is a function that takes $\mathbb{CD}$-linear combinations to linear combinations:

$$\underline{L}(\underline{\alpha}\,\underline{\mathbf{z}} + \underline{\beta}\,\underline{\mathbf{w}}) = \underline{\alpha}\,\underline{L}(\underline{\mathbf{z}}) + \underline{\beta}\,\underline{L}(\underline{\mathbf{w}})\,.$$

If $V$ is an $n$-dimensional complex-dual vector space then a choice of basis defines a $\mathbb{CD}$-linear isomorphism of $V$ with $\mathbb{CD}^n$ that takes $\underline{\mathbf{z}}$, as in (23.1), to the $n$-tuple ($\underline{z}^1$, …, $\underline{z}^n$). If bases $\{\underline{\mathbf{e}}_1, …, \underline{\mathbf{e}}_n\}$ and $\{\underline{\mathbf{f}}_1, …, \underline{\mathbf{f}}_m\}$ are chosen for both $V$ and $W$, respectively, then any complex-dual linear map $\underline{L}$ can be associated with a complex-dual $m \times n$ matrix $\underline{L}_i^a$ by way of:

$$L(\underline{\mathbf{e}}_i) = \underline{\mathbf{f}}_a \, \underline{L}_i^a .$$

The action of $\underline{L}$ on $\underline{\mathbf{z}}$ can be associated with a corresponding action of $\underline{L}_i^a$ on $\underline{z}^i$ by matrix multiplication:

$$\underline{L}(\underline{z}^i) = \underline{L}_i^a \, \underline{z}^i .$$

Like complex-dual vectors, the complex-dual matrix $\underline{L}_i^a$ can decomposed into complex-plus-pure-complex-dual form, real-dual-plus-imaginary-dual form, or even a sum of four real matrices:

$$\underline{L}_i^a = L_i^a + \varepsilon A_i^a = \underline{\mathcal{R}}_i^a + i\underline{\mathcal{I}}_i^a = \mathcal{R}_i^a + \varepsilon \mathcal{A}_i^a + i\mathcal{I}_i^a + i\varepsilon \mathcal{B}_i^a ,$$

in which $L_i^a$ and $A_i^a$ are complex matrices, $\underline{\mathcal{R}}_i^a$ and $\underline{\mathcal{I}}_i^a$ are real-dual matrices, and all of the matrices in the last expression are real.

When expressed in complex-plus-pure-complex-dual form, the product of two complex-dual square matrices $\underline{L}_j^i = L_j^i + \varepsilon A_j^i$ and $\underline{M}_j^i = M_j^i + \varepsilon B_j^i$ takes the form:

$$\underline{L}_k^i \underline{M}_j^k = L_k^i M_j^k + \varepsilon (A_k^i M_j^k + B_k^i L_j^k) .$$

Thus, it behaves like the multiplication of complex square matrices for the complex parts, but has the characteristically more involved form for the pure-complex-dual parts.

If the $\mathbb{CD}$-linear map $\underline{L} : V \to V$ is invertible then so is the matrix $\underline{L}_j^i$, so a matrix $\underline{\tilde{L}}_j^i$ exists such that:

$$\underline{\tilde{L}}_k^i \underline{L}_j^k = \underline{L}_k^i \underline{\tilde{L}}_j^k = \delta_j^i .$$

In complex-plus-pure-complex-dual form, this implies the conditions that the complex part $\tilde{L}_j^i$ of $\underline{\tilde{L}}_j^i$ must be, in fact, the inverse of $L_j^i$, which must then exist, and the pure complex-dual part $\tilde{A}_j^i$ of $\underline{\tilde{L}}_j^i$ must satisfy:

$$\tilde{A}_j^i = -\tilde{L}_k^i A_l^k \tilde{L}_j^l .$$

As before, this places no restriction on $A_j^i$ itself.

The set of all invertible complex dual linear maps, or all invertible complex-dual $n{\times}n$ matrices, for that matter, then forms a group $GL(n; \mathbb{CD})$. As a complex Lie group, it has dimension $2n^2$, and as a real Lie group, it has dimension $4n^2$. It includes $GL(n; \mathbb{C})$ as a subgroup by way of the invertible complex $n{\times}n$ matrices and the complex translation group $\mathbb{C}^{n^2}$ by way of the matrices of the form $I + \varepsilon A$, where we now drop the matrix indices, for brevity.

We can introduce a scalar product on a $\mathbb{CD}$-linear space $V$ as we did for a $\mathbb{D}$-linear one – i.e., a symmetric $\mathbb{CD}$-bilinear functional $<.,.>$ on $V$ that is non-degenerate, in the sense that the map $V \to V^*$, $\underline{\mathbf{v}} \mapsto <\underline{\mathbf{v}},.>$ is a $\mathbb{CD}$-linear isomorphism. In particular, the scalar product now takes its values in $\mathbb{CD}$. When two complex dual vectors $\underline{\mathbf{v}}$ and $\underline{\mathbf{w}}$ are expressed in complex-plus-pure-complex-dual form as $\mathbf{v} + e\mathbf{a}$ and $\mathbf{w} + e\mathbf{b}$, respectively, the scalar product takes the form:

$$< \underline{\mathbf{v}}, \underline{\mathbf{w}} > \; = <\mathbf{v}, \mathbf{w}> + \varepsilon(<\mathbf{v}, \mathbf{b}> + <\mathbf{w}, \mathbf{a}>). \tag{23.2}$$

Thus, orthogonality of $\underline{\mathbf{v}}$ and $\underline{\mathbf{w}}$ would imply two conditions:

$$<\mathbf{v}, \mathbf{w}> = 0, \quad <\mathbf{v}, \mathbf{b}> + <\mathbf{w}, \mathbf{a}> = 0. \tag{23.3}$$

That is, the complex parts would have to be orthogonal in the usual sense, while the pure complex dual parts would have to satisfy a more elaborate condition.

A basis $\{ \underline{\mathbf{e}}_1, \ldots, \underline{\mathbf{e}}_n \}$ for $V$ is (Euclidian) orthonormal for this scalar product if one has:

$$< \underline{\mathbf{e}}_i, \underline{\mathbf{e}}_j > \; = \delta_{ij}. \tag{23.4}$$

Thus the scalar product of two complex dual vectors $\underline{\mathbf{v}} = \underline{v}^i \underline{\mathbf{e}}_i$ and $\underline{\mathbf{w}} = \underline{w}^i \underline{\mathbf{e}}_i$ will take the component form:

$$< \underline{\mathbf{v}}, \underline{\mathbf{w}} > \; = \delta_{ij} \underline{v}^i \underline{w}^j. \tag{23.5}$$

When the components are expressed in complex-plus-pure-complex-dual form as $v^i + \varepsilon a^i$ and $w^i + \varepsilon b^i$, respectively, this takes the form:

$$< \underline{\mathbf{v}}, \underline{\mathbf{w}} > \; = \delta_{ij} v^i w^j + \varepsilon \delta_{ij}(v^i b^j + w^i a^j). \tag{23.6}$$

A $\mathbb{CD}$-linear transformation $\underline{L}: V \to V$ is $\mathbb{CD}$-*orthogonal* iff it preserves the Euclidian scalar product. That is, for every $\underline{\mathbf{v}}$, $\underline{\mathbf{w}} \in V$ one must have:

$$< \underline{L}\mathbf{v}, \underline{L}\mathbf{w} > = < \underline{\mathbf{v}}, \underline{\mathbf{w}} > . \tag{23.7}$$

As usual, the condition on the matrix that represents $\underline{L}$ with respect to some basis (which we shall also represent by $\underline{L}$) is:

$$\underline{L}^{\mathrm{T}}\underline{L} = \underline{L}\underline{L}^{\mathrm{T}} = I; \tag{23.8}$$

i.e.:

$$\tilde{\underline{L}} = \underline{L}^{\mathrm{T}}. \tag{23.9}$$

If $\underline{L}$ is expressed in complex-plus-pure-complex-dual form as $L + \varepsilon A$ then the condition (23.8) takes the form:

$$L^{\mathrm{T}}L = LL^{\mathrm{T}} = I, \qquad A^{\mathrm{T}} = -L^{\mathrm{T}}AL^{\mathrm{T}}. \tag{23.10}$$

Thus, the complex part belongs to $O(n; \mathbb{C}\mathbb{D})$, while the pure dual part satisfies a more involved constraint.

In order to reduce to $SO(n; \mathbb{C}\mathbb{D})$, one must introduce a volume element $\mathcal{V}$ on $V$, which takes the form of a non-zero completely anti-symmetric CD-multilinear functional on $V$:

$$\mathcal{V}(\underline{\mathbf{v}}_1, \cdots, \underline{\mathbf{v}}_n) = \frac{1}{n!} \varepsilon_{i_1 \cdots i_n} v_1^{i_1} \cdots v_n^{i_n} = \det[\underline{\mathbf{v}}_1 | \cdots | \underline{\mathbf{v}}_n] \mathcal{V}. \tag{23.11}$$

Of course, the determinant of the component matrix will take its values in $\mathbb{C}\mathbb{D}$.

Under a $\mathbb{C}\mathbb{D}$-linear transformation $\underline{L}$ one will have:

$$\mathcal{V}(\underline{L}\mathbf{v}_1, \cdots, \underline{L}\mathbf{v}_n) = (\det \underline{L}) \mathcal{V}(\underline{\mathbf{v}}_1, \cdots, \underline{\mathbf{v}}_n). \tag{23.12}$$

Thus, for a volume-preserving $\mathbb{C}\mathbb{D}$-linear transformation, one will have that $\det \underline{L} = 1$. The subgroup of $GL(n; \mathbb{C}\mathbb{D})$ for which this is true will be denoted by $SL(n; \mathbb{C}\mathbb{D})$, and the subgroup of $O(n; \mathbb{C}\mathbb{D})$ will be denoted by $SO(n; \mathbb{C}\mathbb{D})$.

**5. The algebra of complex dual quaternions**.  The algebra $\mathbb{H}_{\mathbb{C}\mathbb{D}}$ of complex dual quaternions is defined the tensor product $\mathbb{H} \otimes \mathbb{C}\mathbb{D}$.  Thus, a typical element $\underline{q} \in \mathbb{H}_{\mathbb{C}\mathbb{D}}$ can be expressed in the form:

$$\underline{q} = \underline{q}^{\mu} \mathbf{e}_{\mu}, \tag{24.1}$$

in which the basis elements $\mathbf{e}_\mu$ are the same as in the real case, but the components $\underline{q}^\mu$ are now complex dual numbers.  One can also regard $\mathbb{H}_{\mathbb{C}\mathbb{D}}$ as $\mathbb{H}_{\mathbb{C}} \otimes \mathbb{D}$ or $\mathbb{H}_{\mathbb{D}} \otimes \mathbb{C}$.  That is, the elements of $\mathbb{H}_{\mathbb{C}\mathbb{D}}$ can be regarded as complex quaternions with dual coefficients or dual quaternions with complex coefficients.  We shall often find that the latter representation is most useful, since one then simply complexifies the corresponding dual expressions.

Hence, if one regards the elements of $\mathbb{H}_{\mathbb{C}\mathbb{D}}$ in the form (24.1) then the only thing that changes in the expression for the product of two complex dual quaternions:

$$\underline{q}\,\underline{q}' = \underline{q}^\mu \underline{q}^\nu \mathbf{e}_\mu \mathbf{e}_\nu \qquad (24.2)$$

is in the way that products of components form, while the products of basis elements are still the same as for real quaternions.

There are various decompositions of complex dual quaternions that become useful, since the number of dimensions in the coefficient algebra has doubled from either $\mathbb{C}$ or $\mathbb{D}$. We simply summarize some of them:

$$\text{real + imaginary:} \qquad \underline{q} = \underline{p} + i\underline{r}\,, \qquad \underline{p},\underline{r} \in \mathbb{H}_{\mathbb{D}}\,,$$

$$\text{scalar + vector:} \qquad \underline{q} = \underline{q}^0 + \mathbf{q}\,, \qquad \underline{q}^0 \in DS\mathbb{H}_{\mathbb{C}\mathbb{D}}\,, \ \mathbf{q} \in DV\mathbb{H}_{\mathbb{C}\mathbb{D}}\,,$$

$$\text{complex + dual:} \qquad \underline{q} = p + \varepsilon r\,, \qquad p, r \in \mathbb{H}_{\mathbb{C}}\,.$$

These direct sum decompositions of the vector space $\mathbb{H}_{\mathbb{C}\mathbb{D}}$ :

$$\mathbb{H}_{\mathbb{C}\mathbb{D}} = \text{Re}(\mathbb{H}_{\mathbb{C}\mathbb{D}}) \oplus \text{Im}(\mathbb{H}_{\mathbb{C}\mathbb{D}}) = DS\mathbb{H}_{\mathbb{C}\mathbb{D}} \oplus DV\mathbb{H}_{\mathbb{C}\mathbb{D}} = C(\mathbb{H}_{\mathbb{C}\mathbb{D}}) \oplus D(\mathbb{H}_{\mathbb{C}\mathbb{D}})$$

define projection operators Re, Im, $DS$, $DV$, $C$, $D$ onto the summands, which can be defined by polarizing the automorphisms *, which is still complex conjugation, dual conjugation:

$$\underline{q}^\circ = \underline{q}^0 - \mathbf{q}\,, \qquad (24.3)$$

and quaternion conjugation:

$$\overline{\underline{q}} = \overline{p} + \varepsilon \overline{r}\,. \qquad (24.4)$$

One then gets:

$$\text{Re}(\underline{q}) = \tfrac{1}{2}(\underline{q} + \underline{q}^*)\,, \qquad \text{Im}(\underline{q}) = \tfrac{1}{2}(\underline{q} - \underline{q}^*)\,, \qquad (24.5)$$

$$DS(\underline{q}) = \tfrac{1}{2}(\underline{q} + \underline{q}^\circ)\,, \qquad DV(\underline{q}) = \tfrac{1}{2}(\underline{q} - \underline{q}^\circ)\,, \qquad (24.6)$$

$$C(\underline{q}) = \tfrac{1}{2}(\underline{q} + \overline{\underline{q}})\,, \qquad D(\underline{q}) = \tfrac{1}{2}(\underline{q} - \overline{\underline{q}})\,. \qquad (24.7)$$

In the three forms that we introduced, the product (24.2) becomes:

$$\underline{q}\,\underline{q}' = \underline{p}\,\underline{p}' - \underline{r}\,\underline{r}' + i(\underline{p}\,\underline{r}' + \underline{r}\,\underline{p}') \tag{24.8}$$

$$\underline{q}\,\underline{q}' = (\underline{q},\underline{q}') + \underline{q}^0\mathbf{q}' + \underline{q}'^0\mathbf{q} + \mathbf{q}\times\underline{\mathbf{q}}', \tag{24.9}$$

$$\underline{q}\,\underline{q}' = pp' + \varepsilon(pr' + rp'). \tag{24.10}$$

into which we have introduced the scalar product:

$$(\underline{q},\underline{q}') = DS(\underline{q}\,\underline{q}') = \underline{q}^0\underline{q}'^0 - \sum_{i=1}^{3}\delta_{ij}\underline{q}^i\underline{q}'^j = (p,p') + \varepsilon[(p,r') + (p',r)], \tag{24.11}$$

and if $\underline{\mathbf{q}} = \mathbf{q} + \varepsilon\mathbf{r}$ and $\underline{\mathbf{q}}' = \mathbf{q}' + \varepsilon\mathbf{r}'$ then one defines the cross product by bilinearity:

$$\underline{\mathbf{q}}\times\underline{\mathbf{q}}' = \tfrac{1}{2}[\underline{\mathbf{q}},\underline{\mathbf{q}}'] = \mathbf{q}\times\mathbf{q}' + \varepsilon(\mathbf{q}\times\mathbf{r}' + \mathbf{r}\times\mathbf{q}'). \tag{24.12}$$

Of course, both of these expressions (24.11) and (24.12) are merely the complexification of their real counterparts.

One can also introduce another isomorphism that corresponds to the adjoint operation for complex quaternions:

$$\underline{q}^{\dagger} = p^{\dagger} + \varepsilon r^{\dagger}. \tag{24.13}$$

By polarization, this defines a decomposition of $\mathbb{H}_{\mathbb{C}\mathbb{D}}$ into a direct sum $H^+ \oplus H^- \oplus \varepsilon H^+ \oplus \varepsilon H^-$ of four-real-dimensional subspaces, two of which are composed of pure dual real quaternions. One can also rearrange this direct product into the form $(H^+ \oplus \varepsilon H^+) \oplus (H^- \oplus \varepsilon H^-)$, which makes $\mathbb{H}_{\mathbb{C}\mathbb{D}}$ the sum of two dour-dimensional dual vectors spaces.

As we see, $\mathbb{H}_{\mathbb{C}\mathbb{D}}$ is still an associative, but not commutative, ring and has a unity element 1 that is still defined by $\mathbf{e}_0$, with a center that is defined by all of the complex dual scalars, which have the form $(q^0 + \varepsilon r^0)\,\mathbf{e}_0$.

Like the algebra of real dual quaternions, the algebra $\mathbb{H}_{\mathbb{C}\mathbb{D}}$ has divisors of zero, and the obvious examples are still the pure complex dual quaternions. Similarly, the only difference between the expression for the inverse of an invertible element $\underline{q} = p + \varepsilon r$:

$$\underline{q}^{-1} = p^{-1} - \varepsilon\,p^{-1}r\,p^{-1} \tag{24.14}$$

and the previous one in the real case is that everything is complex now. In particular, $p$ must not be a null complex quaternion.

The other scalar product that we have been habitually defining is:

$$<\underline{q},\underline{q}'> = DS(\underline{q}\,\overline{\underline{q}}') = \sum_{\mu=1}^{3}\delta_{\mu\nu}\underline{q}^{\mu}\underline{q}'^{\nu}, \tag{24.15}$$

which now takes its values in $\mathbb{CD}$.

If one puts the quaternions involved into quaternion-plus-dual form then the scalar product looks like it did before, but with complex quaternions:

$$< \underline{q}, \underline{q}' > \, = <p, p'> + \varepsilon(<r, p'> + <p, r'>). \tag{24.16}$$

In particular:

$$\| \underline{q} \|^2 = \| p \|^2 + 2\varepsilon <p, r>. \tag{24.17}$$

From (24.17), a null complex dual quaternion must satisfy:

$$\| p \| = 0, \qquad <p, r> = 0. \tag{24.18}$$

Hence, unlike the real dual case, since there non-trivial complex quaternions, there are also non-trivial complex dual quaternions, as well.  However, analogous to the real case, we are now dealing with the complex Study quadric, as the second condition shows.

Similarly, a unit quaternion must satisfy:

$$\| p \| = 1, \qquad <p, r> = 0. \tag{24.19}$$

Therefore, it defines a Cartesian product $(p, r) \in \mathbb{H}_{\mathbb{C}} \times \mathbb{H}_{\mathbb{C}}$ of a unit complex quaternion $p$ and complex quaternion $r$ that is complex-orthogonal to $p$.

If $\underline{q}$ is non-null then $\| \underline{q} \|$ can be factored out and any invertible complex dual quaternion can be expressed as a product:

$$\underline{q} = \| \underline{q} \| \, \underline{u}, \tag{24.20}$$

with the obvious definition for the complex dual unit quaternion $\underline{u}$.

Furthermore, we have a canonical form for $\underline{u}$ that is the complex analogue of the real expression:

$$\underline{u} = \cos \tfrac{1}{2} \underline{\alpha} + \sin \tfrac{1}{2} \underline{\alpha} \mathbf{u}, \tag{24.21}$$

in which the angle $\underline{\alpha}$ is a complex dual number, while $\mathbf{u}$ is a complex dual unit vector.

Since:

$$\| \underline{q} \, \underline{q}' \|^2 = DS(\underline{q} \, \underline{q}' \overline{\underline{q} \, \underline{q}'}) = DS(\underline{q} \, \underline{q}' \, \overline{\underline{q}}' \, \overline{\underline{q}}) = \| \underline{q} \|^2 \| \underline{q}' \|^2, \tag{24.22}$$

i.e.:

$$\| \underline{q} \, \underline{q}' \| = \| \underline{q} \| \, \| \underline{q}' \|, \tag{24.23}$$

one sees that the product of two complex dual null quaternions is a complex dual null quaternion and the product of two complex dual unit quaternions is again a unit quaternion.  Therefore, although null quaternions do not have multiplicative inverses, and therefore do not define a group, nonetheless, the unit quaternions do form a group, since one also has that the inverse of a unit quaternion is a unit quaternion.  This follows from (24.23) when the left-hand side is unity.

Thus, the group $\underline{CQ}^*$ of invertible complex dual quaternions and the group $\underline{Q}^*$ that we introduced previously differs only by the complexification everything. In particular, it contains the subgroup $CQ^*$ of invertible complex quaternions ($r = 0$), which is the set complement of the null hypersurface, and the subgroup of elements of the form $1 + \varepsilon r$, which is then isomorphic to the translation group of $\mathbb{C}^4$. The two groups intersect only at 1 and are both four-complex-dimensional, so the group $\underline{CQ}^*$ has complex dimension 8. From (24.20), $\underline{CQ}^*$ can be expressed as the product group $\mathbb{CD}^* \times \underline{CQ}_1$ of the invertible complex dual numbers with the group of complex dual unit quaternions.

As we shall see, the group $\underline{CQ}_1$ is isomorphic to the semi-direct product $\mathbb{C}^3 \times_s SO(3; \mathbb{C})$, which is the group of complex rigid motions. Hence, it has complex dimension six or real dimension twelve.

The Lie algebra $\mathfrak{iso}(3; \mathbb{C}) = \mathbb{C}^3 \oplus_s \mathfrak{so}(3; \mathbb{C})$ of infinitesimal complex rigid motions is isomorphic to the Lie algebra $\underline{\mathbf{q}}_0^{\mathbb{C}}$ of complex dual vectors, which amounts to the complexification of the corresponding statement for infinitesimal real rigid motions and real dual vectors. One simply repeats the argument that was given above under the assumption that all of the component vectors involved are complex, now.

For complex dual vectors, one has:

$$\underline{\mathbf{q}}\,\underline{\mathbf{q}}' = - <\underline{q}, \underline{q}'> + \underline{\mathbf{q}} \times \underline{\mathbf{q}}', \tag{24.24}$$

$$<\underline{\mathbf{q}}, \underline{\mathbf{q}}'> = <\mathbf{p}, \mathbf{p}'> + \varepsilon(<\mathbf{p}, \mathbf{r}'> + <\mathbf{r}, \mathbf{p}'>), \tag{24.25}$$

$$\|\underline{\mathbf{q}}\|^2 = \|\mathbf{p}\|^2 + 2\varepsilon<\mathbf{p}, \mathbf{r}>, \tag{24.26}$$

which is, of course, merely the complexification of the real situation.

In particular, a null complex dual vector satisfies:

$$\|\mathbf{p}\| = 0, \qquad <\mathbf{p}, \mathbf{r}> = 0, \tag{24.27}$$

which is now non-trivially possible, and a unit complex dual vector satisfies:

$$\|\mathbf{p}\| = 1, \qquad <\mathbf{p}, \mathbf{r}> = 0. \tag{24.28}$$

Thus the set of all unit complex dual vectors becomes the complexification of the real set, or $T(\mathbb{C}S^2)$. It is not, however, a group, which was also true for the real case.

Just as $DV(\mathbb{H}_\mathbb{D})$ represented the real affine space $A^3$, similarly, $DV(\mathbb{H}_{\mathbb{CD}})$ represents the tangent bundle $T(A_\mathbb{C}^3)$ to the complex affine space $A_\mathbb{C}^3$. The complex vector part $\mathbf{v}$ of $\mathbf{v} + \varepsilon \mathbf{a}$ represents the vector part $\mathbf{v}$ of the point $(x, \mathbf{v}) \in T(A_\mathbb{C}^3)$, while the pure complex dual part $\mathbf{a}$ indirectly represents the translation that takes some reference point $O$ to $x$.

In order to find the nilpotents of degree two in $\mathbb{H}_{\mathbb{CD}}$, we start with the expression for $\underline{q}^2$, namely:

$$\underline{q}^2 = p^2 + \varepsilon(pr + rp). \qquad (24.29)$$

Once again, this equation is unchanged from the real dual case, except that now everything is complex, and similarly, for the necessary conditions for $\underline{q}^2$ to vanish, one has:

$$p^2 = 0, \qquad pr + rp = 0.$$

The possibility that $p = 0$ still leads to the fact that the complex dual quaternions of pure complex dual type are all nilpotents, as they were in the real case. However, $p$ is now a complex quaternion, and we saw that $\mathbb{H}_{\mathbb{C}}$ also admits non-trivial nilpotents of degree two. Thus, in addition to the pure dual case, one can consider nilpotents for which $p$ is non-trivial. Recall that for $\mathbb{H}_{\mathbb{C}}$ they took the form $p = \mathbf{p}$ with $<\mathbf{p}, \mathbf{p}> = 0$; in particular, $p^0 = 0$.

Now put everything into scalar-plus-vector form and look at the corresponding form of the latter conditions. After some straightforward calculations, one gets that:

$$pr + rp = 2(r^0 \mathbf{p} - <\mathbf{p}, \mathbf{r}>).$$

The vanishing of the left-hand side would imply that:

$$r^0 \mathbf{p} = 0, \qquad <\mathbf{p}, \mathbf{r}> = 0.$$

If $\mathbf{p} = 0$ then $\underline{q} = \varepsilon r$ is a pure complex dual quaternion.

The case in which $r^0$ vanishes makes $r = \mathbf{r}$ a pure complex quaternion, as well as $p$, although it must be restricted by the last condition above. Thus, such a nilpotent takes the form:

$$\underline{q} = \mathbf{p} + \varepsilon \mathbf{r}, \qquad <\mathbf{p}, \mathbf{r}> = 0. \qquad (24.30)$$

As for the idempotents, if one sets $\underline{q}^2 = \underline{q}$ and expands this into complex-plus-dual form then this gives the following necessary conditions:

$$p^2 = p, \qquad pr + rp = r.$$

Thus, $p$ must be an idempotent in $\mathbb{H}_{\mathbb{C}}$, which we know exist non-trivially. It is clear that setting $p$ to either 0 or 1 would make $r = 0$, which are both trivial cases. Thus, we set $p$ equal to its canonical form:

$$p = \tfrac{1}{2}(1 + i\mathbf{u}), \qquad \| \mathbf{u} \| = 1. \qquad (24.31)$$

If we put $r$ into its scalar-plus-vector form $r^0 + \mathbf{r}$ then the second condition above takes the form:

$$r^0 - i\langle \mathbf{u}, \mathbf{r} \rangle = r^0, \qquad \mathbf{r} + ir^0\,\mathbf{u} = \mathbf{r},$$

which imply:

$$\langle \mathbf{u}, \mathbf{r} \rangle = 0, \qquad r^0\,\mathbf{u} = 0.$$

Hence, either $\mathbf{u} = 0$, which does not lie on the unit sphere and is therefore unacceptable, or $r^0 = 0$. We can then put a non-trivial idempotent $\underline{\iota}$ in $\mathbb{H}_{\mathbb{C}\mathbb{D}}$ into the form:

$$\underline{\iota} = \tfrac{1}{2}(1 + i\mathbf{u}) + \varepsilon\mathbf{r}, \qquad \langle \mathbf{u}, \mathbf{u} \rangle = 1, \qquad \langle \mathbf{u}, \mathbf{r} \rangle = 0. \tag{24.32}$$

Since $\mathbb{H}_{\mathbb{C}\mathbb{D}}$ admits non-trivial idempotents $\underline{\iota}$, it will also admit the non-trivial left and right ideals $\mathcal{I}_l(\underline{\iota})$ and $\mathcal{I}_r(\underline{\iota})$ that are generated by them. As in the complex case, the conjugate $\overline{\underline{\iota}}$ of an idempotent is also idempotent and since it is therefore null, one still has $\underline{\iota}\,\overline{\underline{\iota}}$, which means that $\overline{\underline{\iota}}$ is orthogonal to $\underline{\iota}$ and one can decompose $\mathbb{H}_{\mathbb{C}\mathbb{D}}$ into direct sums $\mathcal{I}_l(\underline{\iota}) \oplus \mathcal{I}_l(\overline{\underline{\iota}})$ and $\mathcal{I}_r(\underline{\iota}) \oplus \mathcal{I}_r(\overline{\underline{\iota}})$, with a corresponding decomposition of the identity operator:

$$I = \underline{\iota} + \overline{\underline{\iota}}. \tag{24.33}$$

The vector subspaces $\mathcal{I}_l(\underline{\iota})$, $\mathcal{I}_l(\overline{\underline{\iota}})$, $\mathcal{I}_r(\underline{\iota})$, $\mathcal{I}_r(\overline{\underline{\iota}})$, which are also subalgebras, then have complex dual dimension two and are invariant subspaces under left and right multiplication, respectively.

**6. The action of the group of complex rigid motions on complex dual quaternions.** The various actions that we introduced in the context of real dual quaternions can all be complexified, which also introduces the possibilities that we introduced for complex quaternions, due to the increased number of automorphisms.

As in the real case, the action:

$$\underline{CQ}_1 \times \mathbb{H}_{\mathbb{C}\mathbb{D}} \to \mathbb{H}_{\mathbb{C}\mathbb{D}}, \qquad (\underline{u}, \underline{q}) \mapsto \underline{u}\,\underline{q}\,\overline{\underline{u}} \tag{25.1}$$

has $DS(\mathbb{H}_{\mathbb{C}\mathbb{D}})$ and $DV(\mathbb{H}_{\mathbb{C}\mathbb{D}})$ for invariant subspaces, so that much has merely been complexified. Similarly, the argument that makes the action isometric for the scalar product $\langle .,. \rangle$ works the same, except that the dual numbers become complex dual ones.

One also sees that both $\underline{u}$ and $-\underline{u}$ produce the same effect on $\underline{q}$, as usual. Thus, when one expresses $\underline{q}$ in component form as $\underline{q}^\mu \mathbf{e}_\mu$, with complex dual coefficients, one can associate the above action of $\underline{u}$ with a 4×4 matrix $\underline{L}^\mu_\nu$ with complex dual coefficients by way of:

$$(\underline{u}\,\underline{q}\,\overline{\underline{u}})^\mu = \underline{L}^\mu_\nu\,\underline{q}^\nu, \tag{25.2}$$

which can also be defined by the corresponding transformation of the frame $\mathbf{e}_\mu$ :

$$\underline{u}\,\mathbf{e}_\mu\,\overline{\underline{u}} \;=\; \underline{L}^\nu_\mu\mathbf{e}_\nu \;. \tag{25.3}$$

Since $DS(\mathbb{H}_{\mathbb{C}\mathbb{D}})$ and $DV(\mathbb{H}_{\mathbb{C}\mathbb{D}})$ are invariant subspaces, the matrix $\underline{L}^\mu_\nu$, in a frame that is adapted to the decomposition, splits into a direct sum of a $1\times1$ – complex dual scalar – matrix and a $3\times3$ complex dual submatrix $\underline{L}^i_j$. We now need to show that $\underline{L}^i_j$ represents a complex rigid motion. However, this follows directly from the fact that $\underline{u}\,q\,\overline{\underline{u}}$ is an isometry of the complex dual scalar product on $\mathbb{H}_{\mathbb{C}\mathbb{D}}$, and thus, of its restriction to $DV(\mathbb{H}_{\mathbb{C}\mathbb{D}})$. Of course, the map $\underline{CQ_1} \to ISO(3;\mathbb{C})$ that takes $\{\underline{u}, -\underline{u}\}$ to $\underline{L}^i_j$ is the two-to-one covering homomorphism $\mathbb{C}^3 \times_s SL(2;\mathbb{C}) \to \mathbb{C}^3 \times_s SO(3;\mathbb{C})$.

**7. The role of complex rigid frames in physics.** Rather than examine the kinematics of complex rigid frames in terms of complex dual quaternions, since no discussion of such matters seems to exist in the physics mainstream, we shall first attempt to motivate the physical significance of such a study. The first issue to address is the physical applicability of the complex affine space $A^3_\mathbb{C}$.

Since the best-established application of the complex vector space $\mathbb{C}^3$ is to electromagnetism, it seems reasonable to extend that discussion to the affine case. Thus, one needs to justify that the concept of the translation of a 2-form or bivector by another 2-form or bivector actually occurs naturally in the theory of electromagnetism. In fact, this is the case.

This is due to the fact that since the Maxwell equation for the electromagnetic field strength 2-form $F$ – namely, $dF = 0$ – is linear and homogeneous in $F$, any general solution to that equation will be determined only up to an additive constant field $F_0$. Of course, imposing boundary or initial-value conditions will eliminate that indeterminacy, but it does exist for the general solutions; i.e., it is a symmetry of the system of equations.

By contrast, the dual equation [8] $\delta\mathfrak{H} = \mathbf{J}$ for the electromagnetic excitation bivector field $\mathfrak{H}$ has that symmetry only in the absence of sources, such as for points outside the support of the source current $\mathbf{J}$.

However, that would suggest that the application of complex dual quaternions to physical models is most natural in the context of the symmetries of the field equations for electromagnetism, while the main focus of the present study was to the role of quaternions in kinematics. Therefore, we shall defer a more detailed study of the physical role of complex dual quaternions to a later paper that will deal with the application of quaternions to physical field theories.

---

[8] Here, we are defining the divergence operator $d$: $\Lambda_k M \to \Lambda_{k-1} M$ to be the adjoint $\#^{-1} \cdot d \cdot \#$ of the exterior derivative operator $d$ with respect to the Poincaré isomorphism #: $\Lambda_k M \to \Lambda^{n-k} M$ that is defined by a choice of volume element on $T(M)$. This definition does, in fact, agree with the usual divergence operator for vector fields.